\documentclass[fleqn,usenatbib,useAMS]{mnras}
\usepackage[utf8]{inputenc}

\usepackage{newtxtext,newtxmath}

\usepackage[T1]{fontenc}
\usepackage{float}
\DeclareRobustCommand{\VAN}[3]{#2}
\let\VANthebibliography\thebibliography
\def\thebibliography{\DeclareRobustCommand{\VAN}[3]{##3}\VANthebibliography}
\usepackage{subfig}

\newcommand{\Eq}[1]{Eq.~(\ref{eq:#1})}

\newcommand{\Eqb}[1]{(Eq.~\ref{eq:#1})}
\newcommand{\Eqnb}[1]{Eq.~\ref{eq:#1}}
\newcommand{\se}[1]{Section \ref{sec:#1}}
\newcommand{\app}[1]{Appendix \ref{app:#1}}
\newcommand{\Fig}[1]{Fig.~\ref{fig:#1}}

\newcommand{\Tab}[1]{Table~\ref{tab:#1}}
\newcommand{\be}{\begin{equation}}
\newcommand{\ee}{\end{equation}}
\newcommand{\bad}{\begin{equation} \begin{aligned}}
\newcommand{\ead}{\end{aligned} \end{equation}}

\newcommand{\rhodm}{\rho_{\rm DM}}
\newcommand{\Mdm}{M_{\rm DM}}
\newcommand{\MBH}{M_{\rm BH}}
\newcommand{\Mvir}{M_{\rm 200}}
\newcommand{\Md}{M_{\rm d}}
\newcommand{\Macc}{\dot{M}}

\newcommand{\Lbol}{L_{\rm bol}}
\newcommand{\Medd}{\dot{M}_{\rm Edd}}
\newcommand{\Rvir}{R_{\rm 200}}

\newcommand{\dd}{\text{d}}

\usepackage{graphicx}	
\usepackage{amsmath}	

\title[AGN winds: variable coupling efficiency]{Simulating AGN wind feedback with variable feedback efficiencies in idealised disc galaxies}

\author[Jinning Liang et al.]{
Jinning Liang,$^{1,2,3}$\thanks{Email: \href{mailto:jnliang25@stu.pku.edu.cn}{jnliang25@stu.pku.edu.cn}}
Cedric G. Lacey,$^{1}$\thanks{Email: \href{mailto:cedric.lacey@durham.ac.uk}{cedric.lacey@durham.ac.uk} (Corresponding author)}
Filip Hu\v{s}ko,$^{4,1}$
Evgenii Chaikin,$^{1,4}$
Sownak Bose$^{1}$
\\
$^{1}$Institute for Computational Cosmology, Department of Physics, University of Durham, South Road, Durham, DH1 3LE, UK\\
$^{2}$Department of Astronomy, School of Physics, Peking University, Beijing 100871, China\\
$^{3}$Kavli Institute for Astronomy and Astrophysics, Peking University, Beijing 100871, China\\
$^{4}$Leiden Observatory, Leiden University, PO Box 9513, 2300 RA Leiden, the Netherlands\\
}

\date{Accepted 2026 November 17. Received 2026 October 08; in original form 2026 March 12}

\pubyear{\the\year{}}

\begin{document}
\label{firstpage}
\pagerange{\pageref{firstpage}--\pageref{lastpage}}
\maketitle

\begin{abstract}
Active Galactic Nucleus (AGN) feedback plays a critical role in galaxy formation and evolution. AGN-driven winds can significantly influence their host galaxies, although the details of their impact remain unclear. In this study, we investigate the feedback effects of AGN winds on idealized disc galaxies using the SWIFT hydrodynamical code with COLIBRE subgrid physics. We implement a new thermal AGN feedback model in which the energy injection coupling efficiency has a power-law dependence on the Eddington ratio of the black hole (BH) accretion rate, motivated by scaling relations for AGN winds from numerical models and observations. We simulate idealised Milky Way-mass galaxies, incorporating a BH, cold gas disc, stellar disc, and hot circumgalactic medium, within a static dark matter halo. We vary the BH mass and the slope and normalisation of the new coupling efficiency model. For a fixed BH mass, we find that while systematic trends with coupling efficiency exist, most galaxy and BH properties show only modest variations. This likely reflects BH self-regulation in the COLIBRE model, which modulates the effects of changes in the feedback efficiency, provided the BH mass is sufficiently high. Key exceptions are the BH accretion rate and mass growth history, and outflow behaviour, where lower coupling efficiencies lead to faster BH growth and weaker outflows, potentially helping to explain the presence of overmassive BHs at high redshifts. Varying the BH mass, however, has a much larger impact, confirming that BH mass remains the primary factor shaping galaxy and BH evolution in our simulations.
\end{abstract}




\section{Introduction}\label{sec:intro}
Active galactic nucleus (AGN) feedback from supermassive black holes (SMBHs) residing at the centres of host galaxies is widely thought to have large effects on galaxy formation and evolution, especially in regulating the growth of massive galaxies, by releasing significant amounts of energy into the surrounding gas \citep{Fabian12}. 
Including AGN feedback in galaxy formation models helps in reproducing the quenching of massive galaxies by suppressing their star formation \citep[e.g.][]{Bower06,Harrison17}. However, while current models accounting for AGN feedback can reproduce key observational relations, such as the $z \approx 0$ galaxy stellar mass function and the black hole mass ($M_{\rm BH}$) - stellar mass ($M_{\rm *}$) relation, their success is based on simplified, phenomenological prescriptions for AGN feedback, with adjustable parameters that are tuned to reproduce these observations \citep[e.g.][]{Schaye15,Pillepich18}. Discrepancies may be found when comparing to some observed properties that are not used in the calibration of the model parameters (e.g. the number density of massive quenched galaxies at high redshift and gas fractions in massive haloes) but are potentially related to AGN feedback \citep[e.g.][]{Crain23,Baker25}. This suggests that the treatment of AGN feedback remains incomplete, potentially due to missing feedback modes and/or gaps in our understanding of the underlying physical mechanisms \citep[e.g.][]{Best12}. This motivates the investigation of more physically motivated models of AGN feedback in numerical simulations of galaxy formation and evolution.

Black holes (BH) grow either by accreting surrounding material (mainly gas) or by mergers with other BHs. Due to non-zero angular momentum of the accreted material, accretion of gas usually proceeds via an accretion disc \citep[e.g.][]{Volonteri10}. The nature of this disc \citep[e.g.][]{SS73,Narayan94,Abramowicz88}, however, depends on the accretion rate normalised to the Eddington accretion rate ($\dot{m}$). 
AGN feedback occurs once a BH accretion disc has formed. This feedback mainly occurs in two main types: \emph{wind feedback} and \emph{jet feedback}, although radiation pressure on dust in the interstellar medium (ISM) may constitute a distinct channel. 
Winds are launched from the accretion disc by a combination of radiation, thermal, and magnetic driving, and can be inferred observationally from ultra-fast outflows (UFOs) in luminous AGN \citep{King15}. They are relatively isotropic and slower than jets \citep{Tombesi10}. Jets, by contrast, are collimated and relativistic, depositing energy out to circumgalactic and intergalactic medium (CGM/IGM) scales \citep{McNamara05,McNamara07}.

Given the uncertainties in how AGN feedback operates, and the importance of AGN feedback to galaxy formation and evolution, many investigations have been conducted over the past two decades \citep[e.g.][]{Fabian12,Best12,King15,Harrison17}. 
Cosmological hydrodynamical simulations generally implement AGN feedback through phenomenological subgrid models, as the AGN accretion disc is unresolved. These models mostly inject thermal or kinetic energy into the surrounding gas. Pure thermal feedback, used in simulations such as EAGLE, ASTRID, and Magneticum \citep{Schaye15,Bird22,Dolag25}, could represent either radiative heating or the thermalization of accretion-disc winds through shocks. Other simulations, including Illustris, Horizon-AGN, IllustrisTNG, SIMBA, and NewHorizon \citep{Vogelsberger14,Dubois14,Nelson18,Pillepich18,Dave19,Dubois21}, incorporate kinetic feedback, in the form of jets or winds, typically in combination with thermal feedback, depending on the accretion rate. Recently, COLIBRE \citep{Schaye25, Husko25} introduced a model that includes accretion discs, with radiative, wind, and jet feedback launched simultaneously with different efficiencies depending on the accretion state. However, the implementation of thermal feedback, particularly from the thin-disc, with constant coupling efficiency, remains phenomenological and motivates our study.

Beyond cosmological simulations, other approaches provide different advantages but at the cost of losing some galaxy-scale context. (Semi-)analytical models of galaxy formation \citep[e.g.][]{Lacey16,Henriques20,Mo24,Chen25} and of AGN accretion disc systems \citep[e.g.][]{Dittmann24,Risaliti10,Quera-Bofarull20,Quera-Bofarull23} are cheap, but capture only limited aspects of BH growth and AGN feedback owing to their simplified physics. Simulations of BHs and their accretion discs resolve additional small-scale physics, including general relativity, radiative transfer, and magnetohydrodynamics \citep{Narayan12,Higginbottom14,Nomura16,Yuan18,Jiang19,Higginbottom24}, but omit the host galaxy, probe only short ($\sim$Myr) timescales, and remain computationally expensive.

In this work, we focus on AGN wind feedback. 
Observations reveal multiple facets of AGN winds, including molecular winds, ionised winds, broad absorption line (BAL) winds, and X-ray absorption line winds (such as ultra-fast outflows, or UFOs), 
with the X-ray and BAL winds observed at small radii likely to be driving the ionised and molecular winds seen further out \citep{Laha,Fiore17}. The most powerful of these are the highly ionised UFOs, which reach velocities $v\sim 0.1-0.3c$ with kinetic luminosities $L_{\rm kin}\sim 0.1\%-10\%\,L_{\rm Edd}$ (or $\sim 1\%-50\%\,\Lbol$), large enough to remove significant amounts of gas from the host galaxy and regulate star formation. They are detected in a substantial fraction of local radio-quiet AGN \citep[e.g.][]{Tombesi10} and in some higher redshift objects \citep[e.g.][]{Chartas09,Lanzuisi12} through blueshifted Fe K-shell absorption \citep{Tombesi15,Gofford15}, indicating an origin close to the accretion disc. A promising driving mechanism is radiation pressure on UV absorption lines \citep{Proga00}, analogous to the winds of O-star photospheres \citep{CAK75}. Such winds are termed UV line-driven winds, and they provide the motivation for the present work.

There have been attempts to model the launching of UV line-driven winds from BH accretion disks using both (semi-)analytical models and hydrodynamical simulations \citep{Nomura16,Risaliti10,Higginbottom14,Quera-Bofarull20,Quera-Bofarull23,Higginbottom24}. One such semi-analytical model is Qwind \citep{Risaliti10,Quera-Bofarull20,Quera-Bofarull23}, which computes the trajectories of gas parcels launched from a standard thin accretion disc \citep{SS73} and illuminated by UV and X-ray flux, accounting for gravity and radiation pressure via the formalism from \cite{CAK75} but neglecting gas pressure and hydrodynamical forces, except during the launching phase. It is therefore far cheaper than full radiation hydrodynamic simulations, allowing a rapid exploration of wind properties over a wide parameter space. From Qwind, \cite{Quera-Bofarull23} find that the mass outflow rate, kinetic luminosity, momentum outflow rate and average outflow velocity at fixed BH mass all depend on the Eddington ratio through roughly power-law relations, consistent with numerical simulations \citep[e.g.][]{Nomura17} . The feedback coupling efficiency, i.e. the fraction of the total radiative luminosity $\Lbol$ transferred to the kinetic luminosity of the AGN wind, is likewise roughly a power law in the Eddington ratio, with only a weak dependence on BH mass over the range $\sim 10^7 - 10^9~{\rm M}_\odot$. These predictions have not yet been incorporated into a hydrodynamical galaxy formation code. Our aim is therefore to implement them as a new subgrid model for AGN feedback in idealised galaxy evolution simulations, in contrast to previous subgrid prescriptions that mostly assume constant coupling efficiencies for AGN feedback, and to investigate the effects on the evolution of galaxies and their SMBHs.

In this paper, we use idealised simulations of isolated MW-mass disk galaxies to study the effects of varying the coupling efficiency of AGN feedback on BH growth and on the properties of the host galaxy. We implement a new model in the thermal AGN feedback module of \texttt{SWIFT} \citep{Schaller24}, in which the coupling efficiency varies as a power law of the Eddington ratio. We run simulations with the subgrid physics of the COLIBRE galaxy formation model \citep{Schaye25}. The \texttt{SWIFT} code and COLIBRE model will be summarised in \se{simulation}, where we also describe the implementation of our new variable coupling efficiency model and the initial conditions (ICs) for all galaxy components in the simulations. \se{result} presents our results and analysis. \se{discussion} compares our results with previous work, as well as discussing the role of the CGM and prospects for future improvements. Finally, \se{conclusion} presents our conclusions. While cosmological parameters are not needed in our idealised runs, we assume the Hubble constant to be $H_0=$ 70.4 km s$^{-1}$ Mpc$^{-1}$ when defining virial mass and virial radius.

\section{Simulation code and initial conditions}\label{sec:simulation}

In this section, we briefly introduce the simulation code and the subgrid physics used to run our simulations. Then, as motivated by the model predictions for UV line-driven winds described in the Introduction \citep{Quera-Bofarull23}, we explain how our new AGN coupling efficiency model is implemented. Finally, we describe the initial setup for our idealised galaxy runs.

\subsection{The \textsc{SWIFT} code and COLIBRE subgrid model}
The simulations in this work use the simulation code {\tt SWIFT}\footnote{\href{www.swiftsim.com}{www.swiftsim.com}} \citep{Schaller24}. {\tt SWIFT} integrates cosmology, gravity, hydrodynamics, and subgrid models for physical processes needed for galaxy formation such as for radiative cooling, star formation, chemistry, and feedback from stars and black holes, using task-based parallelism. The equations of hydrodynamics are solved using the smoothed particle hydrodynamics (SPH) density-energy scheme {\tt SPHENIX} \citep{Borrow22}. 
The COLIBRE \footnote{\href{http://colibre-simulations.org/}{http://colibre-simulations.org/}} galaxy formation model \citep{Schaye25, Chaikin25} is implemented in {\tt SWIFT} and will be used in this work. Below, we summarise the COLIBRE model with an emphasis on the prescriptions for BH physics and AGN feedback.

Radiative cooling and heating follow the \texttt{HYBRID-CHIMES} model \citep{Ploeckinger25}, which combines non-equilibrium abundances of primordial elements computed by \texttt{CHIMES} \citep{Richings14a,Richings14b} with metal heating and cooling rates pre-computed by \texttt{CHIMES} under the assumption of ionization equilibrium. The cooling and heating rates
account for a $z=0$ homogeneous radiation background from distant galaxies
and quasars, and, within the ISM, an interstellar radiation field whose
intensity scales with the local star formation rate surface density. The formation, evolution and destruction of six dust species is modelled following \cite{Trayford25}, with dust abundances self-consistently coupled to the chemistry and cooling. The gravitational instability criterion is used to check whether a given gas particle is star-forming, following \cite{Nobels24}. Star-forming gas particles compute their SFRs according to the Schmidt law \citep{Schmidt59} with a fixed star-formation efficiency per free-fall time-scale of 1 per cent, and are converted into stellar particles stochastically. The implementation of early stellar feedback is detailed in \cite{Benitez-Llambay25} accounting for processes including H~\textsc{ii} regions, stellar winds, and radiation pressure using the Binary Population and Spectral Synthesis (\texttt{BPASS}) tables \citep{Eldridge17, Stanway18}. COLIBRE tracks the abundances of 12 chemical elements: H, He, C, N, O, Ne, Mg, Si, Fe, Sr, Ba, and Eu. The stellar evolution model in COLIBRE is based on Correa et al. (submitted) and includes chemical enrichment channels from core-collapse supernovae (CC SNe), type-Ia supernovae (SNIa), stellar winds, asymptotic giant branch stars, neutron star mergers, common envelope jet supernovae and collapsars.

The CC SNe feedback model in COLIBRE includes energy feedback in kinetic \citep{Chaikin23} and thermal forms (based on \citealt{DallaVecchia12}). SN energy injections are distributed among gas neighbours in a statistically isotropic way, following \cite{Chaikin22}. The total energy from CC SNe available for a stellar particle of initial mass $m_*$ in the time-step from $t$ to $t+\Delta t$ is
\be
\Delta E_{\rm CC SNe}=10^{51}\text{erg} \times f_{\rm E}m_*\int^{m_{\rm d}(t+\Delta t)}_{m_{\rm d}(t)}\Phi(m)\dd m,
\label{eq:E_CCSN}
\ee
where $\Phi(m)$ is the stellar initial mass function assumed to be \cite{Chabrier03}, $m_{\rm d}(t)$ is the mass of the stars that explode as CC SNe at age $t$, and $f_{\rm E}$ is a parameter representing the fraction of the total CC SNe energy in units of $10^{51}$erg, increasing monotonically with stellar birth gas pressure, but capped at 4.0 (see \cite{Schaye25} for details). For CC SNe kinetic feedback, a fraction of $f_{\rm kin}=0.1$ of the total energy is released by kicking pairs of gas particles in random but opposite directions, conserving energy, linear momentum, and angular momentum, with a target velocity of 50 km s$^{-1}$\citep{Chaikin23}. The remainder of the total CC SNe energy, $(1-f_{\rm kin})\Delta E_{\rm CC SNe}$, is released in thermal form. The probability that a gas particle is heated is equal to the ratio of the total energy available for thermal SN feedback and the energy required to increase the temperature of the gas particle by $\Delta T_{\rm SN}$, expressed as 
\be
\Delta E_{\rm heat, SN}= \frac{m_{\rm g} k_{\rm B}\Delta T_{\rm SN}}{\left(\gamma-1\right)\mu m_{\rm p}},
\ee
where $\mu=0.6$ is the mean particle mass for fully ionized gas, $k_{\rm B}$ is the Boltzmann constant, $m_{\rm g}$ is the mass of the gas particle being heated, and $m_{\rm p}$ is the proton mass. 

The temperature increment $\Delta T_{\rm SN}$ increases with gas density as
\be
\Delta T_{\rm SN}=10^{6.5} \text{K} \left(\frac{n_{\rm H,SN}}{n_{\rm H,pivot}}\right)^{2/3},
\label{eq:deltaT_SN}
\ee
but is limited to the range $\Delta T_{\rm SN,min} \leq \Delta T_{\rm SN} \leq \Delta T_{\rm SN,max}$. In equation (\ref{eq:deltaT_SN}), $n_{\rm H,SN}$ is the gas hydrogen number density (assuming a hydrogen mass fraction of $X_{\rm H} = 0.756$), computed using the SPH formalism at the location of the stellar particle at the time of SN feedback, and $n_{\rm H,pivot}$ is a normalization parameter. The strengths of SN and AGN feedback in the COLIBRE simulations at each resolution have been calibrated to the observed $z=0$ galaxy stellar mass function, galaxy size -- stellar mass relation, and the BH mass -- stellar mass relation of massive galaxies \citep{Chaikin25}. The calibration was performed separately at each resolution, resulting in slightly different subgrid parameter values for SN and AGN feedback, including $\Delta T_{\rm SN,min}$, $\Delta T_{\rm SN,max}$, and $n_{\rm H,pivot}$ (see table 1 in \citealt{Schaye25} for all parameters that vary with resolution). Since the mass resolution of the simulations in this work ($m_{\rm g} = 10^5~\mathrm{M_\odot}$)\footnote{
We adopt a gas particle mass resolution of $m_{\rm g}=10^5~\mathrm{M_\odot}$, as this allows us to explore a wide range of parameter variations within a reasonable computational time (2-3 days for one run, $\sim 10^5$ CPU hours for all runs). This resolution was also the fiducial resolution in previous
works presenting simulations of a MW-mass galaxy run with the
\textsc{SWIFT} code \citep[e.g.][]{Chaikin23,Nobels24}. This choice is also close to the resolution of the highest-resolution COLIBRE simulations \citep{Schaye25}.
Accordingly, we use the calibrated m5 COLIBRE parameters, and since the resolutions are sufficiently similar, no further recalibration is required.} is comparable to that of the fiducial COLIBRE cosmological simulations at m5 resolution ($m_{\rm g} = 2.3 \times 10^5~\mathrm{M_\odot}$), we adopt the values of $\Delta T_{\rm SN,min}$, $\Delta T_{\rm SN,max}$, and $n_{\rm H,pivot}$ used for m5 resolution, namely $\Delta T_{\rm SN,min} = 10^7~\mathrm{K}$, $\Delta T_{\rm SN,max} = 10^8~\mathrm{K}$, and $n_{\rm H,pivot} = 1~\mathrm{cm}^{-3}$.

The energy release from SNIa feedback is calculated following the same scheme as for CC SNe feedback but with $f_{\rm E}=1$ and $f_{\rm kin}=0$, and with the integral in \Eq{E_CCSN} replaced by $\int_t^{t+\Delta t}\mathcal{\dot{N}}_{\rm SNIa}(t-t_*)\,\dd t$, where $\mathcal{\dot{N}}_{\rm SNIa}(t-t_*)$ is the rate of SNIa per unit initial stellar mass for a simple stellar population born at time $t_*<t$ (Nobels et al. in preparation).

\subsection{Black holes and AGN feedback}\label{sec:COLIBRE_BH}
The fiducial model of supermassive black hole growth and AGN feedback in COLIBRE is based on \cite{Booth09}. BH particles grow in mass by merging with other BHs (not used in this work) and by accreting gas from its surroundings. In COLIBRE, the Bondi-Hoyle-Lyttleton model \citep{Hoyle39,Bondi44} with a modification accounting for turbulence and vorticity \citep{Krumholz05,Krumholz06} is used to calculate the gas accretion rate limited at 100 times the Eddington rate
\be\label{eq:acc}
\dot{M}_{\rm acc}={\rm min}\{f_{\rm TV}\frac{4\pi G^2 \MBH^2\rho_{\rm gas}}{c_{\rm s}^3},100\dot{M}_{\rm Edd}\},
\ee
where $c_{\rm s}$ is the sound speed of the ambient gas, and $f_{\rm TV}$ is the modification parameter which takes form of  
\be\label{eq:fTV}
f_{\rm TV}=\frac{f_{\rm turb}f_{\rm ang}}{\left(f_{\rm turb}^2+f_{\rm ang}^2\right)^{1/2}}.
\ee
The turbulence limiter is
\be
f_{\rm turb}=\left[\frac{\lambda^2+\mathcal{M}^2}{(1+\mathcal{M}^2)^4}\right]^{1/2},
\ee
with $\lambda=1.1$ \citep{Krumholz06} and where $\mathcal{M}=v/c_{\rm s}$ is the Mach number of the gas. Here, the gas velocity $v$ is calculated as $v^2=v_{\rm BH}^2+\sigma_{\rm turb}^2$, where $v_{\rm BH}$ and $\sigma_{\rm turb}$ are, respectively, the bulk velocity relative to the BH and the velocity dispersion of the ambient gas in the frame of the BH.

The other parameter in \Eq{fTV} is the vorticity limiter $f_{\rm ang}$ \citep{Krumholz05}, calculated as
\be
f_{\rm ang}=0.34f(\omega_\star),
\ee
where $\omega_\star=\omega r_{\rm B}/c_{\rm s}$, $\omega=|\nabla\times \boldsymbol{v}|$ is the vorticity of the ambient gas, $r_{\rm B}=G\MBH/c_{\rm s}^2$ is the Bondi radius, and $f(\omega_\star)=(1+\omega_\star^{0.9})^{-1}$.

In \Eq{acc}, the Eddington rate is given by
\be
\dot{M}_{\rm Edd}=\frac{4\pi G m_{\rm p}\MBH}{\epsilon_{\rm r}\sigma_{\rm T} c},
\ee
where $\sigma_{\rm T}$ is the Thomson cross-section, $c$ is the speed of light, and $G$ is the gravitational constant. The Eddington ratio $\dot{m}$ is therefore defined as $\dot{m}=\Macc/\Medd$. Note that, while \Eq{acc} limits the BH accretion rate to 100 times the Eddington rate, we find that the accretion rate never actually exceeds the Eddington rate in our simulations.

A fraction $\epsilon_{\rm r}=0.1$ of the total mass accreted over time-step $\Delta t$ is converted to radiation and the remainder is added to the mass of the SMBH, with the excess mass nibbled from the surrounding gas following the nibbling algorithm from \cite{Bahe22}. The rate at which the BH grows is 
\be\label{eq:bondi}
\dot{M}_{\rm BH}=(1-\epsilon_{\rm r})\dot{M}_{\rm acc}.
\ee


COLIBRE models AGN feedback by using either a thermal model or a hybrid model that includes both thermal and jet feedback \citep{Husko25}. The hybrid model is not included here because we only want to study feedback by AGN winds. Therefore, we simply introduce the thermal mode adopted in our new subgrid model here, even though it is intended to represent the effect of an accretion disc wind.
This approach is justified owing to the limited resolution of cosmological simulations, which prevents a clear distinction between isotropic thermal and kinetic feedback \citep{Husko24a}. More specifically, real AGN accretion disc winds would be likely to be shocked on subgrid scales, thermalising the wind kinetic energy, so that by the time the outflowing material reaches the resolved scales, it manifests as a hot, expanding bubble, driven primarily by thermal gas pressure. This explains why kinetic feedback in low-resolution cosmological simulations produces results very similar to thermal feedback, as the kicked gas is immediately shocked \citep{Weinberger18,Husko24a}. Idealised calculations of the interaction of fast AGN winds (as expected for UV line-driving) with the surrounding ISM imply that radiative cooling in the wind shocks will typically be weak, so the shocks will be nearly adiabatic \citep{FG12,Costa20}, and almost 100\% of the initial kinetic energy of the accretion disc wind will be converted to thermal energy.

In the thermal mode, the SMBH inject energy at a rate $\dot{E}_{\rm AGN}=\eta\epsilon_{\rm r} \dot{M}_{\rm acc}c^2 $ 
into the surrounding gas with coupling efficiency $\eta$, i.e. the fraction of the radiative luminosity $\epsilon_{\rm r}\Macc c^2$ that is
assumed to couple to the ambient gas, after the required amount of energy has been accumulated in the reservoir of the BH. In this work, we vary the $\eta$ value according to our new model, and set 0.05 as a reference value when using the fiducial COLIBRE prescription for AGN feedback. Note that the value of $\eta$ in the fiducial COLIBRE cosmological simulations \citep{Schaye25} depends on the mass resolution. Our choice of 0.05 is to match the similar mass resolution in the fiducial runs, i.e. $10^5$ M$_\odot$. In the fiducial COLIBRE AGN thermal feedback model, $\eta$ is a constant, independent of the BH mass or accretion rate, while in this work, we allow $\eta$ to vary according to a specific model. 

 \begin{table}
\centering
\caption{Coupling efficiencies $\eta$, and the corresponding normalisations $N_\eta$ and slopes $\alpha_\eta$, adopted for the Eddington ratio $\dot{m}$-dependent model in this work. We run fiducial constant and fiducial variable coupling efficiency models for all BH masses, while the variations of normalisations $N_\eta$ and slopes $\alpha_\eta$ are only tested for $\MBH=10^6,10^7,10^8$ M$_\odot$.}
{\fontsize{7.5}{8.5}\selectfont	
\begin{tabular}{lclclcc}
\hline
$\eta$                             & $N_\eta$                      & $\alpha_\eta$       & \multicolumn{4}{c}{Comments}                                            \\ \hline
$N_\eta\dot{m}^{\alpha_\eta}$     & 39.81                         & 2.6                & \multicolumn{4}{c}{Fiducial variable, }                                   \\
     &                          &                 & \multicolumn{4}{c}{all BH masses}                                   \\ \\
0.05                               & NA                            & NA                 & \multicolumn{4}{c}{Fiducial constant,}\\  
                              &                             &                  & \multicolumn{4}{c}{all BH masses}                                   \\ \\
$N_\eta\dot{m}^{\alpha_\eta}$     & 10, 300, 3000      & 2.6                & \multicolumn{4}{c}{Variation for variable $\eta$, }    
                   \\
     &       &                 & \multicolumn{4}{c}{$\MBH=10^6, 10^7, 10^8$ M$_\odot$}    
                   \\ \\
$N_\eta\dot{m}^{\alpha_\eta}$     & $39.81\times 10^{\alpha_\eta-2.6}$ & 0.5,1.5,3   & \multicolumn{4}{c}{Variation for variable $\eta$, }                       \\
     &  &    & \multicolumn{4}{c}{$\MBH=10^6, 10^7, 10^8$ M$_\odot$}                       \\ \hline
\end{tabular}}
\label{tab:parameters2}
\end{table}

Gas particles surrounding the BH are heated by AGN feedback by a temperature increment $\Delta T_{\rm AGN}$ corresponding to an energy
\be
\Delta E_{\rm heat, AGN}= \frac{m_{\rm g} k_{\rm B}\Delta T_{\rm AGN}}{\left(\gamma-1\right)\mu m_{\rm p}}.
\ee
The model accumulates $\dot{E}_{\rm AGN} \, \Delta t$ from multiple time-steps in an energy reservoir until $E_{\rm BH}\geq \Delta E_{\rm heat, AGN}$, where $\Delta t$ is a single BH time-step. If this condition is satisfied, the SMBH will heat up the nearest surrounding particle(s) by $\Delta T_{\rm AGN}$.

\begin{table*}
\centering	
\caption{Parameters for initialisations of DM halo, cold gas disc, stellar disc, hot CGM and SMBH used in this work. The parameters for all components except the mass of SMBH are the same for all simulations used in this work. The CGM mass listed in the table is measured within $R_{200}$.}
\begin{tabular}{ccllclclcclcc}
\hline
Component     & \multicolumn{12}{c}{Parameters}                                                                                                                                                                                                                                     \\ \hline
DM halo       & \multicolumn{3}{c}{Profile}                            & \multicolumn{2}{c}{$M_{200}/$M$_\odot$}                  & \multicolumn{2}{c}{$R_{200}$/kpc}                 & $r_{\rm s, Hern}$/kpc     & \multicolumn{2}{c}{$\lambda$}                  \\
              & \multicolumn{3}{c}{Hernquist}                          & \multicolumn{2}{c}{$1.37\times 10^{12}$}               & \multicolumn{2}{c}{228}                              & 52.13                  & \multicolumn{2}{c}{0.033}                        \\ \hline
Cold gas disc & \multicolumn{3}{c}{Profile}                            & \multicolumn{2}{c}{$M_{\rm d, gas}/$M$_\odot$}           & \multicolumn{2}{c}{$R_{0,\rm gas}$/kpc}           & $z_{0,\rm gas}$/kpc & \multicolumn{2}{c}{$Z/Z_\odot$} & \multicolumn{2}{c}{$T_{0,\rm gas}$/K} \\
              & \multicolumn{3}{c}{Exponential}                        & \multicolumn{2}{c}{$5.48\times10^9$}                   & \multicolumn{2}{c}{4.3}                           & 0.43                & \multicolumn{2}{c}{1}           & \multicolumn{2}{c}{$10^4$}            \\ \hline
Stellar disc  & \multicolumn{3}{c}{Profile}                            & \multicolumn{2}{c}{$M_{\rm d, *}/$M$_\odot$}             & \multicolumn{2}{c}{$R_{0,*}$/kpc}                 & $z_{0,*}$/kpc       & \multicolumn{2}{c}{}            & \multicolumn{2}{c}{}                  \\
              & \multicolumn{3}{c}{Exponential}                        & \multicolumn{2}{c}{$4.93\times10^{10}$}               & \multicolumn{2}{c}{4.3}                           & Hydro Equilibrium   & \multicolumn{2}{c}{}            &                   &                   \\ \hline
Hot CGM       & \multicolumn{3}{c}{Profile}                            & \multicolumn{2}{c}{$M_{\rm CGM}/$M$_\odot$}              & \multicolumn{2}{c}{$r_0$/kpc}                     & $T_0$/K             & \multicolumn{2}{c}{$Z/Z_\odot$} & \multicolumn{2}{c}{$\lambda^\prime$}  \\
              & \multicolumn{3}{c}{Numerical}                          & \multicolumn{2}{c}{$9.3\times 10^{10}$}                & \multicolumn{2}{c}{3.5}                           & $10^6$              & \multicolumn{2}{c}{0.1}         & \multicolumn{2}{c}{0.05}              \\ \hline
SMBH       & \multicolumn{3}{c}{}                            & \multicolumn{2}{c}{$\MBH/$M$_\odot$}               \\
              & \multicolumn{3}{c}{}                          & \multicolumn{2}{c}{$0, 10^6, 4\times 10^6, 10^7, 10^8, 10^9$}         \\ \hline

\end{tabular}
\label{tab:parameters}
\end{table*}

Note that larger values for $\Delta T_{\rm AGN}$ will make individual feedback events more energetic and more intermittent. The fiducial COLIBRE cosmological simulations use an AGN heating temperature that depends linearly on the BH mass. However, in this work we adopt a fixed $\Delta T_{\rm AGN} = 10^{9}~\mathrm{K}$ because we want to isolate the effects of a variable coupling efficiency and therefore we do not want an additional dependence of $\Delta T_{\rm AGN}$ on the BH mass. Furthermore, \cite{Chaikin25} showed that COLIBRE simulations with $\Delta T_{\rm AGN} = 10^9~\mathrm{K}$ reproduce the observational data similarly well to those with the fiducial variable $\Delta T_{\rm AGN}$ in the implementation of AGN feedback (see figures~14 and 15 in \citeauthor{Chaikin25}~\citeyear{Chaikin25}). 

Finally, the effects of dynamical friction on the BH, i.e. BH repositioning, are not modelled in our idealised simulations, and instead the BH is pinned to the centre of the box.

\subsection{New AGN feedback model with varying coupling efficiency}\label{sec:implementation}

Our new coupling efficiency model is motivated by the predictions of Qwind, described in \se{intro}. Qwind outputs several wind properties, and \cite{Quera-Bofarull23} find that these properties depend on $\dot{m}$ approximately as power laws at fixed BH mass, over the radiatively efficient thin disc regime ($0.01 < \dot{m} < 1$) in which UFOs are thought to be launched. The scaling of most relevance here is $\eta = L_{\rm kin}/L_{\rm bol} \propto \dot{m}^{2.6}$, where $\eta$ is the coupling efficiency of AGN feedback, i.e. the fraction of the radiative luminosity that couples to the kinetic power of the wind. The dependence of $\eta$ on BH mass is predicted to be weak over the range $10^7-10^9$~M$_\odot$, so we neglect it here. This behaviour contrasts with that assumed in most cosmological simulations, which assume $\eta$ to be constant and calibrate its value to reproduce the observed $M_{\rm BH}$--$M_*$ relation, for example obtaining 0.15 \citep{Schaye15}, 0.1 \citep{Weinberger18} or 0.05 \citep[the value used in COLIBRE runs at a mass resolution of $\sim 10^5$~M$_\odot$;][]{Schaye25}. Although the slope and normalisation of the Qwind relation depend on the details and parameters of the accretion disc wind model, the existence of a correlation between $\eta$ and $\dot{m}$ appears robust.

Therefore, we implement a new model for the coupling efficiency in COLIBRE. Instead of directly adopting the normalisation and slope of the power law from Qwind, we keep these two as free parameters. This is due to the potential dependence of these parameters on the specifics of the line-driven wind model. 
Note that the normalization is more uncertain than the slope in Qwind, as it depends on the inner wind-launching radius of the accretion disc, which is not determined by the model. The variable coupling efficiency is therefore assumed to have the form 
\be\label{eq:new_eta}
	\eta=\text{min}\{N_{\eta}\dot{m}^{\alpha_\eta},1\},
\ee
where $N_\eta$ is the normalisation and $\alpha_\eta$ is the slope. Here we cap this efficiency at 100\%. Observational data on UFOs \citep{Mestici24} imply that $\eta \approx 0.1$ for $\dot{m} \approx 0.1$, while constraining the dependence on $\dot{m}$ more weakly. 
For our fiducial variable coupling efficiency model, we adopt $\alpha_\eta=2.6$ based on the Qwind model, but normalise to $\eta=0.1$ for $\dot{m}=0.1$, based on observations, which yields $N_\eta = 39.81$ based on \Eq{new_eta}. We also explore a wider range of dependencies by considering variations of the slope $\alpha_\eta$ and normalisation $N_\eta$ around this fiducial model, in one of two ways. The first method varies the normalisation ($N_{\eta}=10,300,3000$) while keeping the slope fixed ($\alpha_\eta=2.6$). The other method varies the slope ($\alpha_\eta=0.5,1.5,3$) but fixes $\eta=0.1$ when $\dot{m}=0.1$, which yields $N_{\eta}=39.81\times 10^{\alpha_{\eta}-2.6}$. All coupling efficiencies and their parameters used in this work are listed in \Tab{parameters2}. We note that although our variable coupling efficiency model is inspired by the UV line-driven wind model Qwind, we do not attempt to reproduce the predictions of that model in detail. Rather, our goal is to investigate the effects of a variable coupling efficiency depending on the Eddington ratio. To this end, we allow the Eddington ratio to take values both below and above the typical range expected for UV line-driven winds.

In the standard version of COLIBRE with a constant coupling efficiency, the time-step for a BH is limited to be no larger than $\Delta t=\Delta E_{\rm heat, AGN}/\dot{E}_{\rm AGN}$. In addition to the limiter based on the BH accretion rate, the BH time-step is always limited by gravity, and to be at most 4 times larger than the time-steps of the BH's gas neighbours.
We note that a shorter time-step is necessary when calculating BH quantities using this variable coupling efficiency model, compared to the standard constant efficiency. If this constraint on the time-step is not imposed, then the calculation of the energy injection by AGN feedback becomes inaccurate. Specifically, the AGN energy input saved in the output files may be inconsistent with the energy calculated from a time integral over the accretion rate ($\int \eta \epsilon_{\rm r}\dot{M}_{\rm acc}c^2\dd t$), since the effect of fluctuations of the accretion rate and variable coupling efficiency cannot be fully captured in a large time interval.
Therefore, to address these issues, we modify the BH time-step to have a maximum value $\Delta t^\prime=\Delta E_{\rm heat, AGN}/\dot{E}^\prime$, where $\dot{E}^\prime=\epsilon_{\rm r}\dot{M}_{\rm Edd}c^2$, corresponding to a high heating rate.

\subsection{Initial conditions}

\begin{figure*}
    \centering
    \includegraphics[width=1\textwidth]{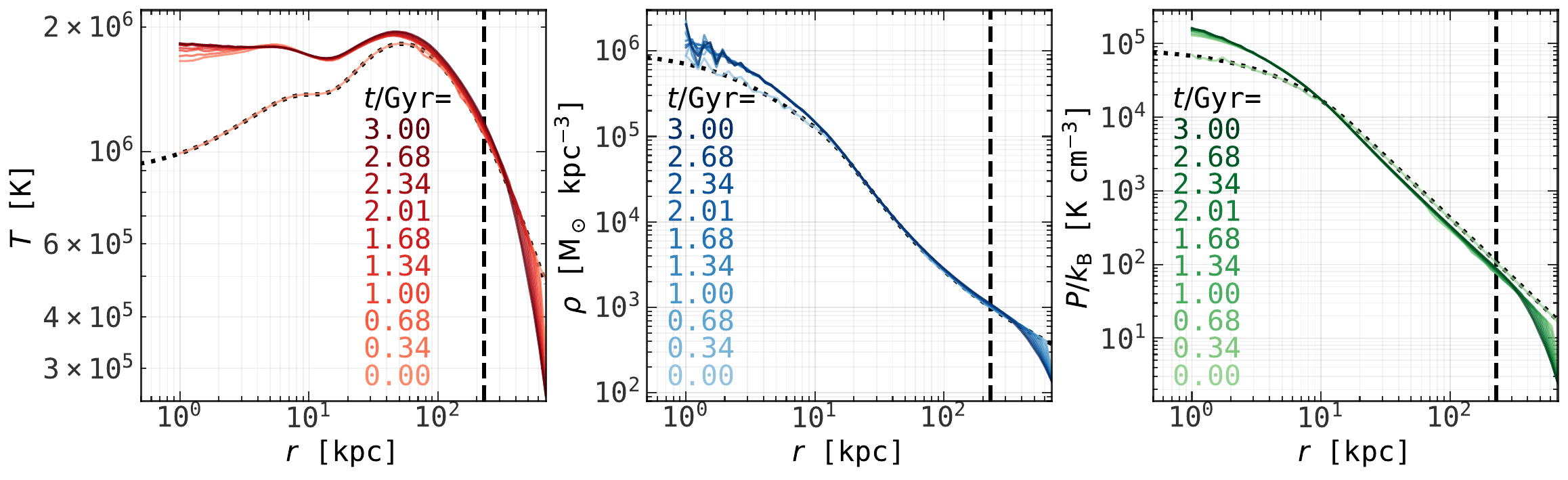}
    \caption{
    Initial temperature (left panel), density (middle panel), and pressure (right panel) profiles of the CGM in this work, using the initialisation method from \protect\cite{Nobels22} modified for the inclusion of a stellar disc (see text). The black dotted lines indicate the initial profiles, while the vertical black dashed lines show the radius $R_{\rm 200}$. Coloured solid lines represent the profile in the hydro- and gravity-only simulation at different times, as indicated in the panels, with darker colours corresponding to progressively later times. Equilibrium within $2R_{200}$ is reached already after 0.3 Gyr, whereas the gas profiles at radii beyond $2R_{200}$ continue to evolve at later times.}
\label{fig:CGM_init}
\end{figure*}

We simulate an idealised disc galaxy consisting of an exponential disc of stars and cold gas (with no stellar spheroid) embedded in a hot CGM inside a static DM halo potential. All particles in the
simulation (gas, stars, and the BH) follow the same SPH neighbour-finding algorithm to identify their gas neighbours, using a quartic spline kernel. The target particle smoothing length is set to
$1.2348$ times the local inter-particle separation, which, for a quartic spline, yields an expected number of $\approx 64.9$ gas neighbours
within the kernel. A minimum smoothing length of $2 \times 10^{-6}~\mathrm{pc}$ is imposed, although it is never reached,  while the maximum allowed smoothing length is set to $50~\mathrm{kpc}$. Gravitational forces are softened
using a Plummer-equivalent gravitational softening length of $200~\mathrm{pc}$.

The parameters of the simulated galaxy, including the temperatures and metallicities for the gas, are chosen to be similar to those for the Milky Way. The initialisation of the stellar and gas discs and DM halo potential follows \cite{Nobels24}, which is based in turn on the {\tt MAKENEWDISK} code from \cite{Springel05}, while the distribution of the hot CGM is modelled in a similar way to \cite{Nobels22}. Note that we used the ICs from \cite{Nobels24} which contain the stellar and gas discs and were generated by them. We describe the procedures in detail below, but with one important change to the CGM initialisation that we allow in our initial conditions for the gravitational potential being non-spherical due to the stellar disc. Specifically, we allow an initial dynamical relaxation phase for the CGM in the non-spherical potential (see below), which is a new feature for \texttt{SWIFT} simulations. All parameters are summarised in \Tab{parameters}.

\subsubsection{Workflow for generating initial conditions and achieving equilibrium}

Below we outline the procedure used to generate the ICs for our simulations and to evolve them to a state of equilibrium:

\begin{itemize}
	\item Initialise the background analytical DM potential with a Hernquist profile \Eqb{Hernquist}, using the virial mass $\Mvir$, and NFW concentration $c$. 
    The Hernquist scale radius $r_{\rm s, Hern}$ is derived using \Eq{get_rs}.
	\item  Initialise the stellar and cold gas discs using \Eq{rho_disc}. Inputs include the stellar disc mass $M_{\rm d,*}$, cold gas disc mass $M_{\rm d,gas}$, initial cold gas temperature and metallicity. The scale radius $R_0$ for both discs is computed following \cite{MMW98} from $\Mvir$, $\lambda$ and $c$. The stellar disc scale height $z_0$ is fixed to 10 per cent of $R_0$ while the scale height of the cold gas disc is determined by vertical hydrostatic equilibrium. Velocity dispersions and rotational velocities are calculated from moments of the collisionless Boltzmann equations,  following the procedure in \citet{Springel05}.
	\item Initialise the CGM with a temperature profile \Eqb{Ttot} and solve for its density profile numerically, assuming spherical symmetry. The enclosed mass $M_{\rm en}$ here includes only the DM and stellar disc. Required inputs for the CGM are the scale radius $r_0$, central temperature $T_0$, total mass, initial metallicity and the spin parameter $\lambda^\prime$. The CGM rotation velocity is set by its angular momentum \Eqb{CGMAM}, using $\Mvir$, $M_{\rm en}$ and $\lambda^\prime$.
	\item Evolve the CGM with the DM and stellar disc for 3 Gyr under gravity and hydrodynamics only, no longer assuming spherical symmetry, to reach dynamical equilibrium. From the final snapshot, extract the CGM distribution, replace the evolved stellar disc with the initial one, and add the cold gas disc to obtain the ICs for all of the runs.
	\item Enable all subgrid physics and evolve for 0.3 Gyr with no BH to stabilise the star formation rate.
	\item Insert a central BH with the target mass (except for the SN-only run), apply different coupling efficiency models, and evolve for another 3 Gyr.
\end{itemize}
All parameters that need to be input are listed in \Tab{parameters}. We detail these procedures in the following sections.

\subsubsection{Dark matter halo}

The DM halo can be characterized by halo mass and concentration. The halo mass $\Mvir$ is defined as the mass within the virial radius $\Rvir$, the radius within which the average density of the halo is 200 times the critical density of the universe, i.e. $M_{\rm 200}=4\pi\Delta_{\rm c}\rho_{\rm c} R_{\rm 200}^3/3$, where $\Delta_{\rm c}=200$ is the overdensity factor, and $\rho_{\rm c}=3H_0^2/8\pi G$ is the critical density. 
In all runs, we choose a halo virial mass $\Mvir=1.37\times 10^{12}$ M$_\odot$. 

In {\tt MAKENEWDISK} \citep{Springel05}, a static dark matter halo, i.e. an analytical potential representing the dark matter halo, is used. Based on cosmological N-body simulations of the DM evolution, DM density profiles are well fit by Navarro, Frenk \& White profiles \citep[hereafter NFW]{Navarro97}. However, {\tt MAKENEWDISK} assumes a Hernquist profile \citep{Hernquist90} instead, for computational convenience. To be consistent with \cite{Nobels24}, we also use a Hernquist profile for the halo potential. To find the Hernquist profile that best matches the nominal NFW profile, we require the central parts of the two DM density profiles to be equal. By doing so, we then only need to specify the virial mass and halo concentration for the NFW profile, from which we can derive the corresponding parameters for the Hernquist profile to generate initial conditions and run the simulations. The procedure is outlined below.

The NFW profile takes the form
\be
\rho_{\rm DM, NFW}\left(r\right)=\frac{\Mvir}{4\pi f(c)}\frac{1}{r \left(r_{\rm s, NFW}+r\right)^2},
\ee
where $r_{\rm s,NFW}$ is the scale radius of the NFW halo, the halo concentration is defined as $c=\Rvir/r_{\rm s,NFW}$ and $f(c)=\ln(1+c)-c/(1+c)$. We choose $c=9$ in all runs.

The Hernquist profile is defined as
\be
\rho_{\rm DM, Hern} \left( r \right)=\frac{\Mdm}{2\pi}\frac{r_{\rm s, Hern}}{r(r+r_{\rm s, Hern})^3},
\ee
where $\Mdm$ is the total mass of DM obtained by integrating $\rhodm$ to infinity, and $r_{\rm s, Hern}$ is the scale radius in the Hernquist profile. The total DM mass can be connected with the virial mass by $\Mvir=\int_0^{\Rvir}\rhodm 4\pi r^2\dd r$. We assume that the virial mass (and the corresponding virial radius) are the same for the two profiles. Then one can rewrite the Hernquist profile as
\be\label{eq:Hernquist}
\rho_{\rm DM, Hern} \left( r \right)=\frac{M_{\rm 200} (\Rvir + r_{\rm s, Hern})^2}{2\pi \Rvir^2}  \frac{r_{\rm s, Hern}}{r (r + r_{\rm s, Hern})^3}.
\ee

Next, we match the Hernquist profile with the NFW profile to determine $r_{\rm s, Hern}$, forcing the two density profiles to match at small radii (i.e. $\rho_{\rm DM, Hern}(r) = \rho_{\rm DM, NFW}(r)$ for $r \ll r_{\rm s,Hern}$, see \cite{Nobels24} for more details), which yields
\be\label{eq:get_rs}
r_{\rm s, Hern}=\frac{b+\sqrt{b}}{1-b}\Rvir,
\ee
where $b=2\left( \ln(1 + c) - \frac{c}{1 + c} \right)/c^2$.

The angular momentum of the DM halo is initialised by the dimensionless spin parameter $\lambda$ \citep{Peebles69}, which will be used to determine the disc scale radius (see the next subsection). In all runs, the DM halo spin parameter is set to $\lambda=0.033$.

\subsubsection{Stellar disc and cold gas disc}
The profiles of the stellar disc and cold gas disc are given by

\be\label{eq:rho_disc}
\rho_{\rm d}\left(R,z\right)=\frac{\Md}{4\pi z_0 R_0^2}\text{sech}^2\left(\frac{z}{2z_0}\right)\exp\left(-\frac{R}{R_0}\right),
\ee
 where the total disc mass is set to 4 per cent of the virial mass, $\Md=5.48\times10^{10}$ M$_\odot$. The cold gas disc mass is 10 per cent of the total disc mass, $M_{\rm d, gas}=5.48\times10^9 $ M$_\odot$ while the remainder is in the stellar disc, corresponding to $M_{\rm d, *}=4.932\times10^{10} $ M$_\odot$. $R_0$ and $z_0$ are the scale radius and the scale height of the disc, respectively. We follow the scenario in \cite{MMW98}, which relates $R_0$ to $\lambda$ depending on $R_{\rm 200}$ and $c$, and solves for $R_0$ numerically, assuming an NFW profile. This yields a disc scale radius $R_0=4.3$ kpc. Although both the stellar and cold gas discs have the same scale radius set as above, their scale heights are different. The stellar scale height is fixed to 10 per cent of disc scale radius, $z_0=0.43$ kpc, while the scale height of the gas is determined by vertical hydrostatic equilibrium \citep{Springel05,Nobels24}. 

 The stellar velocity distribution at each position can in principle be determined by solving the collisionless Boltzmann equation based on the density profile, while the rotational velocity distribution follows the procedure in \cite{Springel05}. Note that our velocity initialisations for the discs are adopted from \cite{Nobels24} and we add a CGM manually later. Therefore, the total potential used to set up the stellar velocity distribution consists only of a static DM halo, a stellar disc and a cold gas disc.

  The particle masses for the stellar disc and the cold gas disc are both
set to $10^5~\mathrm{M_\odot}$. Unlike newly formed stellar particles, stellar particles from the initial conditions do not contribute to metal
enrichment or energy feedback during the simulation. We assume a solar metallicity (${\rm Z}_\odot=0.0133$) for the initial cold gas disc and that its initial temperature is $10^4$ K. For simplicity, the initial dust mass fraction of the cold gas disc is set to 0, but dust is produced by newly formed stars and subsequently grows in mass in the ISM.

\subsubsection{Circumgalactic medium}

From observations, there is clear evidence that our MW and other similar galaxies are surrounded by a hot CGM \citep{Tumlinson17}. Additionally, the cold gas disc is replenished by the CGM, part of which may later be accreted onto the BH. It is therefore desirable to include a CGM component in our idealised galaxy setup, alongside the DM halo, stellar disc, and cold gas disc. However, it is not easy to set up an exact dynamical equilibrium for all components at the beginning of the simulation, due to the gravitational potential of the disc not being spherically symmetric. To resolve this issue, we adopt the formalism in \cite{Nobels22} to first construct a system with a spherical CGM in dynamical equilibrium.
In their work, dynamical equilibrium was achieved in the gravitational potential of a static DM halo and stellar components following a spherical distribution. Instead, our stellar distribution is a cylindrically symmetric exponential disc, but we treat the mass distribution of the disc as spherical only for calculating its gravitational potential when setting up the initial spherical CGM distribution, and not afterwards. The initial temperature, density, and pressure profile are shown in \Fig{CGM_init}. We present the mathematical formalism for setting up these profiles below.

The CGM is assumed to initially be in hydrostatic equilibrium in a spherically symmetric potential, with thermal pressure balancing gravity,
\be
\frac{\dd P}{\dd r}=-\frac{GM_{\rm en}\left(r\right)\rho\left(r\right)}{r^2},
\ee
where $M_{\rm en}(r)$ is the mass of the DM halo and stellar disc enclosed within a sphere of radius $r$. Since the density of the CGM should be quite low compared to the DM, its contribution to the enclosed mass is neglected. The equation above can be rewritten as 
\be\label{eq:CGMhydro}
\frac{\dd \ln P}{\dd \ln r}=-\gamma\frac{v_{\rm c}^2}{c_{\rm s}^2},
\ee
where $c_{\rm s}=\sqrt{\gamma P/\rho}$ is the local sound speed, $\gamma=5/3$ is the adiabatic index of the gas, and $v_{\rm c}=\sqrt{GM_{\rm en}(r)/r}$ is the circular velocity. If we assume that $c_{\rm s}=v_{\rm c}$\footnote{It is expected that hot halo gas is shock-heated on infall, converting kinetic energy set by the halo potential into thermal energy, and is subsequently supported against gravity by thermal pressure, implying that $c_{\rm s} \sim v_{\rm c}$ \citep{Stern19, Nobels22}.}, the condition of  hydrostatic equilibrium implies that
\be\label{eq:CGMhydro_equi}
P=P_0\left(\frac{r}{r_0}\right)^{-\gamma},
\ee
where $P_0$ is a normalisation constant. One can then obtain the temperature profile using $c_{\rm s}=\sqrt{k_{\rm B}\gamma T/\mu m_{\rm p}}=v_{\rm c}$, which takes the form
\be\label{eq:Tcirc}
T_{\rm circ}=\frac{\mu m_{\rm p}}{k_{\rm B}\gamma}\frac{GM_{\rm en}\left(r\right)}{r}.
\ee
However, the assumption $v_{\rm c} = c_{\rm s}$ is not sufficient in the galaxy centre, where $v_{\rm c}$ approaches zero, leading to $T \to 0$, which is unrealistic. Therefore, one needs to modify the temperature profile by imposing a temperature floor
\be\label{eq:Ttot}
T_{\rm tot}=T_{\rm circ}+\frac{T_0}{1+\exp\left(\frac{r-2r_0}{r_0}\right)},
\ee
with a central temperature $T_0$ and a scale radius $r_0$ being two free parameters. With this temperature profile, one can solve \Eq{CGMhydro} numerically. One can then obtain the corresponding density profile from $n(r)=P(r)/(k_{\rm B}T(r))$ and place CGM particles accordingly. In this model, $T_0$, $r_0$ and the total CGM mass within $R_{500}$ are free parameters, where $R_{500}=159$ kpc is the radius within which the average density of the halo is 500 times the critical density of the universe. In all runs, $T_0$ is set to $10^6$ K while $r_0$ is set to 3.5 kpc. The smallest distance between a CGM
particle and the centre is 0.244 kpc, while the 0.1th percentile is 6.11 kpc. The total CGM mass within $R_{200}$ ($R_{500}$) is $M_{\rm CGM}=9.3\times 10^{10} $ M$_\odot$ ($5\times 10^{10} $ M$_\odot$), which is 6.8 (4.5) per cent of $M_{\rm 200}$ ($M_{\rm 500}$), where $M_{500}$ is the halo mass within $R_{500}$. Our estimated CGM mass fraction lies within the observationally constrained range for MW-mass galaxies (with stellar masses $M_*\sim 10^{10.5}$ M$_\odot$), which is approximately 4.8 - 16\% \citep{Tumlinson17}.

 The CGM is assumed to initially be rotating around the disc axis, although the CGM rotational velocity is ignored when setting up the initial CGM density profile, which is spherical. The magnitude of the rotation velocity $v_{\rm rot}$ is based on the CGM specific angular momentum distribution ($j_{\rm CGM}$) and a spherical radial dependence ($v_{\rm rot}=j_{\rm CGM}/r$) within $R_{\rm 200}$, but with a slightly different spin parameter for the gas ($\lambda^\prime = 0.05$ different from the DM spin $\lambda=0.033$). The radial profile of $j_{\rm CGM}$ is obtained by using the fitting functions from \cite{Bullock01}, which give
\be\label{eq:CGMAM}
j_{\rm CGM}=j_0\frac{M_{\rm en}(r)}{M_{200}}\frac{1}{\mu_\lambda^\prime-M_{\rm en}(r)/M_{200}},	
\ee
where the first parameter $\mu_\lambda^\prime$ is set to 1.25 and the second parameter $j_0$ is
\be
j_0=\frac{\sqrt{2}V_{200}R_{200}\lambda^\prime}{-\mu_\lambda^\prime\ln({1-\mu_\lambda^\prime}^{-1})-1},
\ee
where $V_{200}=\sqrt{GM_{200}/R_{200}}$ is the virial velocity. Note that different from \cite{Bullock01}, the enclosed mass here is for stars and DM instead of DM only. The scheme for setting up the initial rotational velocity of the CGM is identical to that in \cite{Nobels22}, except that the stellar density profile is that of an exponential disc instead of a spheroid. 

The CGM is divided into high- and low-resolution regions at a radius of
$250~\mathrm{kpc}$, while the CGM properties are measured at radii well within the high-resolution region. Within the high-resolution region, the gas particle
mass is $10^5~\mathrm{M_\odot}$, the same as that of the cold gas disc
particles. In the low-resolution region, the mass resolution decreases monotonically with increasing galactocentric distance, with particle
masses ranging from $10^5~\mathrm{M_\odot}$ to $7.5 \times 10^5~\mathrm{M_\odot}$ (at 1875 kpc). Specifically, the dependence of particle mass on
distance is linear, which expressed as $m_{\rm gas, low}=m_{\rm gas}(r/250~\mathrm{kpc})$, where $m_{\rm gas, low}$ is the low resolution particle mass. 
The initial metallicity of the CGM is 0.1 times solar metallicity
, which is reasonable based on observations of star-forming galaxies of MW mass \citep[e.g.][]{Prochaska17}. The initial dust mass fraction of the CGM is 0, the same as for the cold gas disc.

\subsubsection{Achieving dynamical equilibrium for the circumgalactic medium}

To place CGM particles and achieve dynamical equilibrium with the other components, we first initialise a stellar disc and a spherical CGM distribution within a static DM halo, as described above. We then simulate the system with hydrodynamics and gravity (now using the real non-spherical gravitational potential) for 3 Gyr, turning off stellar feedback, AGN feedback, gas cooling, and star formation. From the last snapshot (3 Gyr), we extract the CGM distribution, add a black hole of variable mass, replace the time-evolved stellar disc with the initial one, i.e. before running for 3 Gyr, and introduce a cold gas disc to construct the ICs for our idealised simulations.

 These steps ensure that the system (including CGM) reaches a good approximation to dynamical equilibrium before turning on all of the subgrid gas physics for cooling, star formation, SN and AGN feedback. This setup is then used as the initial state at time $t=0$ for all of our simulations including a black hole. If the CGM were added directly without allowing it to dynamically relax, the system would not be fully in dynamical equilibrium at the start. Running the simulation with only gravity and hydrodynamics helps to achieve a dynamical equilibrium state for the initial conditions. We initially exclude the cold gas disc during the CGM dynamical relaxation calculation to prevent energy exchange between it and the hot CGM, which would otherwise heat up the gas disc and prevent it from remaining cold in the ICs. Additionally, the stellar disc dominates the disc potential, meaning the absence of the cold gas disc during the gravity and hydrodynamics only simulation has a minimal effect on the system achieving dynamical equilibrium.
 
 The evolution of the density, temperature and pressure profiles during the CGM dynamical relaxation calculation is shown by coloured lines in \Fig{CGM_init}. As can be seen, these profiles evolve away from the initial distribution as soon as the simulation begins, but reach dynamical equilibrium very soon (within $\sim$0.3 Gyr), after which they remain stable until the end of the simulation ($t = 3$ Gyr). All of the profiles become higher at small radii compared to the initial state as the gas contracts toward the centre due to the potential of the disc.

\begin{figure}
    \centering
    \includegraphics[width=1\columnwidth]{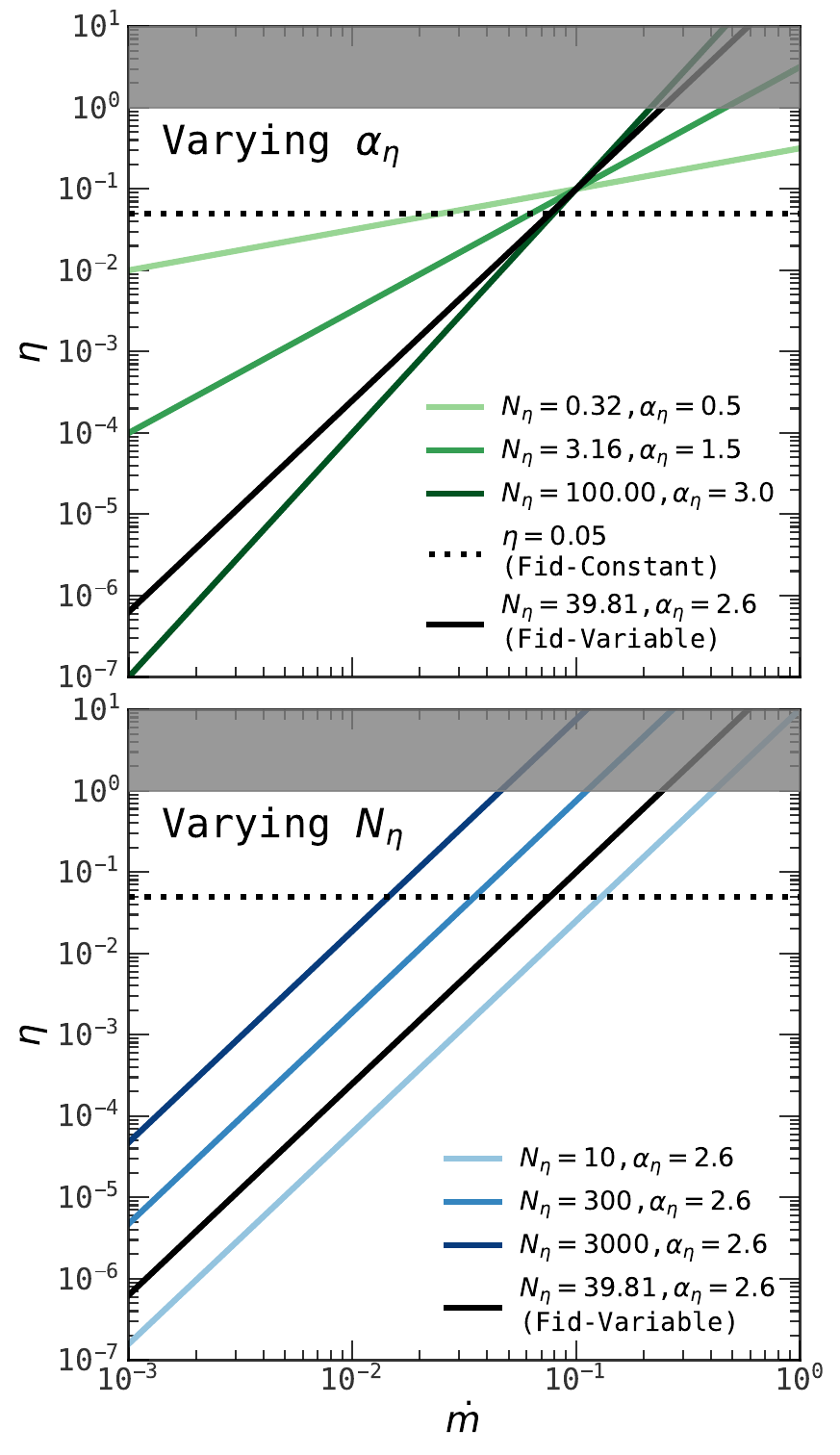}
    \caption{The coupling efficiency $\eta$ as function of the Eddington ratio $\dot{m}$, using our variable coupling efficiency model ($\eta=N_\eta\dot{m}^{\alpha_\eta}$) where different colours correspond to different combinations of the model parameters. We show the variations of $\alpha_\eta$ (varying $N_\eta$ as $39.81\times 10^{\alpha_\eta-2.6}$) in the upper panel while the variations of $N_\eta$ (keeping $\alpha_\eta$ fixed) are shown in the lower panel. Dark colors correspond to larger values of the parameter being varied. The parameters for the fiducial variable model are $N_\eta=39.81$ and $\alpha_\eta=2.6$, included as black lines in both panels. The gray shaded area represents the ceiling set on the coupling efficiency. All $\eta$ in this region will be set to unity. The horizontal lines at $\eta=0.05$ stand for the fiducial-constant value.}
    \label{fig:eta_mdot}
\end{figure}

\subsubsection{Supermassive black hole}

For all simulations, the setup for the components described above (DM halo, stellar disc, cold gas disc and hot CGM) remains the same. The only exception is in \se{CGMfree}, where we investigate how including the CGM affects our results, by comparing simulations with and without the CGM. We pin the single SMBH to the galaxy centre, with its coordinates coinciding with
the minimum of the potential, and vary the initial mass across different values in different simulations: $M_{\rm BH}=(10^6, 4\times10^6, 10^7, 10^8, 10^9)$ M$_\odot$.
The choice of $4\times10^6 $ M$_\odot$ is to match the observationally inferred BH mass for the MW \citep{MWBHmass}, while the other BH masses cover the range of observed BH masses for spiral and S0 galaxies \citep[$1\times 10^6 $ M$_\odot$ - $9\times 10^8 $ M$_\odot$, table 3 in][]{KH13}. 

We investigate two alternatives for the coupling efficiency: one is a constant efficiency, $\eta=0.05$ (fiducial constant or Fid-Constant), and the other is a variable coupling efficiency that depends on $\dot{m}$. The latter includes the fiducial variable coupling efficiency (Fid-Variable), and also coupling efficiencies with varying parameters as described in \se{implementation} (run for the cases $\MBH=10^6,10^7,10^8 $ M$_\odot$). In \Fig{eta_mdot}, we show how the coupling efficiency $\eta$ changes with the Eddington fraction $\dot{m}$  for all values of $N_\eta$ and $\alpha_\eta$ used in these two sets of simulations. For the cases varying $\alpha_\eta$ while fixing $\eta=0.1$ for $\dot{m}=0.1$, all lines cross the point (0.1,0.1) in $\dot{m}-\eta$ plane. Note that in all cases $\eta$ is capped at unity, as marked by the grey shaded area. A higher $N_\eta$ or a smaller $\alpha_\eta$ typically lead to a larger value of $\eta$, given that $\dot{m}$ is smaller than 0.1 for nearly all the time in our idealised simulations. Therefore, a monotonic trend between $\eta$ and $N_\eta$ or $\alpha_\eta$ always holds. The relationships between galaxy or BH properties and the parameters in our variable coupling efficiency model can thus be understood in terms of how their corresponding values of $\eta$ influence galaxy or BH properties.

\begin{figure*}
    \centering
    \includegraphics[width=1\textwidth]{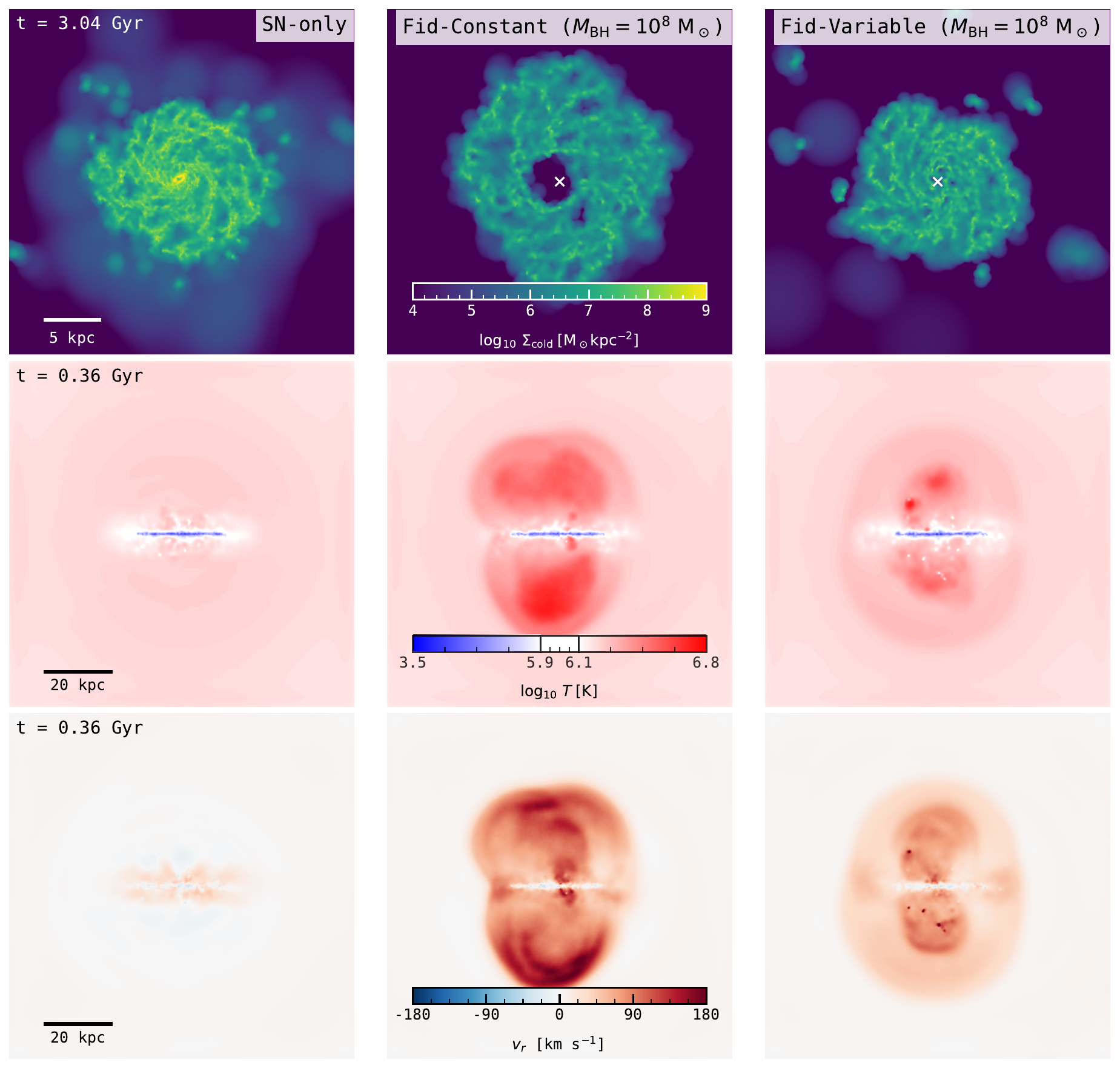}
    \caption{Cold gas surface density, gas temperature and gas radial velocity in different runs. From left to right, the columns represent the SN-only run, the run with $\MBH = 10^8$ M$_\odot$ using the fiducial constant coupling efficiency model, and the run with $\MBH = 10^8$ M$_\odot$ using the fiducial variable coupling efficiency model, respectively.
    \emph{Top row}: Face-on cool ($T\leq 8000$ K) gas surface density maps, each 30 kpc wide including all particles in projection, at $t=3.04$ Gyr. A 5 kpc scale bar and a colour bar indicating cold gas surface density are shown in the left and middle panels respectively. The white `X' in the second and third panels shows the position of the central BH.
 \emph{Middle row}: Edge-on mass-weighted, projection-averaged temperature maps, each 100 kpc wide including all particles in depth, at $t=0.36$ Gyr. A 20 kpc scale bar and a colour bar indicating gas temperature are shown in the left and middle panels respectively.  
 \emph{Bottom row}: Edge-on mass-weighted, projection-averaged radial velocity maps, each 100 kpc wide including all particles in depth, at $t=0.36$ Gyr. A 20 kpc scale bar and a colour bar indicating gas radial velocity are shown in the left and middle panels respectively. 
 }
    \label{fig:maps}
\end{figure*} 

We find that, while not shown here, in the SN-only run, the star formation rate increases during the first $\sim$0.3 Gyr, after which it stabilizes at a level of $\sim 2.5$ M$_\odot$ yr$^{-1}$. This behaviour arises because at the start of the simulation, SN feedback is not yet in equilibrium with star formation and cooling processes. To avoid confusion and to separate the effects of AGN feedback from SN feedback, we evolve the system from $t=0$ to $t=0.3$ Gyr with the BH mass set to zero. After this period, we set the BH mass at the galaxy centre to an initial mass of $M_{\rm BH} = 10^6$, $4\times10^6$, $10^7$, $10^8$, or $10^9\ $ M$_\odot$. Due to the absence of the BH during the first 0.3 Gyr, a significant amount of gas builds up in the galaxy centre. At this point, changing the BH mass suddenly leads to a temporarily high BH accretion rate (due to the high gas density). 
As a result, this can lead to the accretion of an unrealistically large amount of gas and, consequently, unrealistically strong AGN feedback during this first time-step. To prevent this artificial behaviour, BH accretion is disabled during the first time-step of the BH after it is placed into the simulation. After this time-step, we cap the BH accretion rate at the Eddington limit, until the BH accretion rate drops below the cap. After this, this temporary cap is removed. This acts as an additional safety check.

\section{Results}\label{sec:result}

\begin{figure*}
    \centering
    \includegraphics[width=0.75\textwidth]{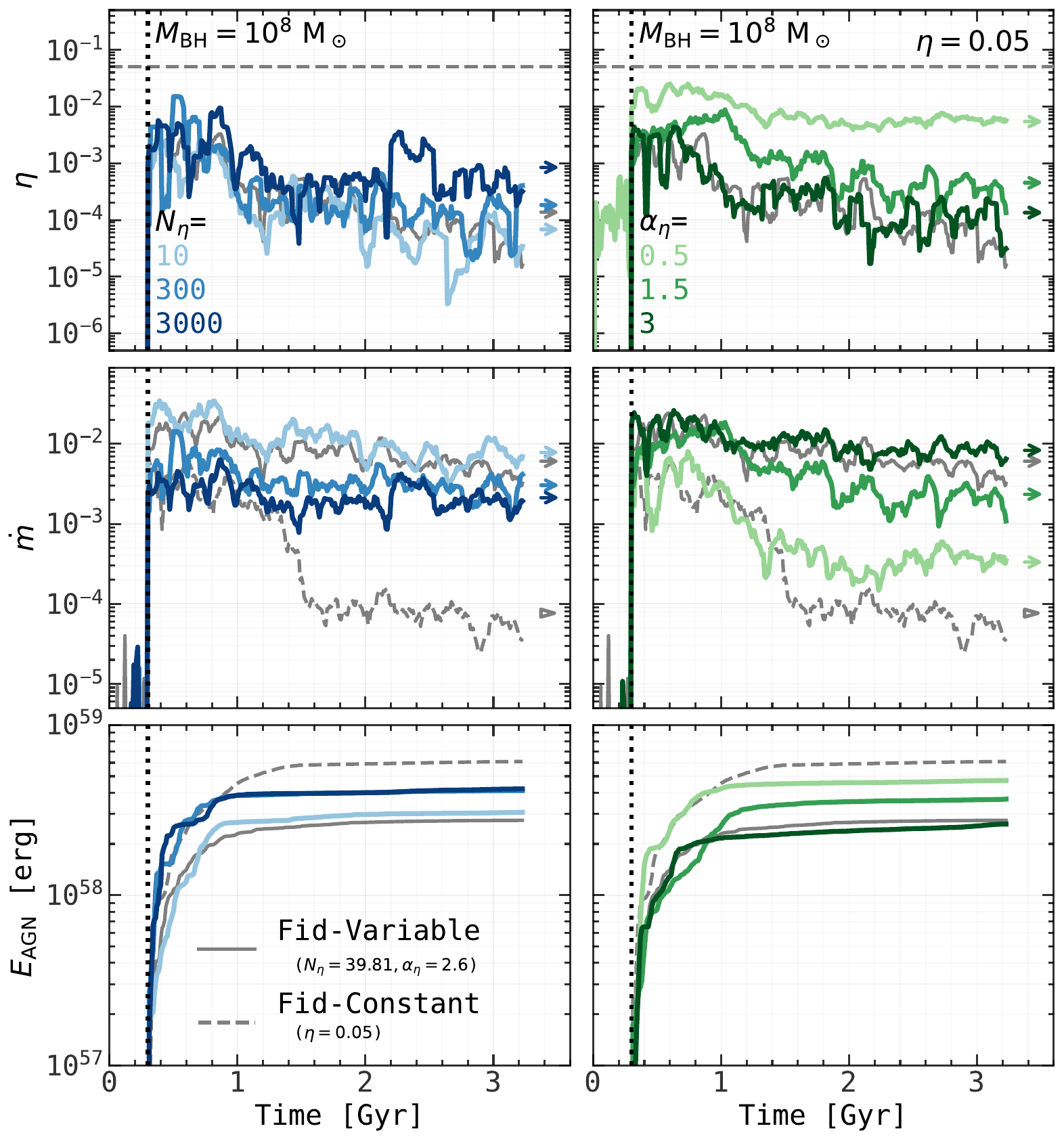}
    \caption{Effects of parameter variations in the variable coupling efficiency model on AGN properties. We show the  coupling efficiencies $\eta$ (first row), Eddington-normalised accretion rates $\dot{m}$ (second row), and cumulative AGN energy injection $E_{\rm AGN}$ (third row), for different times in runs with $\MBH=10^8$ M$_\odot$ using the variable coupling efficiency model with varying normalisations (first column) and varying slopes (second column). Variations in normalisation (slope) are represented by blue (green) colours, with darker shades corresponding to larger parameter values. The arrows or wedges in the right of the panels indicate the mean values averaged over the final 1.5 Gyr, with arrows denoting the fiducial variable coupling efficiency model and wedges denoting the fiducial constant coupling efficiency model. The values plotted are all averaged over 0.09 Gyr time intervals. The fiducial variable coupling efficiency model (solid lines) and fiducial constant coupling efficiency model (dashed lines) are shown by gray lines. The vertical dotted lines represent the time at which the BH is placed into the simulations. On average, higher values of $\eta$ lead to lower $\dot{m}$ and higher $E_{\rm AGN}$.}
    \label{fig:AGNprop_vary}
\end{figure*}

\begin{figure*}
    \centering
    \includegraphics[width=0.8\textwidth]{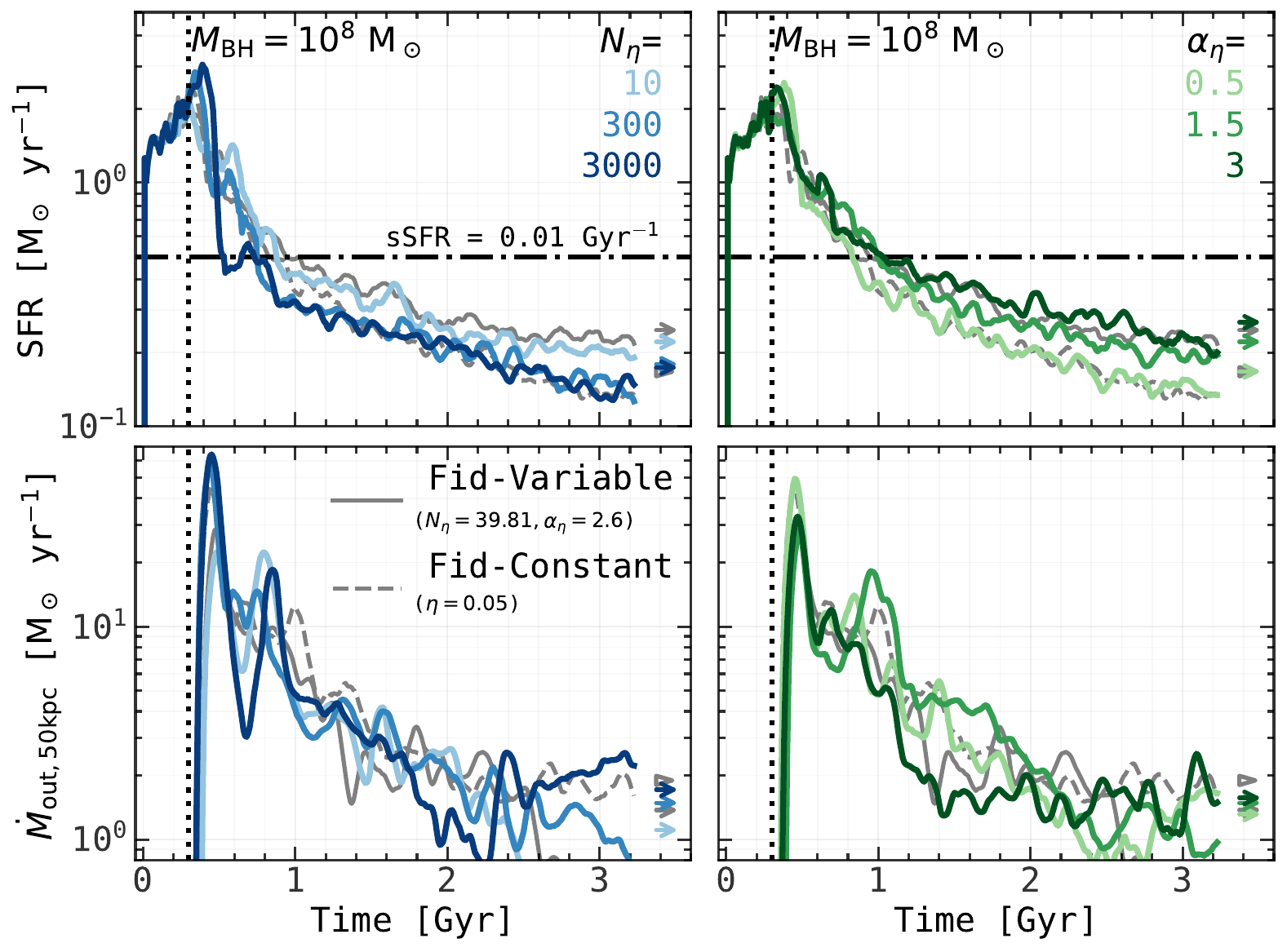}
    \caption{Effects of parameter variations in the variable coupling efficiency model on galaxy properties. We show the evolution of the galaxy SFR (first row) and the mass outflow rate $\dot{M}_{\rm out, 50~kpc}$ measured at a radius of 50 kpc (second row). The mass outflow rate is calculated for gas that is moving outwards with radial velocity $v_r > 40~{\rm km~s^{-1}}$. The black horizontal dot-dashed lines in the first row mark the quenching threshold of sSFR = 0.01 Gyr$^{-1}$ (or $\mathrm{SFR} \approx 0.49~\mathrm{M_\odot} ~\text{yr}^{-1}$, given that the stellar masses in these simulations change by less than 3 per cent).  The values plotted are all smoothed over 0.09 Gyr time intervals. Colours and linestyles remain the same as in \Fig{AGNprop_vary}. On average, larger values of $\eta$ result in lower SFRs and higher mass outflow rates.
    }
    \label{fig:Galaxyprop_vary}
\end{figure*}

In this section, we present the results concerning both BH and galaxy properties for different coupling efficiency models, using the ICs and workflow described above and evolving the systems for 3.3 Gyr. 

We first qualitatively compare the strength of feedback effects in a run including only SN feedback, and runs also including AGN feedback for a BH of mass $\MBH=10^8 $ M$_\odot$, using both coupling efficiency models.  \Fig{maps} shows images of the gas distribution at two different times in the simulations. In the upper row, we show the face-on cold gas ($T\leq 8000$ K) density maps at $t=3.04$ Gyr. The SN-only run shows prominent and dense spiral arms, as no AGN feedback is present to expel or heat up the cold gas. In contrast, the run with $\MBH=10^8 $ M$_\odot$\footnote{Movies showing the evolution of cold gas surface density and gas temperature maps for the fiducial runs and the SN-only run can be seen in \href{https://github.com/JinningLianggithub/VariableAGN_movie/}{https://github.com/JinningLianggithub/VariableAGN\_movie/}. Simulations with $\MBH\leq 10^7 $ M$_\odot$ produce temperature and density structures similar to the SN-only case while simulations with a higher BH mass are similar to the $\MBH=10^8$ M$_\odot$ cases but exhibit an even more pronounced hot gas cavity in the constant coupling efficiency case.} using the fiducial constant coupling efficiency model exhibits a central cavity in the cold gas disc, indicating recent gas expulsion by strong AGN feedback. The average gas density at late times is lower in the BH runs, as is seen later in \Fig{outflow_prof_mass}. The variable coupling efficiency run with the same $\MBH$ retains more central cold gas, suggesting a more moderate impact of AGN feedback. This is also evident from its lower AGN energy injection compared to the constant coupling efficiency run with the same $\MBH$, shown in \Fig{AGNprop_mass2}. In the middle row, we show edge-on temperature maps shortly after the BH is introduced at $t=0.36$ Gyr. The SN-only run lacks a hot gas ($T\sim10^7$ K) bubble in the polar directions, indicating that SN feedback alone is insufficient to heat significant volumes of gas to high temperatures. Notably, more warm gas ($T\sim10^6$ K) cools and falls back onto the cold gas disc in the SN-only run, as indicated by small white gas parcels, which will be further discussed in \se{outflow_infall}. In contrast, the AGN feedback runs produce large hot gas bubbles, particularly in the constant coupling efficiency case, where feedback heats the gas and drives powerful outflows along the polar axis. Finally, in the bottom row, we show edge-on radial velocity maps at the same time as in the second row. The SN-only run shows multiple outflow and inflow streams with small velocity magnitudes within a radius of $\sim$20 kpc. In contrast, the AGN runs show much larger outflow velocities, easily propagating to CGM scales. Moreover, the outflow velocity in the fiducial variable coupling efficiency model is smaller than in the fiducial constant coupling efficiency model. Note that, although we present only two snapshots as representative examples, these findings hold more generally for the duration of the simulations.


In the following subsections, we further explore the influence of our new variable coupling efficiency model. We will first demonstrate model differences in general BH properties, then for general galaxy properties, as functions of time. 
We begin by looking at the effects of varying the parameters in our new variable coupling efficiency model in \se{general_vary}, followed by comparing the fiducial variable and constant coupling efficiency models for different BH masses in \se{general_mass}. 

Following this, to further explore the impact of AGN feedback on the gas, we investigate the mass outflow profiles at different times across various models and BH masses in \se{outflow_infall}.
We also study evolution of total gas density profiles. Finally, we show the phase-space diagram of gas temperature vs. gas density for different models and BH masses in \se{phase}. 

 When evaluating parameter variations within the variable coupling efficiency model, we restrict our analysis to $M_{\rm BH} = 10^8$ M$_\odot$. Corresponding results for $M_{\rm BH} = 10^6$ M$_\odot$ and $M_{\rm BH} = 10^7$ M$_\odot$ are provided in \app{AGN_galaxy_prop}. For all of the figures below, instantaneous quantities, e.g. SFR and outflow rates, are averaged over 0.09 Gyr time intervals, centred at the time of interest.

\begin{figure*}
    \centering
    \includegraphics[width=0.8\textwidth]{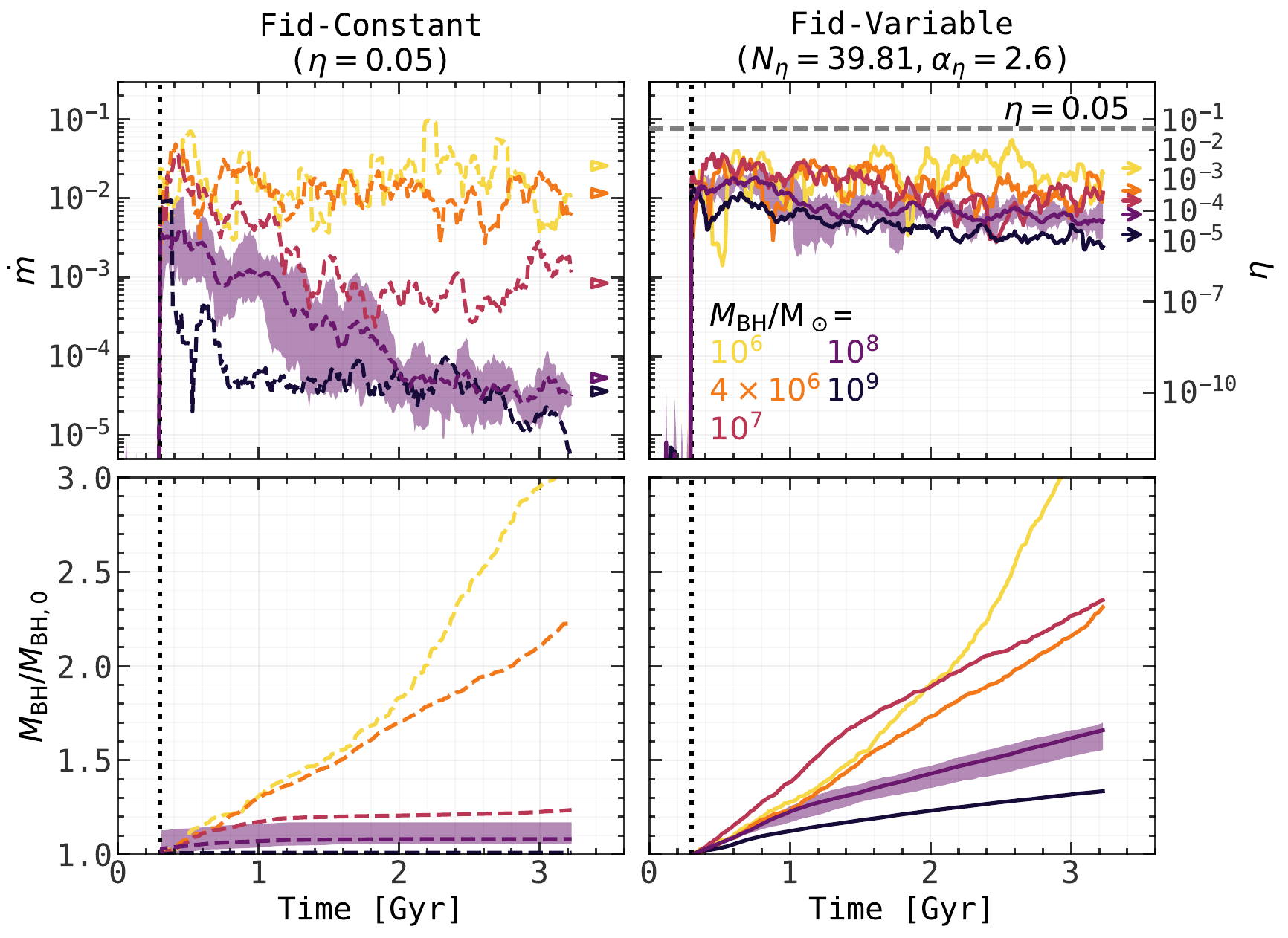}
    \caption{Eddington-normalised accretion rates (first row), and the BH masses normalised by the BH initial masses (second row) are shown at different times for the fiducial constant coupling efficiency model (dashed lines, first column) and the fiducial variable coupling efficiency model (solid lines, second column), for different values of BH mass. The corresponding coupling efficiencies, $\eta$, for the fiducial variable model ($\eta=N_\eta\dot{m}^{\alpha_\eta}$ with $\alpha_\eta= 2.6$ and $N_\eta = 39.81$) are shown on the secondary y-axis in the upper-right panel. The fiducial constant coupling efficiency ($\eta=0.05$) is indicated by a grey horizontal line. Different values of $M_{\rm BH}/$M$_\odot$ are represented by different colours: $10^6$ (yellow), $4\times 10^6$ (orange), $10^7$ (brown), $10^8$ (dark purple), $10^9$ (dark blue). The values plotted are all smoothed over 0.09 Gyr time intervals. The arrows or wedges in the right of the panels stand for the mean values averaged over the last 1.5 Gyr. The vertical dotted lines represent the time when the BH is placed into the simulation domain, i.e. $t=0.3$ Gyr. The purple shaded regions indicate the full run-to-run min--max scatter for the representative $\MBH=10^8~{\rm M}_\odot$ cases. For a given BH mass, the fiducial variable coupling efficiency model leads to smaller $\eta$, larger $\dot{m}$, and faster BH growth  compared to the fiducial constant one.}
    \label{fig:AGNprop_mass1}
\end{figure*}

\subsection{Effects of varying AGN feedback efficiency parameters}\label{sec:general_vary}
\subsubsection{AGN properties}
\Fig{AGNprop_vary} shows AGN properties including the coupling efficiency $\eta$, Eddington ratio $\dot{m}$ and cumulative AGN energy injection $E_{\rm AGN}$ for different AGN feedback efficiency  parameter variations. We first focus on the differences in $\eta$ in the first row. When examining parameter variations while fixing the BH mass at $M_{\rm BH}=10^8 $ M$_\odot$, we find that $\eta$ increases with higher normalisation $N_\eta$ or lower slope $\alpha_\eta$ (given that  $\dot{m}<0.1$ for most of the time), consistent with \Fig{eta_mdot}. 
The fiducial variable model produces intermediate $\eta$ values, as both its $N_\eta$ and $\alpha_\eta$ parameters lie between the extremes tested in the parameter variations. 
In particular, all cases show much smaller values of $\eta$ compared to the fiducial constant coupling efficiency ($\eta=0.05$).

 Next, we compare differences in Eddington-normalised accretion rate $\dot{m}$ in the second row of \Fig{AGNprop_vary}. We find that lower slope $\alpha_\eta$ or higher normalisation $N_\eta$ lead to lower $\dot{m}$ given the same BH mass $\MBH$. Since $\MBH$ does not change much (we find that the maximum change in $\MBH$ is no more than a factor of two), the variation in $\dot{m}\propto \dot{M}_{\rm acc}/\MBH\propto f_{\rm TV}\MBH\rho_{\rm gas}/c_{\rm s}^3$ is dominated by the variation in gas properties, i.e. $f_{\rm TV}\rho_{\rm gas}/c_{\rm s}^3$ (see \Eqnb{acc}). If the coupling efficiency $\eta$ is higher, as a result of $N_\eta$ increasing or $\alpha_\eta$ decreasing, the feedback will be stronger. As a result, more gas is expelled and the gas density $\rho_{\rm gas}$ is smaller, which leads to lower $\dot{m}$. Although the fiducial variable coupling efficiency model lies between these variations, the fiducial constant coupling efficiency model shows the lowest $\dot{m}$ because of its larger $\eta$ and therefore stronger feedback. In particular, $\dot{m}$ only drops over time from its peak value by $\lesssim$0.7 dex for the fiducial variable model, compared to $\sim3$ dex in the constant efficiency model.
As a consequence, models with higher accretion rates, e.g. the variable coupling efficiency models, particularly ones with smaller normalisation $N_\eta$ or larger slope $\alpha_\eta$, lead to faster BH growth.

In the last row of \Fig{AGNprop_vary}, we compare the cumulative AGN energy injection $E_{\rm AGN}$. We find that a lower slope $\alpha_\eta$ or a higher normalisation $N_\eta$ leads to higher $E_{\rm AGN}$ given the same $\MBH$, which can be understood from the scaling relation $E_{\rm AGN} \propto \int \eta \dot{m}  \dd t\propto \int N_\eta \dot{m}^{\alpha_\eta+1}\dd t$. When varying $\alpha_\eta$, the scaling relation becomes $E_{\rm AGN} \propto \int 10^{\alpha_\eta} \dot{m}^{\alpha_\eta + 1} \dd t$, which implies that $E_{\rm AGN}$ decreases with increasing $\alpha_\eta$, provided that $\dot{m} < 0.1$. In contrast, when varying $N_\eta$, the relation becomes $E_{\rm AGN} \propto \int N_\eta \dot{m}^{3.6} \dd t$, indicating that $E_{\rm AGN}$ increases with increasing $N_\eta$. However, the differences between the different variations are quite small, no more than 0.3 dex. In particular, the case with $\alpha=0.5$ is very close to the case with $\alpha=1.5$ at later times. This is explained by the first two rows of \Fig{AGNprop_vary}, which display the trends of  $\eta$ and  $\dot{m}$. As one can see, lower  $\alpha_\eta$ or higher  $N_\eta$ lead to lower $\dot{m}$ and higher  $\eta$ at the same $\MBH$. The differences in $\dot{m}$ and $\eta$ tend to compensate, resulting in similar values of $E_{\rm AGN}$. Such compensation highlights the tendency for the AGN energy injection to self-regulate. Here, self-regulation means that, for a fixed galaxy and halo system, the self-regulated $\dot{M}$ and $E_{\rm AGN}$ are less sensitive to $\MBH$ and $\eta$, resulting in a smaller spread across $\MBH$ than in the constant-efficiency model. Therefore, $\dot{m}$ adjusts to produce similar AGN energy injection in runs with different coupling efficiencies \citep{Booth09}.

\subsubsection{Galaxy properties}\label{sec:galaxy_vary}

In \Fig{Galaxyprop_vary} we show the effects of varying the parameters in the variable coupling efficiency on the galaxy SFR and the mass outflow rate measured at a radius of 50 kpc, $\dot{M}_{\rm out, 50 ~kpc}$. 
The SFR provides the most direct evidence of how AGN feedback influences galaxy growth. 
As shown in the first row of \Fig{Galaxyprop_vary}, a lower slope $\alpha_\eta$ or a larger normalisation $N_\eta$ lead to stronger suppression of star formation. This trend is consistent with the AGN energy differences shown in \Fig{AGNprop_vary}, where lower $\alpha_\eta$ or larger $N_\eta$ yield higher $\eta$ and larger AGN energy injection. The trends in stellar mass growth closely follow the trend of the SFR, i.e. higher $\eta$ results in slower stellar mass buildup. 

\begin{figure*}
    \centering
    \includegraphics[width=0.8\textwidth]{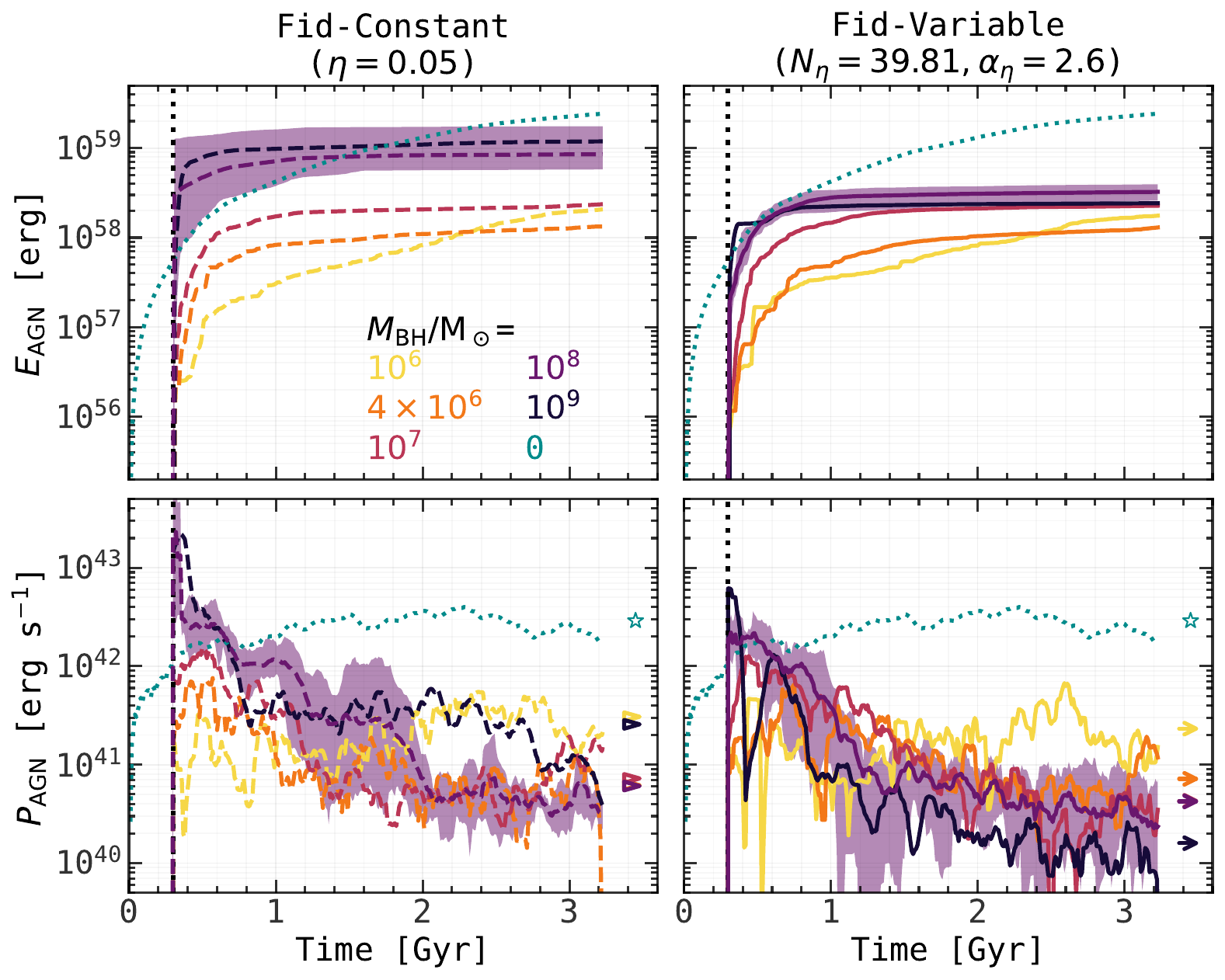}
    \caption{As \Fig{AGNprop_mass1} but showing the cumulative AGN energy injection $E_{\rm AGN}$ (first row) and instantaneous AGN energy injection rate $P_{\rm AGN}$ (second row) versus time. As a reference, we also show by dotted dark cyan lines the SN feedback energy injection and feedback powers from the SN-only run (i.e. $M_{\rm BH}=0$ M$_\odot$). This figure illustrates that the fiducial variable coupling efficiency model leads to slightly smaller $E_{\rm AGN}$ and $P_{\rm AGN}$ compared to the fiducial constant one given the same BH mass. The SN-only run shows much higher feedback energy injection and power at late times. The smaller differences in $E_{\rm AGN}$ in the variable coupling efficiency model ($\approx$0.4 dex compared to $\approx$0.9 dex in the fiducial constant coupling efficiency model) indicate better BH self-regulation than in the fiducial constant efficiency model. Note that the SN energy injection also differs between the AGN runs. }
    \label{fig:AGNprop_mass2}
\end{figure*}

A galaxy at redshift $z\sim0$ is usually defined to be quenched when its specific SFR (${\rm sSFR}={\rm SFR}/M_*$) drops below $10^{-11}$~yr$^{-1}$ \citep[e.g.][]{Wetzel13}, which corresponds approximately to an SFR of $0.49$~M$_\odot$~yr$^{-1}$ when using the initial stellar mass in our setup, given that the stellar mass does not change much over the duration of these simulations. The quenching time scale, which we define as the time from when the BH mass is inserted at 0.3~Gyr up to when the SFR drops below the threshold (horizontal dotted line), spans the range from 0.5 (0.3) Gyr to 0.7 (0.5) Gyr for $\alpha_\eta$ ($N_\eta$)-variations, respectively.

The mass outflow rate $\dot{M}_{\rm out}$ is another clear indicator of the AGN feedback strength. Stronger AGN feedback will push more gas outward in the ISM and into the CGM with faster velocities. In our analysis, $\dot{M}_{\rm out}$ is defined as the mass outflow rate through a sphere, calculated as follows, by considering all gas particles with a positive radial velocity that is larger than a threshold value $V_{\rm th}$, within a spherical shell of width $\Delta r$,
\be\label{eq:Mout}
\dot{M}_{\rm out}=\sum_i^{v_{r,i}>V_{\rm th}}\frac{m_iv_{r,i}}{\Delta r},
\ee
where $m_{i}$ is the mass of the $i$-th gas particle and $V_{\rm th}$ is chosen as a quarter of the maximum of the circular velocity curve of the DM halo, which is 40 km s$^{-1}$ \citep[e.g.][]{Mitchell20}. The velocity threshold is introduced because gas particles moving with small radial velocities (positive or negative) typically represent sloshing motions rather than systematic outflows or inflows, so that including them tends to add noise to the outflow rate (and profiles below). Moreover, in the SN-only run and in the run with $\MBH = 10^6~{\rm M}_\odot$, some gas motions on CGM scales are not driven by AGN or SN feedback. Excluding these particles allows us to better isolate and analyze the effects of feedback.  

In the second row of \Fig{Galaxyprop_vary}, we show the mass outflow rate measured at a radius of 50 kpc (CGM scale), $\dot{M}_{\rm out, 50~ kpc}$ as a function of time. The shell width $\Delta r$ is set to 1 kpc, and we neglect times when the shell contains fewer than 10 particles. When comparing different parameter variations for the variable coupling efficiency models, we find that a higher normalisation $N_\eta$ or a smaller slope $\alpha_\eta$ lead to higher mass outflow rates $\dot{M}_{\rm out, 50 ~kpc}$, indicating stronger AGN feedback is more effective at driving gas into CGM scales. This is clear at all times, although large fluctuations are seen in some cases at late times. We will further explore outflow behaviours in \se{outflow_infall}.

\subsection{Effects of varying black hole mass}\label{sec:general_mass}

\subsubsection{AGN properties}
\begin{figure*}
    \centering
    \includegraphics[width=0.75\textwidth]{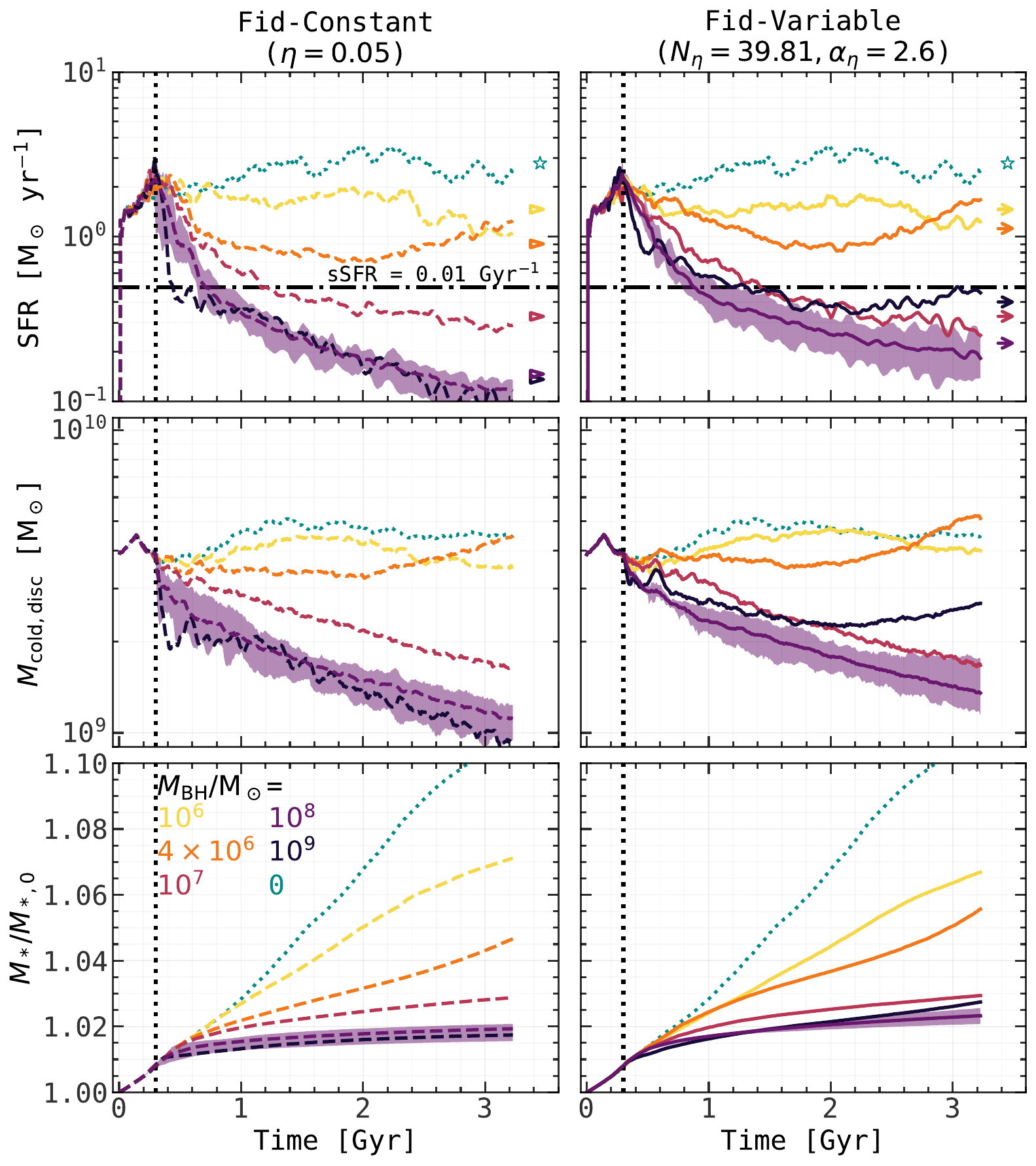}
    \caption{Evolution of the galaxy SFR (first row), cold gas mass in the disc (second row), and stellar mass growth (third row). The dark cyan dotted lines show the results from the SN-only run, while the dark cyan stars in the right of the panels indicate the mean values of the SFR in the SN-only run over the last 1.5 Gyr. The black horizon dot-dashed lines in the first row mark the quenching threshold (sSFR$=0.01$ Gyr$^{-1}$ or SFR$\approx $0.49 M$_\odot$ yr$^{-1}$) (given that the stellar masses for these runs change by less than 10\%). Everything else remains the same as \Fig{AGNprop_mass1}. For a given BH mass, the evolution of the SFR, cold disc gas mass and stellar mass
    are similar in the variable and constant coupling efficiency models.}
    \label{fig:galaxyprop_mass1}
\end{figure*}

\begin{figure*}
    \centering
    \includegraphics[width=0.8\textwidth]{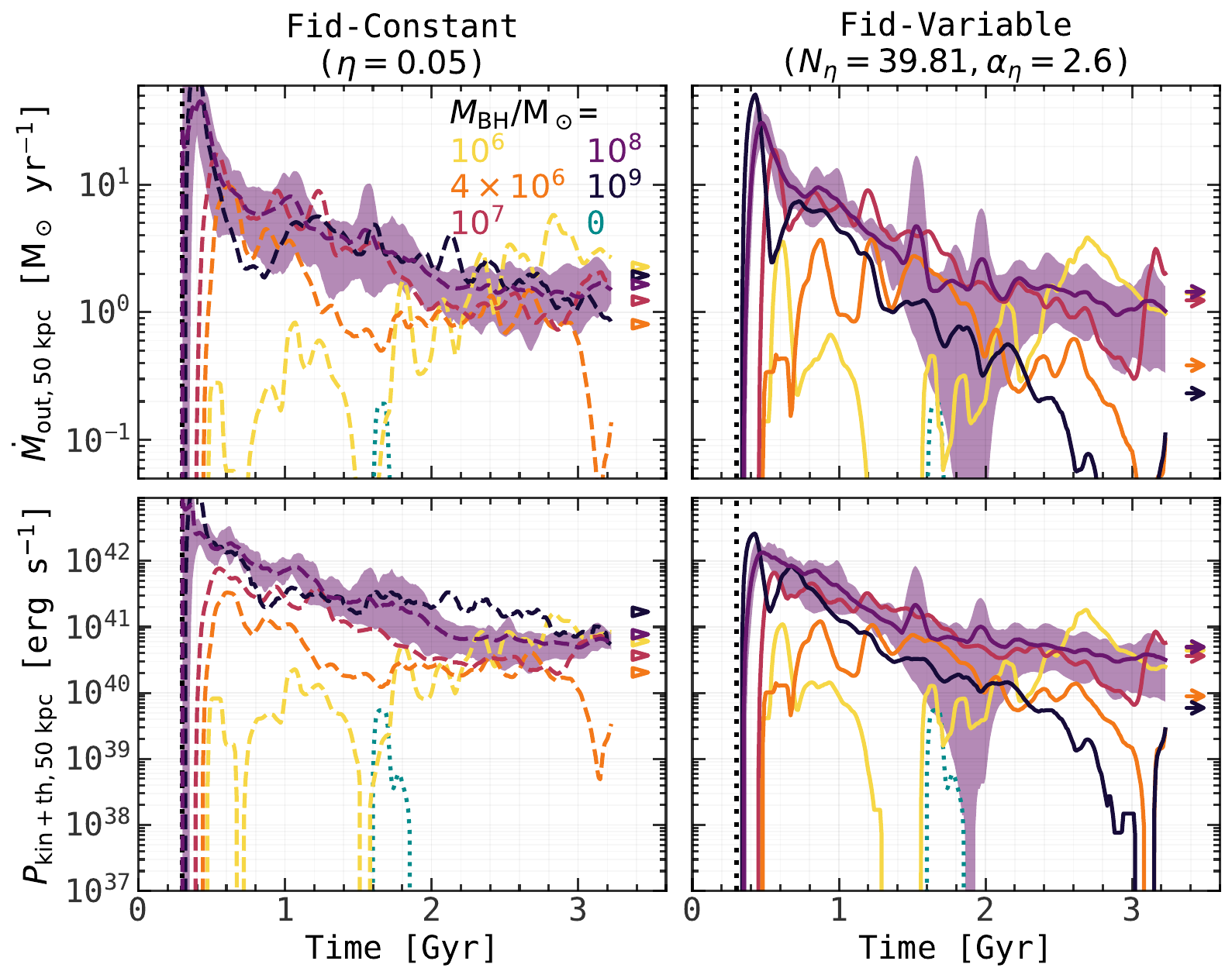}
    \caption{As \Fig{AGNprop_mass1}, but showing the mass outflow rate (\Eqnb{Mout}, top row), and energy outflow rate (\Eqnb{Pout}, bottom row), measured at a radius of 50 kpc. Both outflow rates are calculated for gas that is moving outwards with radial velocity $v_r > 40~{\rm km~s^{-1}}$. The fiducial constant and variable efficiency models produce similar mass and energy outflow rates at late times, except for the $M_{\rm BH} = 10^9~\mathrm{M_\odot}$ case where the variable coupling efficiency yields systematically lower outflow rates.
    }
    \label{fig:galaxyprop_mass2}
\end{figure*}

In this section, we show how various predicted properties depend on BH mass, $\MBH$, for both coupling efficiency models. We performed five repeated simulations with $\MBH=10^8~\mathrm{M}_\odot$ for both coupling efficiency models to quantify the run-to-run min–max scatter, which is shown as shaded regions in \Fig{AGNprop_mass1}–\Fig{galaxyprop_mass2}.
The evolution of AGN properties, including the coupling efficiency $\eta$, Eddington-normalised accretion rate $\dot{m}$, and the increase in BH mass relative to its initial value, $\MBH/M_{\rm BH,0}$, are shown in \Fig{AGNprop_mass1}. We first focus on the differences in $\dot{m}$, shown in the first row of \Fig{AGNprop_mass1}. The trends with $\dot{m}$ differ markedly between the fiducial variable and the constant efficiency models. Although both coupling efficiency models show that higher BH masses generally lead to lower $\dot{m}$, the variation in $\dot{m}$ across different $\MBH$ values is significantly smaller in the variable coupling efficiency model compared to the fiducial constant coupling efficiency model. Moreover, the constant efficiency cases exhibit much lower values of $\dot{m}$ and steeper declines in $\dot{m}$ at late times for $\MBH \geq 10^7$ M$_\odot$, with drops exceeding 1-2 orders of magnitude. Note that higher BH masses generally lead to lower raw accretion rates $ \dot{M}_{\rm acc}$, with the exception of the case with $M_{\rm BH}=10^9~{\rm M}_\odot$ using fiducial constant coupling efficiency. The fiducial constant coupling efficiency cases have accretion rates ranging from $2\times 10^{-4}$ to $2\times 10^{-3}$ M$_\odot$ yr$^{-1}$, while the fiducial variable coupling efficiency cases show values from $3\times 10^{-3}$ to $2\times 10^{-2}$ M$_\odot$ yr$^{-1}$.

Similar to the analysis in the previous section, the trends in the two fiducial coupling efficiency models can be understood as resulting from AGN feedback self-regulation. After placing a BH and turning on AGN feedback at $t=0.3$ Gyr, more massive BHs drive more powerful outflows, reducing gas densities and increasing gas velocities. These effects ultimately cause a drop in $\dot{m}$ with increasing $\MBH$. In the fiducial constant efficiency case with $\eta = 0.05$, the persistent strong feedback prevents efficient gas replenishment, causing $\dot{m}$ to decline by more than an order of magnitude.  In contrast, the variable coupling efficiency models provide stronger self-regulation because $\eta$ decreases when $\dot{m}$ drops due to gas depletion. As shown by the double y-axis in the upper right panel of \Fig{AGNprop_mass1}, the lower values of $\eta$ (much smaller than 0.05) allow more gas to fall back and fuel the BH, leading to an accretion rate that varies less with time and is higher at late times. This is the most prominent and important feature in our new model, reflecting better BH self-regulation. In particular, $\eta$ is always lower than in the fiducial constant efficiency model, and higher $\MBH$ leads to lower $\eta$.

In the second row of \Fig{AGNprop_mass1}, we show the increase in $\MBH$, relative to its initial value. We find that low-mass BHs generally tend to grow more compared to high-mass BHs in terms of $\MBH/M_{\rm BH,0}$. One exception is the $M_{\rm BH}=10^7$ M$_\odot$ case using the fiducial variable coupling efficiency model, which shows faster growth compared to the $M_{\rm BH}=4\times 10^6$ M$_\odot$ case using the same coupling efficiency model. This is potentially due to run-to-run scatter, as illustrated for the $M_{\rm BH}=10^8$ M$_\odot$ case by purple shaded bands.
We also examine the values of the mass increments. For both of the fiducial coupling efficiency models, the mass accreted by the BH increases with the BH mass. In the $\MBH = 10^9$ M$_\odot$ case, the increment is $3.36 \times 10^{8}~ \mathrm{M}_\odot$ ($0.94 \times 10^{7}~ \mathrm{M}_\odot$), while in the $\MBH = 10^6$ M$_\odot$ case, the increment is $2.29 \times 10^{6}~ \mathrm{M}_\odot$ ($2.08 \times 10^{6}~ \mathrm{M}_\odot$) using the fiducial variable (constant) coupling efficiency model. The higher late-time $\dot{m}$ leads to faster BH growth for all $\MBH$, in the fiducial variable coupling efficiency model. For all runs, the BH mass increases by no more than a factor $\sim 3$. 
Cosmological simulations might show larger differences due to more realistic accretion rates. Note that if the gas properties in the Bondi--Hoyle accretion formula \Eq{acc} were to remain unchanged, BHs would grow even faster than exponentially. 
At the other extreme, very massive BHs can quench star formation in their host galaxies, causing their growth to saturate. The case of $\MBH = 4 \times 10^6$ M$_\odot$ with the fiducial variable coupling efficiency model appears to lie between these two regimes, highlighting the mass at which self-regulation becomes important in this model.

In \Fig{AGNprop_mass2}, we compare the cumulative AGN energy injection $E_{\rm AGN}$ and AGN feedback power $P_{\rm AGN}$ for different BH masses, with the cumulative SN energy injection and feedback power in the SN-only run also shown as a reference. Note that the SN energy injection and feedback powers shown here, as well as in figures below, represent the combined contributions from both CC SNe and SNIa. In the first row, one can see that for both coupling efficiency models, low mass BHs inject energy gradually over time, while high mass BHs tend to inject a large amount of energy at the beginning after the BH is inserted. Higher BH masses lead to higher cumulative energy injection values in both models. This behaviour can be understood from the equation $E_{\rm AGN}=\int \epsilon_{\rm r}\eta\dot{M}_{\rm acc}c^2\dd t$. In the variable coupling efficiency case, this scales as $E_{\rm AGN}\propto \int N_\eta \dot{m}^{\alpha_\eta+1}\MBH \dd t$, whereas in the constant coupling efficiency case this simplifies to $E_{\rm AGN}\propto \int 0.05\dot{m}\MBH \dd t$. Although the $\dot{m}^{\alpha_\eta+1}$ or $\dot{m}$ factor tends to decrease when $\MBH$ is increased, the variation in $\MBH$ has the larger effect on $E_{\rm AGN}$. 
We also see that the energy released by SN feedback in the SN-only run is larger than the AGN feedback energy injection at late times. Note that the SN energy injection also differs between the AGN runs, since SN and AGN feedback compete with one another. This will be discussed in \se{powercomparison}.

However, comparing the two fiducial models, $E_{\rm AGN}$ at late times varies less with $\MBH$ in the variable coupling efficiency model due to the nonlinear dependence on $\dot{m}$ (as $\dot{m}^{\alpha_\eta +1}$). This dependence compensates the changes in $\MBH$ more than in the constant coupling efficiency model.
Physically, this reflects more effective self-regulation in the variable coupling efficiency model, where the feedback strength adjusts more dynamically in response to the local gas conditions and accretion rate.

In the evolution of AGN power, $P_{\rm AGN}$, shown in the second row of \Fig{AGNprop_mass2}, both coupling efficiency models show that the $\MBH=10^6$ M$_\odot$ cases exhibit a stable or slightly increasing trend of power with time at early times, while in the larger $\MBH$ cases $P_{\rm AGN}$ declines after an initial peak. This contrast arises because high-mass BHs typically trigger strong early AGN feedback, rapidly expelling surrounding gas and thus limiting later accretion. In contrast, low-mass black holes generate weaker feedback events initially, allowing them to retain more of their surrounding gas. This sustained gas supply supports fast BH growth and, gradually, stronger AGN feedback over time.

Comparing the two coupling efficiency models, higher $\MBH$ leads to lower AGN power and a sharper drop in the AGN power at late times for the fiducial variable coupling efficiency model. In contrast, the fiducial constant coupling efficiency model shows the $\MBH =  10^9 $ M$_\odot$ case tends to exhibit higher AGN power at late times compared to low-mass BHs. At early times, the two fiducial coupling efficiency models show similar powers for the same BH mass. At late times, the variable coupling efficiency model produces comparable or slightly higher power for low-mass BHs but lower power for high-mass BHs.


\subsubsection{Galaxy properties}

Next we compare several galaxy properties for the two fiducial coupling efficiency models. \Fig{galaxyprop_mass1} shows the evolution of the SFR, the cold gas mass in the disc, and the stellar mass relative to its initial value, $M_*/M_{*,0}$. In the first row of \Fig{galaxyprop_mass1}, the SN-only run shows a modest increase in the SFR with time, while AGN runs show stronger suppression of the SFR for larger $\MBH$, consistent with the AGN energy injection trend in \Fig{AGNprop_mass2}. The $\MBH=10^9$ M$_\odot$ variable coupling efficiency model shows a late-time SFR increase, likely due to run-ro-run stochastic variations. At 3.3 Gyr, the fiducial constant coupling efficiency model  yields smaller SFR values for $M_{\rm BH}\geq 10^7$ M$_\odot$, compared to the fiducial variable coupling efficiency model. The cases of lower $M_{\rm BH}$ remain star-forming  even after 3 Gyr of AGN feedback, although the SFR is lower than in the SN-only run. In contrast, for intermediate to high mass BH cases ($\MBH\geq 10^7 $ M$_\odot$), the galaxy is quenched at 3.3 Gyr due to more powerful and efficient AGN feedback. 

The quenching time scales (see definition in \se{galaxy_vary}) for $\MBH=10^7, 10^8, 10^9 $ M$_\odot$ are approximately 1.1 (0.9), 0.7 (0.5), 0.9 (0.1) Gyr, respectively, for the fiducial variable (constant) coupling efficiency model. The shorter quenching times and faster SFR decline in the constant coupling efficiency model indicate its stronger AGN feedback at early times.

In the second row of \Fig{galaxyprop_mass1}, we show the evolution of the cold gas mass in the disc, where the disc is defined as a cylinder with a radius of 10 kpc and a height (full thickness) of 2 kpc. Cold gas is identified as gas with a temperature lower than $10^4$ K\footnote{The chosen temperature threshold reflects the initial conditions, where the gas consists of a cold disc ($T = 10^{4}$ K) and a hot CGM ($T \gtrsim 10^{6}$ K). As the simulation evolves, however, the gas exhibits a broader temperature distribution, as shown in \se{phase}. It is therefore common to classify gas with $T \lesssim 8000$ K as cool gas (without geometric constraints) and gas with $8000~ \text{K} \lesssim T \lesssim 10^{6} ~ \text{K}$ as warm gas (including both the warm CGM and warm ISM), which differs from our definition here.}. The trends follow those of the SFR, since star-forming gas mainly consists of cold gas in the disc. As a consequence of star formation, the stellar mass increases over time, as shown in the third row of \Fig{galaxyprop_mass1}, following similar trends to those seen for the SFR, i.e. lower BH masses (or the absence of a BH) result in a more rapid stellar mass buildup. When comparing the two coupling efficiency models, we find that the stellar mass growth is similar. 
In particular, the differences in stellar mass growth among all runs are within a factor of 1.1 (or a factor of 1.07 excluding the SN-only run).

\Fig{galaxyprop_mass2} shows the evolution of the mass outflow rate, $\dot{M}_{\rm out, 50 ~kpc}$, and the total (kinetic and thermal) energy outflow rate, $P_{\rm kin+th, 50~kpc}$, measured at a radius of 50 kpc.
In our analysis, $P_{\rm kin+th, 50~kpc}$ is defined as $P_{\rm kin,50~kpc} + P_{\rm th,50~kpc}$, where $P_{\rm kin,50~kpc}$ and $P_{\rm th,50~kpc}$ represent the kinetic and thermal outflow powers respectively. These two quantities are calculated as follows, by considering the same particles and the same shell width as for $\dot{M}_{\rm out, 50 ~kpc}$ in \Eq{Mout},
\be\label{eq:Pout}
P_{\rm kin}=\sum_i^{v_{r,i}>V_{\rm th}}\frac{\frac{1}{2} m_i V_{i}^2 \, v_{r,i}}{\Delta r},
\ee
and 
\be\label{eq:Pth}
P_{\rm th}=\sum_i^{v_{r,i}>V_{\rm th}}\frac{m_iu_iv_{r,i}}{\Delta r},
\ee
where $V_i$ is the total velocity magnitude of the gas particle, $u_i$ is its specific internal energy and $V_{\rm th}$ is the same as defined in \Eq{Mout}.

In \Fig{galaxyprop_mass2}, the SN-only run shows a brief spike in $\dot{M}_{\rm out, 50 ~kpc}$ and $P_{\rm kin+th, 50~kpc}$  at $\sim$1.7 Gyr. At other times, there are no fast outflowing ($v_{r,i}>V_{\rm th}$) gas particles at 50~kpc. This indicates that the SN feedback has only a weak and transient effect in pushing gas to CGM scales. 
All other runs show that both $\dot{M}_{\rm out, 50 ~kpc}$ and $P_{\rm kin+th, 50~kpc}$ increase at early times, with a steeper rise for higher $\MBH$, followed by a decline at later times for runs with $\MBH\geq 10^6~{\rm M}_\odot$. The initial increase indicates strong AGN driven outflows while the later decrease results from reduced gas supply. For $\MBH=10^6$ M$_\odot$, both quantities increase more gradually and reach similar late-time levels, indicating that low-mass BHs require longer times to affect the CGM due to weaker feedback. This indicates that low-mass BHs need a long time to influence gas at large radii, i.e. at CGM scales, due to weak AGN feedback.

The mass and energy outflow rates at 50 kpc for lower-mass BHs ($\MBH\leq4\times 10^6$ M$_\odot$) exhibit more episodic fluctuations compared to high-mass BHs, with a characteristic timescale of approximately 0.2-0.4 Gyr. This behaviour can be understood as the effect of a galaxy fountain process: AGN feedback launches winds that push gas out to the CGM scale. 
Before the next powerful AGN episode occurs, the gas begins to fall back and is re-accreted by the galaxy leading to a decline in the outflow rates.
This signal is eliminated in the high BH mass regime, where stronger AGN-driven outflows can expel gas beyond 50 kpc, resulting in a smaller fraction of the gas falling back and fueling further strong feedback. This is what can also be seen in \Fig{maps}.

\begin{figure*}
    \centering
    \includegraphics[width=1.0\textwidth]{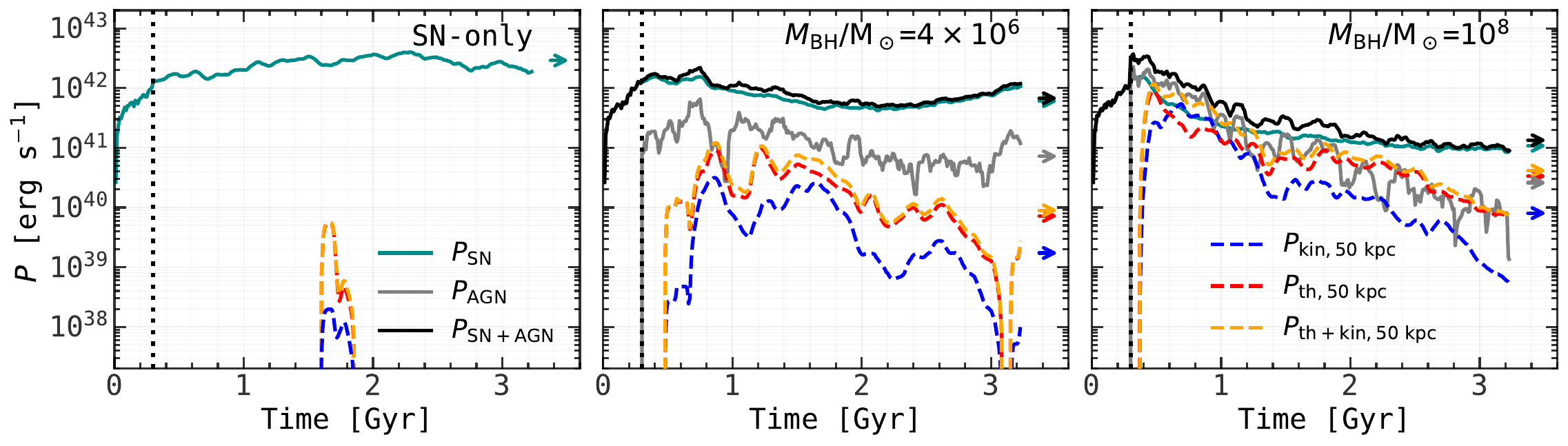}
    \caption{Different powers, $P$, as a function of time for the SN-only run (left panel) and for runs with $\MBH/\text{M}_\odot = 4\times 10^6$ (middle panel) and $10^8$ (right panel). Six types of power are shown: gas kinetic outflow power, $P_{\rm kin, 50~kpc}$ (blue lines), gas thermal outflow power, $P_{\rm th, 50~kpc}$ (red lines), total gas outflow power, $P_{\rm kin+th, 50~kpc}$ (orange lines), SN feedback power, $P_{\rm SN}$ (dark cyan lines), AGN feedback power, $P_{\rm AGN}$ (gray lines), and total feedback power, $P_{\rm SN+AGN}$ (black lines). All gas outflow powers are measured at a radius of 50~kpc, for gas with outward radial velocity $v_r>40~{\rm km~s^{-1}}$. The values plotted are all smoothed over 0.09~Gyr time intervals. The arrows at the right of each panel indicate the mean values averaged over the last 1.5~Gyr. All simulations shown use the fiducial variable coupling efficiency model. The powers in simulations with the fiducial constant coupling efficiency model are similar to the fiducial variable coupling efficiency model.}
    \label{fig:powers}
\end{figure*}

\begin{figure}
    \centering
    \includegraphics[width=1.\columnwidth]{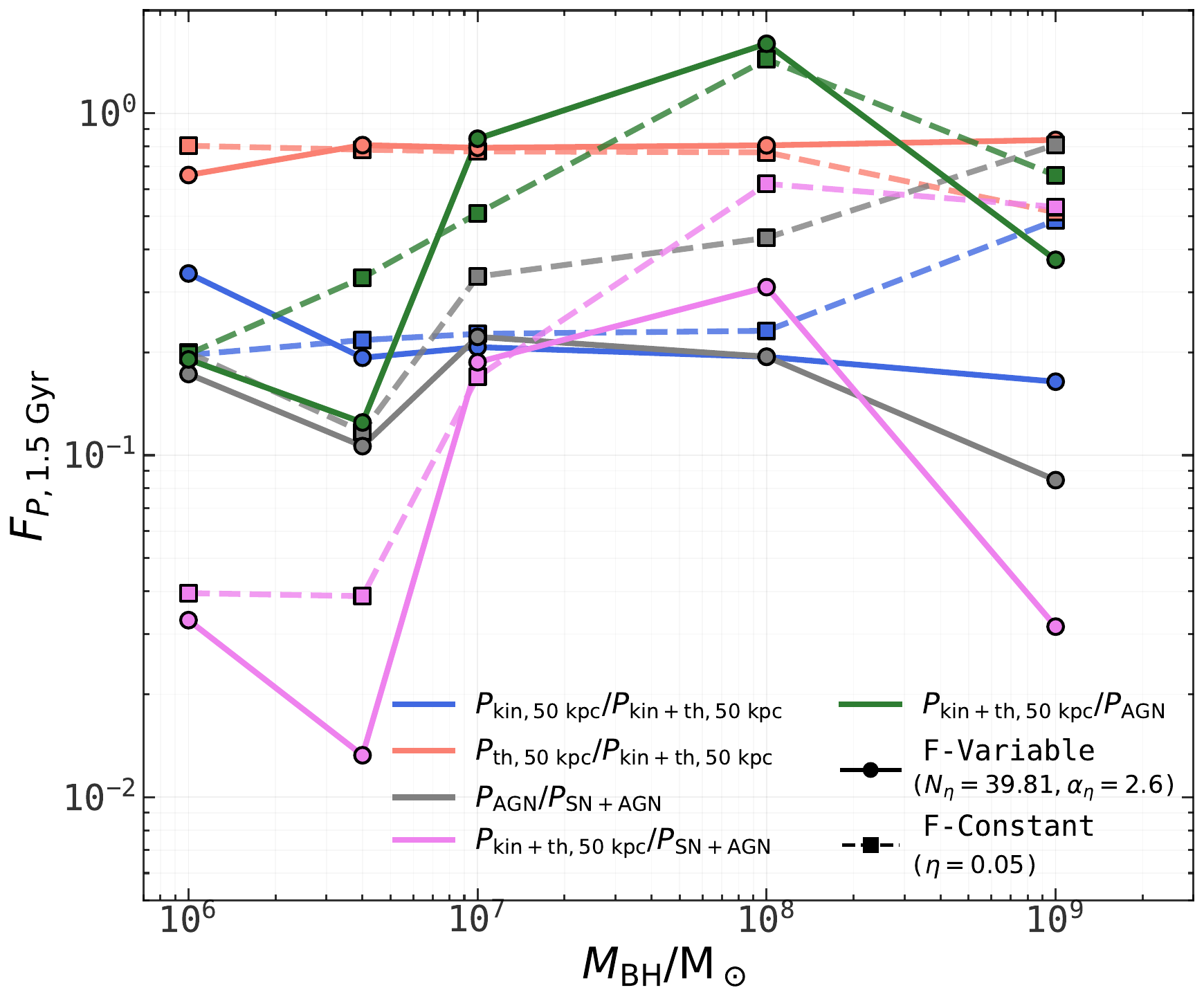}
    \caption{Various ratios of injection and outflow powers, computed as ratios of mean powers averaged over the last 1.5 Gyr ($F_{P, \rm 1.5~Gyr}$), as a function of BH mass for the fiducial variable coupling efficiency model (solid lines) and the fiducial constant coupling efficiency model (dashed lines). Five different power ratios are presented: the ratio of gas kinetic to total gas outflow power, $P_{\rm kin, 50~kpc}/P_{\rm kin+th, 50~kpc}$ (blue lines), the ratio of gas thermal to total gas outflow power, $P_{\rm th, 50~kpc}/P_{\rm kin+th, 50~kpc}$ (red lines), the ratio of AGN feedback to total feedback power, $P_{\rm AGN}/P_{\rm SN+AGN}$ (gray lines), the ratio of total gas outflow to total feedback power, $P_{\rm kin+th, 50~kpc}/P_{\rm SN+AGN}$ (pink lines), and the ratio of total gas outflow to AGN feedback power, $P_{\rm kin+th, 50~kpc}/P_{\rm AGN}$ (green lines). All gas outflow powers are measured at a radius of 50 kpc. }
    \label{fig:powers_ratios}
\end{figure}

\subsubsection{Comparisons of feedback and outflow powers}\label{sec:powercomparison}

To further investigate how the different outflow and feedback powers evolve and relate to each other, we show the various powers as functions of time in \Fig{powers}, and ratios of powers as functions of BH mass in \Fig{powers_ratios}.

\Fig{powers} shows six different powers, namely kinetic ($P_{\rm kin, 50~kpc}$), thermal ($P_{\rm th, 50~kpc}$) and total ($P_{\rm kin+th, 50~kpc}$) outflow powers, all measured at a radius of 50 kpc, and also injection powers for SN feedback ($P_{\rm SN}$), AGN ($P_{\rm AGN}$), and their sum ($P_{\rm SN+AGN}$).
Here, we focus on runs with the fiducial variable coupling efficiency model, while the results from runs with the fiducial constant coupling efficiency are very similar.
When examining the gas outflow powers, thermal energy dominates the total outflow power for all BH masses ($\MBH = 0, 4\times10^6, 10^8$ M$_\odot$). In the SN-only run, the outflow powers show only a brief spike at $\sim$1.7 Gyr, with thermal power exceeding kinetic power by 1--2 orders of magnitude.  In runs with AGN feedback, both kinetic and thermal outflow powers are substantially larger than in the SN-only case, but the kinetic power still contributes only $\sim$20-30 per cent of the total outflow power. For the feedback injection powers, SN feedback dominates in low-mass BH runs, while for high mass BHs, AGN feedback dominates only at early times before SN feedback becomes dominant again. Interestingly, the AGN feedback power never becomes completely dominant, even for the massive BHs.

Comparing powers across different $\MBH$ in \Fig{powers}, the SN-only run shows a relatively stable injection power, while the BH runs exhibit an early increase in power followed by a later decline. Despite a $\sim0.6$ dex higher total injection power, the low-mass BH case produces $\sim0.4$ dex lower total outflow power than the high-mass BH case. The SN-only run also has higher injection power than the BH runs despite lacking AGN feedback. This is because AGN feedback more efficiently expels cold gas, suppressing the SFR and subsequently reducing both SN and AGN feedback, whereas SN feedback provides a more continuous regulation of gas cooling.

\Fig{powers_ratios} shows five different ratios of powers as functions of BH mass: (i) the ratio of kinetic to total outflow power, $P_{\rm kin, 50~kpc}/P_{\rm kin+th, 50~kpc}$; (ii) thermal to total outflow power, $P_{\rm th, 50~kpc}/P_{\rm kin+th, 50~kpc}$; (iii) AGN feedback injection power to total feedback injection power, $P_{\rm AGN}/P_{\rm SN+AGN}$; (iv) total gas outflow power to total feedback injection power, $P_{\rm kin+th, 50~kpc}/P_{\rm SN+AGN}$; and (v) total gas outflow power to AGN feedback injection power, $P_{\rm kin+th, 50~kpc}/P_{\rm AGN}$.

The SN-only run is not shown since at late times both kinetic and thermal power are zero. In runs with BHs up to $\MBH = 10^8$ M$_\odot$, $P_{\rm th, 50~kpc}/P_{\rm kin+th, 50~kpc}$ stays around 0.7–0.8, similar to  values for the Sedov-Taylor blast wave solution \citep{Sedov46,Taylor50}. For the fiducial constant coupling efficiency model, $P_{\rm kin+th, 50~kpc}/P_{\rm SN+AGN}$ and $P_{\rm kin+th, 50~kpc}/P_{\rm AGN}$ increase with $\MBH$, peaking around $\MBH \sim 10^{8}$ M$_\odot$, then dropping for $\MBH=10^9$ M$_\odot$. The fiducial variable coupling efficiency model shows similar drops for the $4\times10^6$ and $10^9$ M$_\odot$ cases. The ratio $P_{\rm kin+th,50~kpc}/P_{\rm AGN}$ can exceed unity for $\MBH=10^8$ M$_\odot$, likely because both SN and AGN feedback contribute to the outflows. Finally, the AGN fraction of the total injected feedback power increases with $\MBH$ in the constant coupling efficiency model.

\begin{figure*}
    \centering
    \includegraphics[width=0.8\textwidth]{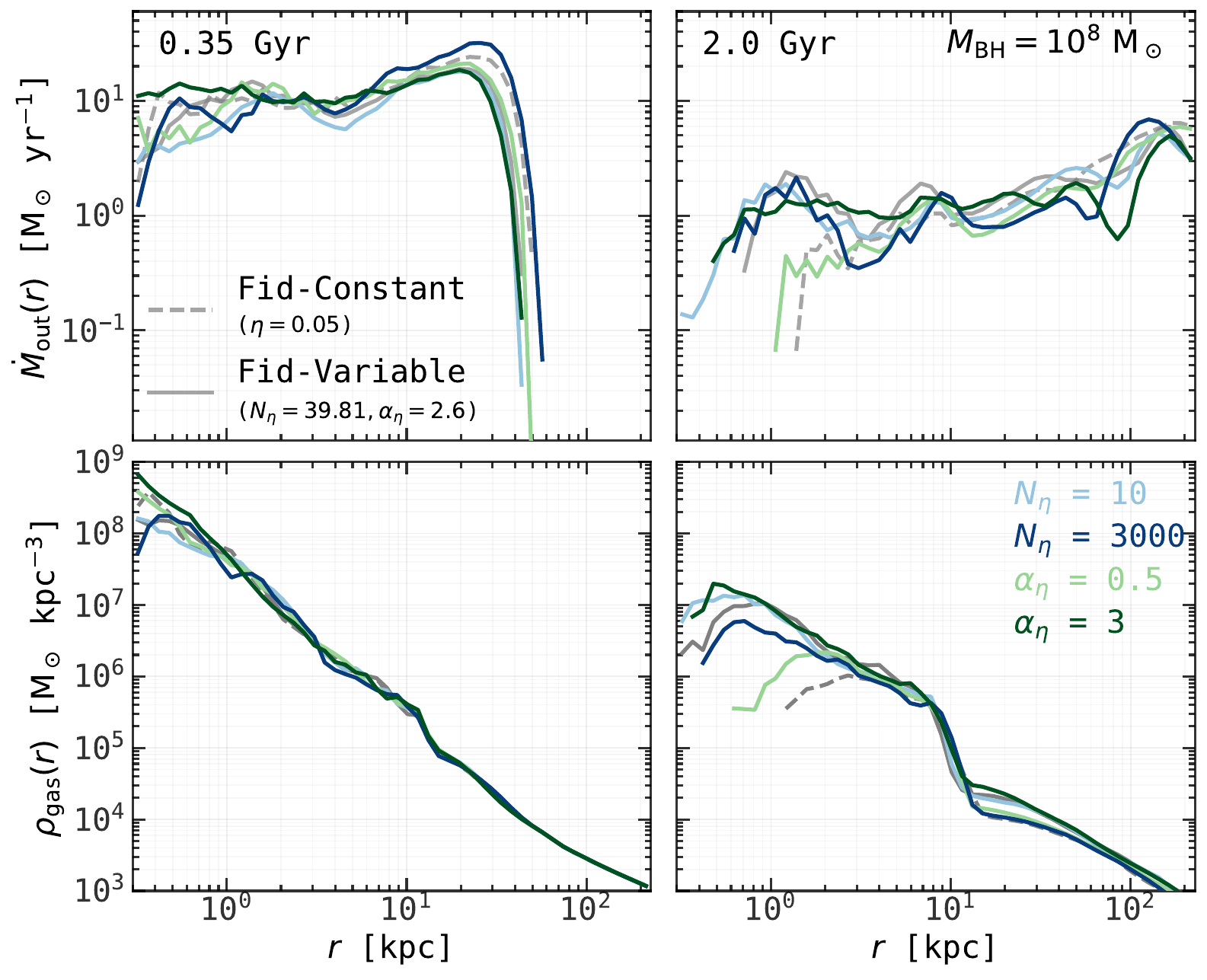}
    \caption{Mass outflow rate profiles (top row) and total gas density profiles (bottom row) at $t=$ 0.35 Gyr (left column) and $t=$ 2.0 Gyr (right column), for runs with $\MBH = 10^8 $ M$_\odot$ using four different parameter variations of the variable coupling efficiency model ($N_\eta = 10$, $3000$ and $\alpha_\eta = 0.5$, $3$). The results are averaged  over a 0.09~Gyr interval centred at each time. For comparison, the fiducial constant (gray dashed lines) and variable (gray solid lines) coupling efficiency models are also included. Linestyles and colours remain the same as in \Fig{AGNprop_vary}. 
    }
    \label{fig:outflow_prof_vary}
\end{figure*}


\subsection{Gas outflow and inflow at different radii}\label{sec:outflow_infall}

In this section, we further investigate the effects of SN and AGN feedback by examining mass outflow rate and density profiles using simulations with $\MBH=10^8$ M$_\odot$. We divide the radial range from 0.3 kpc to the virial radius $\Rvir$ (228 kpc) into 50 logarithmically spaced bins, ignoring bins with gas particle number smaller than 10. At each radius, we calculate the mass outflow rate using \Eq{Mout}, summing over all gas particles with positive radial velocities within a spherical shell of radius $r$ and thickness $\Delta r$.
The same binning strategy is applied to the gas density profiles. To better capture both outflow and inflow behaviours, as seen in a galactic fountain, we also compute the net mass outflow rates, which include all gas particles regardless of the sign of their radial velocities but still requiring their magnitude to be larger than the velocity threshold defined above:
\be
\dot{M}_{\rm net}=\sum_i^{|v_{r,i}|>V_{\rm th}}\frac{m_iv_{r,i}}{\Delta r}.
\ee
In addition, we plot the net mass outflow rates in the polar ($\theta<60^{\circ}$ and $\theta>120^{\circ}$) and equatorial ($85^{\circ}<\theta<95^{\circ}$) regions. This is motivated by what we saw in \Fig{maps}, that gas is mainly blown out in the polar directions. We note that there is an additional intermediate region ($60^{\circ}<\theta<85^{\circ}$ and $95^{\circ}<\theta<120^{\circ}$), mainly composed of gas in the galactic fountain, which is not discussed in detail here. However, it also contributes to the global net mass outflow rate.

\subsubsection{Effects of parameter variations in variable feedback efficiency model}

\begin{figure*}
    \centering
    \includegraphics[width=0.9\textwidth]{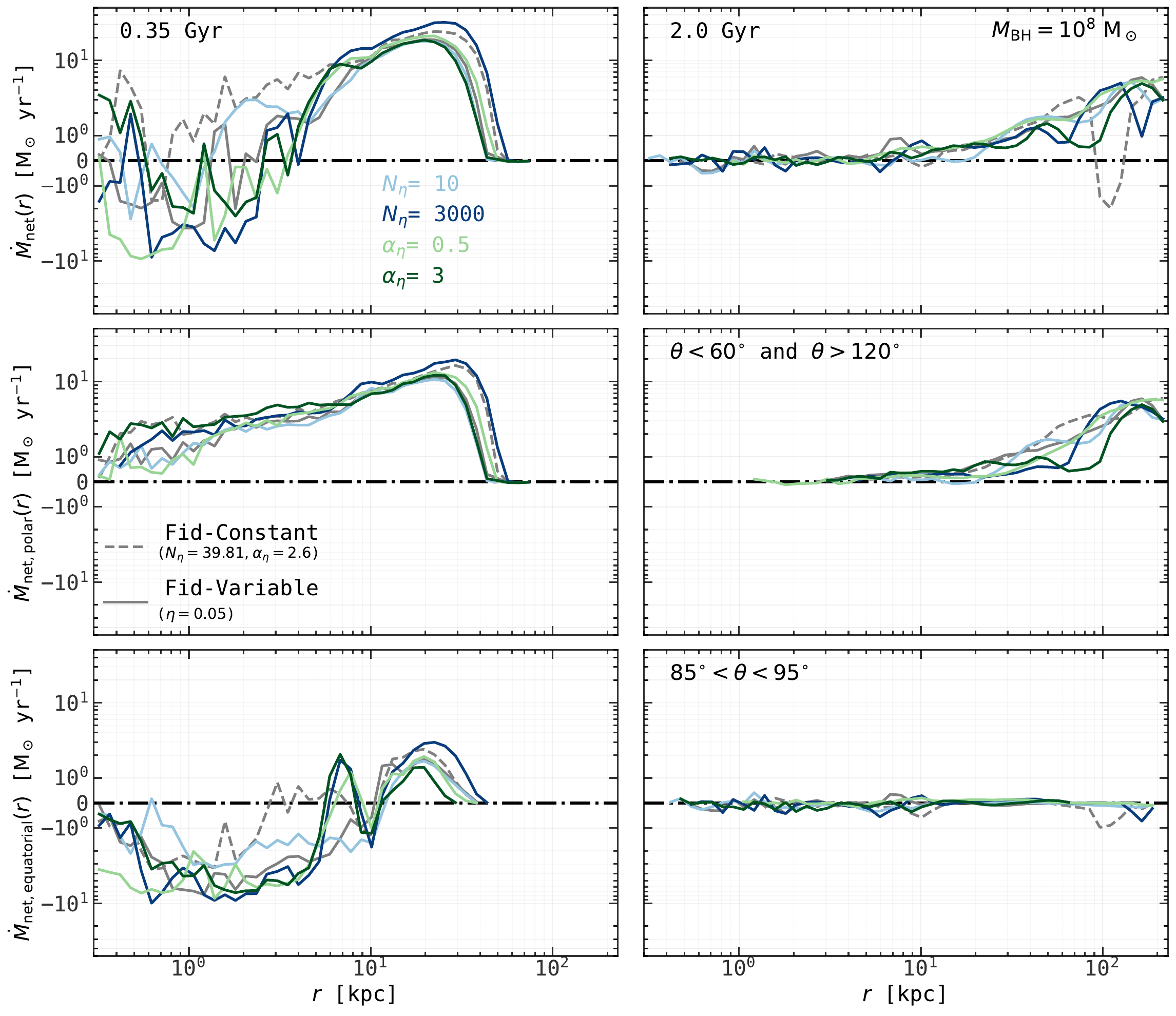}
    \caption{Radial profiles of net mass outflow rate at 0.35 Gyr (left column) and 2.0 Gyr (right column). 
    Net mass outflow rates are calculated for gas with radial velocity $|v_r| > 40~{\rm km~s^{-1}}$, and are averaged over a 0.09 Gyr interval centred at the times of interest. The top row shows outflow rates integrated over all angles, while the middle and bottom rows show outflow rates integrated over the polar ($\theta<60^{\circ}$ and $\theta>120^{\circ}$) and equatorial ($85^{\circ}<\theta<95^{\circ}$) regions respectively. 
    Runs with $\MBH = 10^8 $ M$_\odot$ using four different parameter variations in the variable coupling efficiency model ($N_\eta = 10$, $3000$ and $\alpha_\eta = 0.5$, $3$) are presented. For comparison, the fiducial constant and variable coupling efficiency models are also included. Linestyles and colours remain the same as in \Fig{AGNprop_vary}. 
    } 
    \label{fig:netoutflow_prof_vary}
\end{figure*}

We show in \Fig{outflow_prof_vary} the mass outflow rate profiles, $\dot{M}_{\rm out}$, and density profiles, $\rho_{\rm gas}$, at an early time (0.35 Gyr) and a late time (2.0 Gyr) for four different parameter variations (normalisations $N_\eta=10,3000$ and slopes $\alpha_\eta=0.5,3$), with the fiducial models shown as references. We find that a smaller $\alpha_\eta$ or a larger $N_\eta$ (which lead to higher coupling efficiency $\eta$)  generally produce higher outflow rates at CGM scales at early times, due to stronger AGN feedback driving gas into the CGM (see \Fig{AGNprop_vary}), although these differences are relatively small. At later times, all models show a roughly monotonic trend of the mass outflow rate increasing with radius. The decline in the centre can be explained by the second row of \Fig{outflow_prof_vary} where these runs show lower central densities at late times. The decline in the central outflow rate is thus explained by the depletion of gas in the central region, as strong AGN feedback expels the surrounding gas. 


\begin{figure*}
    \centering
    \includegraphics[width=0.8\textwidth]{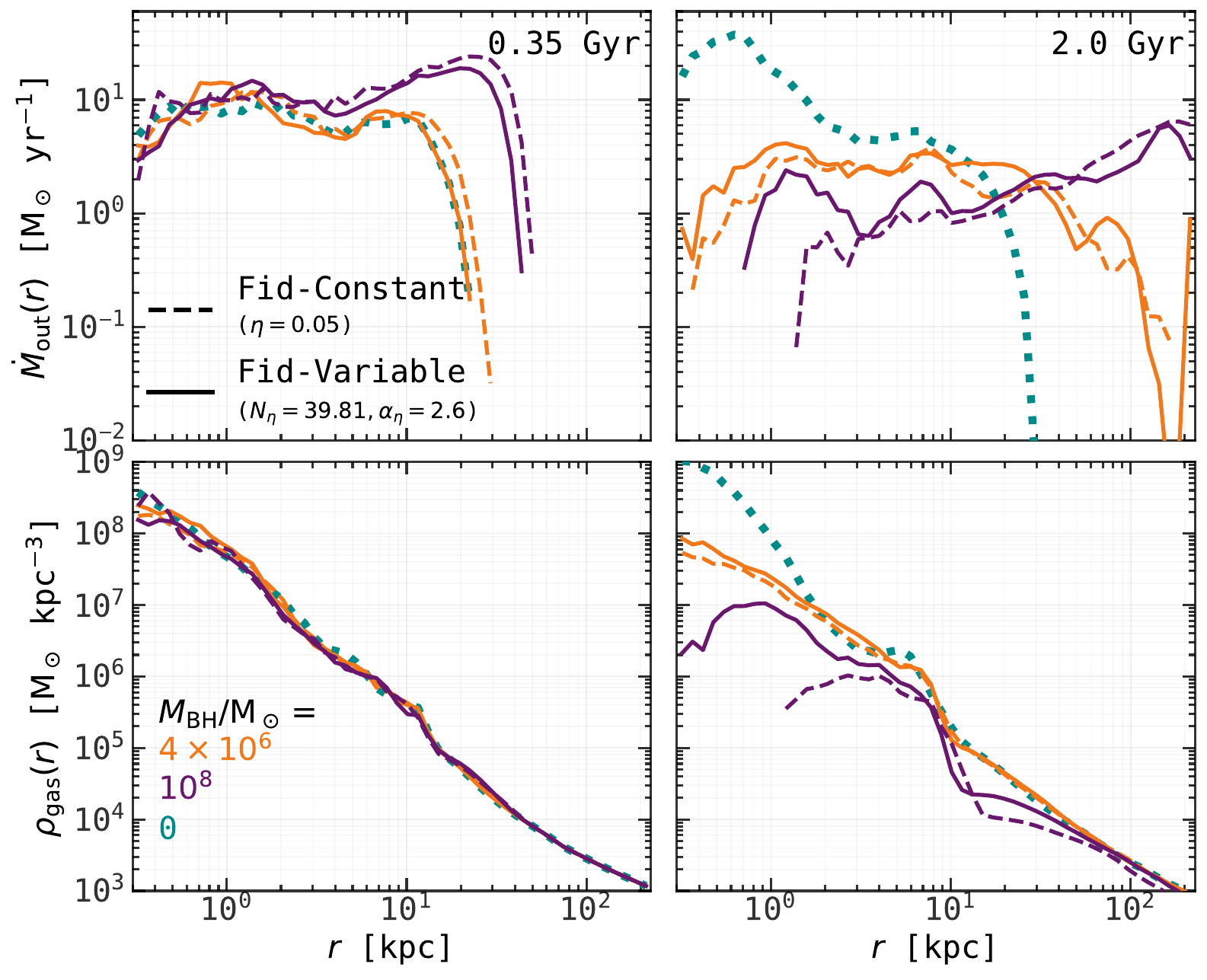}
    \caption{As \Fig{outflow_prof_vary} but showing the SN-only run and runs with $\MBH = 4 \times 10^6 $ M$_\odot$ and $10^8 $ M$_\odot$ using either the fiducial constant or fiducial variable coupling efficiency models. Linestyles and colours remain the same as in \Fig{AGNprop_mass1}. 
    }
    \label{fig:outflow_prof_mass}
\end{figure*}

In \Fig{netoutflow_prof_vary}, we show the net mass outflow rate profiles $\dot{M}_{\rm net}$ for the same parameter variations and at the same times as in \Fig{outflow_prof_vary}. At 0.35 Gyr, simulations with higher coupling efficiency $\eta$ (larger normalisation $N_\eta$ or smaller slope $\alpha_\eta$) exhibit stronger, outflow-dominated net rates at CGM scales in all directions, consistent with the fiducial models for the same BH mass. At smaller radii, increasing $N_\eta$ or decreasing $\alpha_\eta$ also slightly enhances the polar outflow rate, while the equatorial and total net outflow rates exhibit pronounced radial fluctuations, with the total rates frequently oscillating around zero. Such fluctuations are a signature of a galactic fountain effect, driven by AGN or SN feedback. This process expels hot gas bubbles in a polar direction, while cooler, denser gas clumps precipitate back towards the disc or galactic centre, producing outflows in the polar regions and inflows in the equatorial plane, with the dominance alternating from one shell to another. Furthermore, the small equatorial bump at $\sim10$ kpc coincides with the disc edge and is mainly a geometry effect, as the conical equatorial region begins to include outflowing fountain gas at large radii.

At late times (2.0 Gyr), the outflow has propagated to the virial radius, $\Rvir$, which remains untouched (zero net outflow rates) at early times. At large radii, the net outflow is dominated by the polar direction, while its lower magnitude at $r\lesssim40$ kpc reflects weaker AGN feedback. At small radii, the net outflow rate approaches zero in all directions, consistent with the central gas depletion seen in \Fig{outflow_prof_vary}.  Notably, different variations produce similar results at late times at late times, although stronger outflows at large radii are still produced by higher coupling efficiencies $\eta$ (larger normalisations $N_\eta$ or smaller slopes $\alpha_\eta$).

\subsubsection{Comparison of constant and variable efficiency models with different BH masses}

We now turn to a comparison of the two coupling efficiency models with different $\MBH$, including also the SN-only run. In \Fig{outflow_prof_mass}, we show the mass outflow rate profiles at $t=0.35$ and $t=2.0$ Gyr in the top row. At 0.35 Gyr, it is clear that the SN-only run produces similar mass outflow rates to the runs with $\MBH=4\times 10^6$ M$_\odot$ using both fiducial coupling efficiency models. Among the AGN runs, the $\MBH=10^8$ M$_\odot$ case shows larger outflow rates than the $\MBH=4\times 10^6$ M$_\odot$ case, indicating that stronger AGN feedback drives gas out to CGM scales earlier.

At 2.0 Gyr, runs with $\MBH=10^8$ M$_\odot$ show monotonically increasing radial trends of mass outflow rates extending to the virial radius, similar to the behaviour found in \Fig{outflow_prof_vary}. However, the mass outflow rate in the SN-only run and runs with $\MBH=4\times 10^6$ M$_\odot$ shows a monotonically decreasing radial trend. The outflow in the SN-only run stops at $\sim$ 25 kpc, confirming that SN feedback alone cannot transport large amounts of gas beyond $\sim 25$ kpc while the mass outflow rate in runs with $\MBH=4\times 10^6$ M$_\odot$ drops dramatically at $r\gtrsim 80-100$ kpc. 
In the inner region of the galaxy ($r \lesssim 3$ kpc), however, the trend with BH mass is reversed, where higher BH masses correspond to lower outflow rates. This suggests that strong AGN feedback expels central gas and suppresses further activity, whereas weaker feedback is more sustainable. This is also reflected in \Fig{outflow_prof_mass}: at 0.35 Gyr, all runs exhibit similar density profiles, but by late times larger BH masses yield significantly lower central gas densities.

Comparing the two fiducial coupling efficiency models, we find that the latter exhibits a slightly higher outflow rate at CGM scales shortly after the BH is turned on, but substantially lower outflow rates in the central region at late times. Again, this suggests that the constant coupling efficiency model produces stronger but less sustained feedback, in contrast to the variable coupling efficiency model, which enhances BH self-regulation.

\begin{figure*}
    \centering
    \includegraphics[width=0.9\textwidth]{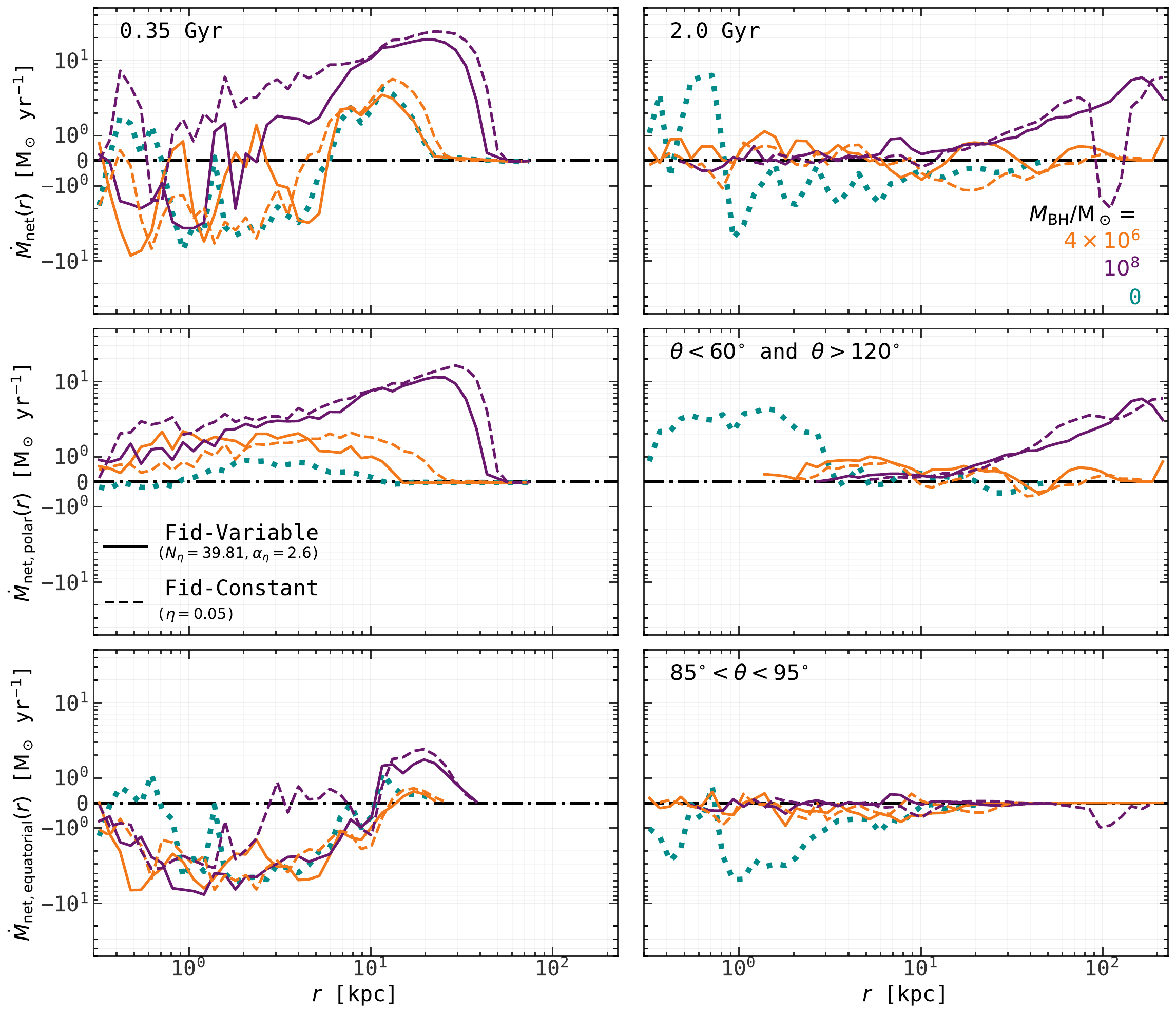}
    \caption{As \Fig{netoutflow_prof_vary}, but showing the SN-only run and runs with $\MBH = 4 \times 10^6 $ M$_\odot$ and $10^8 $ M$_\odot$ using either the fiducial constant or fiducial variable coupling efficiency models. Linestyles and colours remain the same as in \Fig{AGNprop_mass1}.
    }
    \label{fig:netoutflow_prof_mass}
\end{figure*}

In \Fig{netoutflow_prof_mass}, we compare the fiducial models with different $\MBH$ in terms of the net mass outflow rate $\dot{M}_{\rm net}$ and also the polar and equatorial rates. At 0.35~Gyr, the top left panel shows that the outflow dominates between $\sim 2$ (6)~kpc and $\sim 50$ (30)~kpc for the runs with $\MBH=10^8$ ($4\times 10^6$) M$_\odot$, where $\dot{M}_{\rm net}>0$. The lower panels reveal a clear trend in polar and equatorial components: more massive BHs drive stronger polar outflows at early times.
For comparison, outflows powered by SN feedback alone are much weaker and are confined to radii of $r\sim$ 3 kpc. This highlights that strong, large-scale polar outflows are primarily driven by AGN feedback.

At small radii ($r\lesssim$ 2 kpc), nearly all runs with AGN feedback show strong radial fluctuations in the net mass flow rate, consistent with the galactic fountain signatures discussed previously.  In all runs, inflow dominates the inner regions in the equatorial zone. At radii $r\gtrsim$ 10 kpc, all runs exhibit a bump in the equatorial net mass outflow rate. As explained earlier, this is a geometrical artefact of our how we make our measurement.

At late times, as shown in the right column of \Fig{netoutflow_prof_mass}, the SN-only run still exhibits polar outflows at small radii ($r\lesssim 4$ kpc), while its equatorial inflow persists across most radii, except in the outermost regions of the galaxy ($r\gtrsim 10$ kpc). In contrast, the net mass outflow rates in the AGN runs fall to near zero in the inner regions, indicating that AGN feedback is weaker and less frequent at this stage. Notably, the polar direction net outflow rates in AGN runs are lower than in the SN-only case at small radii. This suggests that the small net outflow rate is because of the depletion of central gas, which is also evident in \Fig{outflow_prof_mass}. At larger radii, all AGN runs display similar, positive values of $\dot{M}_{\rm net}$, indicating the outflowing gas propagating to larger radii.

When comparing different fiducial coupling efficiency models at early times, the fiducial variable coupling efficiency model shows weaker AGN feedback, with lower $\dot{M}_{\rm net}$ at CGM scales and more negative central values than the fiducial constant efficiency model. At later times, the latter model shows slightly higher net polar outflow rates than the former at large radii, given the same $\MBH$.

In summary, the figures presented in this section demonstrate that stronger AGN feedback, due to larger $\eta$ or higher $\MBH$, leads to higher outflow or net outflow rates at CGM scales at early times, primarily in the polar direction, with some contribution from the equatorial direction. At small radii, weaker feedback results in inflow-dominated flows, and galactic fountain effects are seen at these radii. By late times, both outflow and net outflow rates near the galactic centre decline to near zero due to central gas depletion, while outflows propagate to larger radii with reduced magnitude. 

\begin{figure*}
    \centering
    \includegraphics[width=1\textwidth]{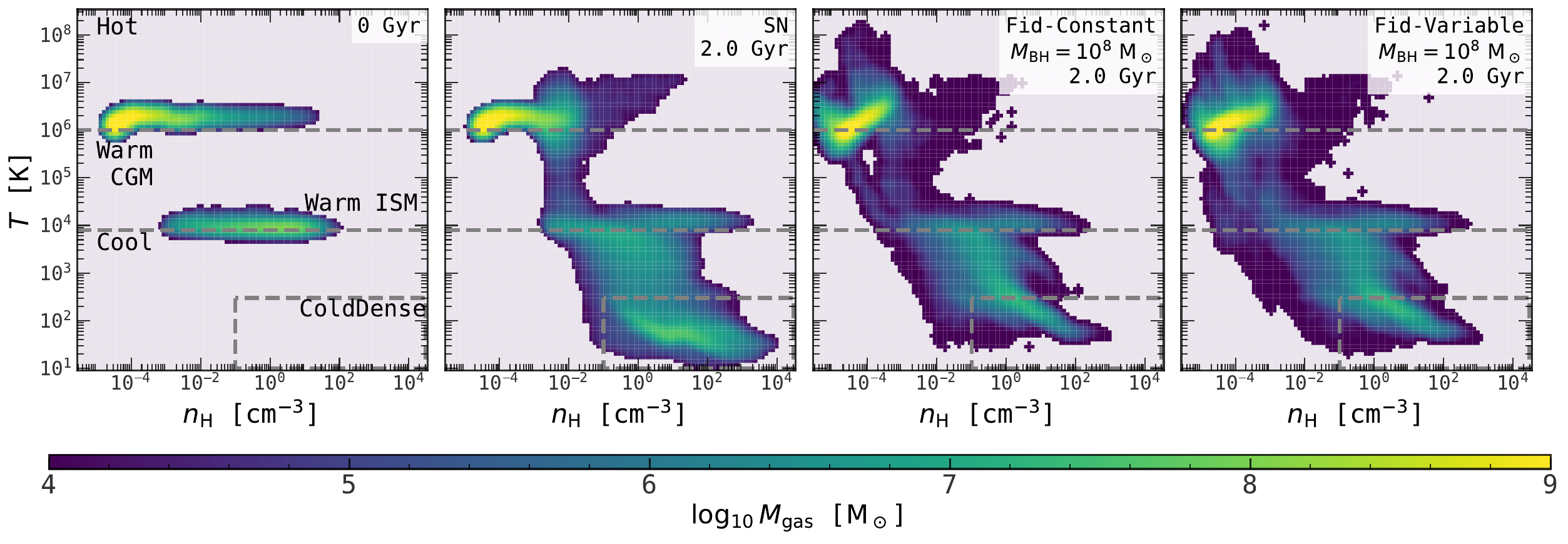}
    \caption{Phase diagram of the gas temperature against the hydrogen number density, colour-coded by the gas mass. The first panel shows results at the initial time, while the remainder show results at 2.0 Gyr, with data averaged over a 0.09 Gyr interval around each time. From left to right, the panels show: the initial conditions; the SN-only run; the $\MBH=10^8 $ M$_\odot$ case using either the fiducial constant or fiducial variable coupling efficiency models. The gray dashed lines divide the gas into hot ($T>10^6$ K), warm ($8\times 10^3~{\rm K}<T\leq10^6$ K), cool ($T\leq 8\times 10^3 $ K), and cold dense ($T\leq 300$ K and $n_{\rm H}\geq 10^{-1}$ cm$^{-3}$) gas. Within the warm region, the horizontal band around $10^4$ K corresponds to the warm ISM, while the remaining warm gas is classified as warm CGM.  This figure indicates that stronger AGN feedback leads to more warm gas and a greater amount of gas heated to higher temperatures and expelled into the CGM.}
    \label{fig:space_SN_4e6_1e8}
\end{figure*}

\subsection{Gas phase diagram}\label{sec:phase}

In \Fig{space_SN_4e6_1e8}, we compare the impact of AGN feedback on the gas phase diagram (i.e. the distribution of gas in density and temperature, including all gas particles in the simulation), between runs with and without a BH, and, for the runs with a BH, between different coupling efficiency models. Here, the hydrogen number density $n_{\rm H}$ is converted from mass density by assuming fixed hydrogen mass fraction $X_{\rm H}=0.756$. The phase diagram can be broadly divided into four regions: the cold and dense gas regime ($T \lesssim 3\times 10^2$ K and $n_{\rm H} \gtrsim 10^{-1}$ cm$^{-3}$), the cool gas regime ($T \lesssim 8\times 10^3$ K, excluding the cold and dense gas), the hot gas regime ($T \gtrsim 10^6$ K, mostly in the CGM but some in the disc), and the warm gas that bridges the hot and cold regimes. The warm gas can be further split into warm ISM and warm CGM. Due to the implementation of a multiphase interstellar medium, cold gas temperatures can reach as low as $T \simeq 10$ K. The interpretation of gas properties in different regimes is discussed in \citet{Schaye25}.

At 2.0 Gyr, we find that runs with AGN feedback result in a shorter tail in the cold dense region of the phase diagram, with gas reaching lower maximum densities, compared to the SN-only run. This indicates that AGN feedback can directly affect the ISM of a galaxy disc. The horizontal band around $T \simeq 10^4$ K also becomes broader in temperature and lower in density in the runs with AGN feedback. Above this floor, the AGN runs show more warm, diffuse gas that bridges the cool and hot phases. In the CGM regime, we observe that AGN feedback produces a steeper slope in temperature versus density compared to SN feedback, indicating more evolution away from the initial conditions.


\section{Discussion}\label{sec:discussion}
	
\subsection{Comparison with previous work}
In this section, we discuss the common features and differences between our results and previous works that also investigated AGN wind feedback in idealised disc galaxies using simulations \citep{Torrey20,Costa20,Farcy25}.

\citet{Torrey20} implemented a phenomenological AGN wind feedback model motivated by observations of BAL outflows, and studied its effects in simulations of isolated disc galaxies, with total mass similar to or slightly smaller than the MW. They used the \texttt{GIZMO} hydrodynamics code \citep{Hopkins15} with the fiducial FIRE-2 subgrid physics model \citep{FIRE}. Their ICs include stellar ($2.9-6.2 \times 10^{10} $ M$_\odot$) and gas discs (gas comprising 20\% of the disc mass), central SMBHs ($\MBH = 10^8 $ M$_\odot$ in the fiducial case), stellar bulges, live DM halos (collisionless particles instead of a static potential), but no CGM. Their gas mass resolution ($2$-$5\times 10^3 $ M$_\odot$) is significantly higher than ours. They ran their simulations for only 50 Myr with AGN feedback turned on, which is much shorter than our simulation time. BH accretion is fixed at the Eddington rate, and BAL winds are launched with mass flux $\dot{M}_{\rm BAL} = \eta_{\rm BAL} \dot{M}$. Feedback is modelled as isotropic kinetic energy injection with energy rate $\dot{E}_{\rm BAL} = \eta_{\rm BAL} \dot{M} v_{\rm BAL}^2 / 2$, where the BAL mass-loading factor, $\eta_{\rm BAL}$, and the assumed wind velocity, $v_{\rm BAL}$, are free parameters. In contrast, our AGN feedback energy injection rate is $\dot{E}_{\rm AGN} = \epsilon_{\rm r} \eta \dot{M}_{\rm acc} c^2$. Using their fiducial values ($\eta_{\rm BAL} = 1$, $v_{\rm BAL} = 0.1c$) for a case with $\MBH=10^8$ M$_\odot$, their energy injection rate is $5.7\times 10^{44}$ erg s$^{-1}$, which is at least $\sim 190$ times higher than our fiducial constant coupling efficiency model and $\sim 290$ times higher than our fiducial variable coupling efficiency model for $\MBH=10^8$ M$_\odot$ (see \Fig{AGNprop_mass2}). The BH accretion rate is likewise a much smaller fraction of the Eddington rate in our simulations (at most $\sim0.02$ for $\MBH=10^8$ M$_\odot$). 
Despite the differences in ICs and feedback energy, we find some notable similarities and differences with their results.

\cite{Torrey20} found that the central gas cavity grows larger for faster BAL winds or higher BH masses. The latter is consistent with our results in \Fig{outflow_prof_mass}. However, our cavity sizes are smaller due to our weaker feedback. In their simulations, AGN feedback triggers a rapid drop in SFR and quenches the galaxy within 50 Myr, whereas their SN-only run maintains a stable SFR. Our SN-only run shows a slight increase in SFR with time, but our SFR is slightly lower than theirs, and the SFR in our AGN runs takes longer to decline. Their higher peak SFR likely stems from the higher disc mass and gas fraction, while the rise in our SN-only run is potentially due to the cooling of the CGM (see \Fig{galaxyprop_noCGM}) which is absent in their simulations. Notably, unlike their results, our simulations show minimal long-term differences between the $\MBH = 10^8 $ M$_\odot$ and $10^9 $ M$_\odot$ cases. This likely arises because our BH accretion is self-regulated via the Bondi-Hoyle accretion rate, whereas theirs is fixed at the Eddington rate, preventing self-regulation by construction. Their phase diagrams further show that AGN feedback heats gas to $\sim 10^{10}$ K and drives outflows up to $10^4$ km s$^{-1}$, depleting nearly all of the cold gas,
highlighting the strength of their kinetic feedback model.

\citet{Costa20} implemented a kinetic AGN feedback model based on small-scale ultrafast winds using the moving-mesh hydrodynamics code \texttt{AREPO} \citep{AREPO}. SN feedback is not modelled explicitly, but cooling and star formation are included. Very differently from ours, their idealised simulations were initialised with a rotating spherical gas cloud following the DM halo density profile, without requiring dynamical equilibrium.  A part of the gas collapses into a central disc, while the remainder of the cloud acts as a CGM. The assumed initial spherical gas fraction is 0.17, which is significantly higher than our total gas fraction, 0.072. Their fiducial mass resolution ($1.6 \times 10^5 $ M$_\odot$) is comparable to ours. They ran their simulations for 0.5 Gyr, with the AGN turned on after 0.15 Gyr. Instead of setting an initial BH mass $\MBH$, they fixed the AGN bolometric luminosity $L_{\rm AGN}$ ($10^{45}$-$5\times10^{47}$ erg s$^{-1}$). Their feedback energy injection rate is defined as $\dot{E}_{\rm w} = \tau\beta L_{\rm AGN}/2$, where $\tau$ is the ratio of the wind momentum injection rate to $L_{\rm AGN}/c$, and $\beta$ is the ratio between wind velocity and the speed of light, both of which are free parameters. The value of $\tau$ is fixed to 1 and the fiducial value of $\beta$ is set as 0.017. Since their AGN feedback is not turned on for all the time, but instead has a duty cycle $\eta_{\rm duty}$, the time-averaged feedback energy injection rate can be calculated as $\eta_{\rm duty}\dot{E}_{\rm w}$. In their work, $\eta_{\rm duty}L_{\rm AGN}$ is set to $5\times 10^{44}$ erg s$^{-1}$ for different runs and the time-averaged feedback energy injection rate is therefore $\sim 4.3\times 10^{42}$ erg s$^{-1}$ using their fiducial $\beta=0.017$. Compared to what we show in \Fig{AGNprop_mass2}, this is at least $\sim 1.4$ times higher than our fiducial models at early times. Our simulations at later times show larger differences compared to theirs.

The simulations by \cite{Costa20} produced a $\sim 40$ times higher SFR peak at 0.1 Gyr, followed by a decline, whereas our models reach a lower peak at a later time. Differences in peak SFR and timing likely stem from differing ICs (e.g., no initial disc) or subgrid physics (e.g., different cooling and no SN feedback).  
However, the decline in their no-AGN runs suggests weaker halo gas cooling at later times, likely due to the lack of SN feedback. Their AGN runs show stronger SFR suppression after a few hundred Myr, indicating more powerful feedback than ours. Despite these differences, both studies agree that stronger AGN energy input leads to greater and faster SFR suppression, while weaker feedback produces results closer to the no-AGN case.

\begin{figure*}
    \centering
    \includegraphics[width=1\textwidth]{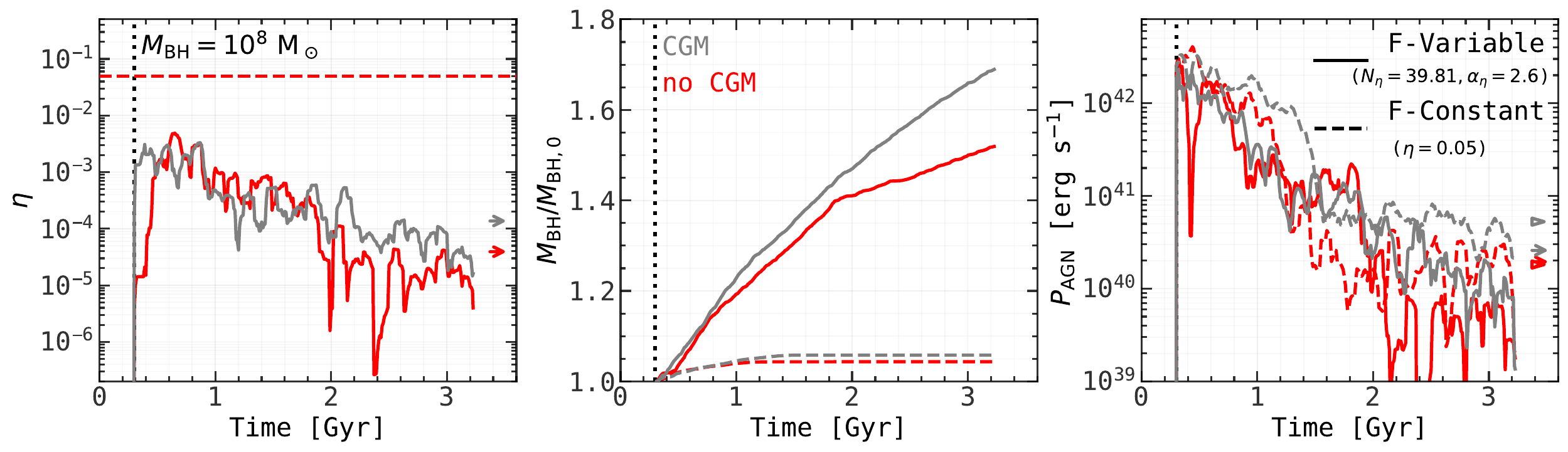}
    \caption{Coupling efficiencies, BH mass growth, and AGN feedback power as functions of time for $\MBH = 10^8 $ M$_\odot$, comparing simulations with and without a CGM. Solid (dashed) lines represent simulations using the fiducial variable (constant) coupling efficiency model. Simulations with a CGM are shown in grey, and those without a CGM are shown in red.  The lines plotted are averaged over 0.09 Gyr. Arrows or wedges on the right side of the panels indicate the mean values averaged over the last 1.5 Gyr.
    }
    \label{fig:AGNprop_noCGM}
\end{figure*}

 \citet{Farcy25} introduced a new AGN feedback scheme, \texttt{Mistral}, implemented in \texttt{AREPO} \citep{AREPO2,AREPO}. This model includes a continuous mode, where gas particles are radially kicked 
at every time-step, and a stochastic mode, where gas particles within the BH smoothing length are kicked along the galaxy spin axis at very high velocity. Unlike the previous two studies, they test their model using both idealised simulations and cosmological zoom-in simulations. Their idealised simulations had a setup similar to ours, including a live DM halo (different from our static DM halo), a stellar disc ($5.7\times 10^{10} $ M$_\odot$), a cold gas disc ($4.2\times 10^{9} $ M$_\odot$), a bulge, a CGM (modelled with a $\beta$ profile of scale radius 20.4~kpc, with $\beta=0.5$, and mass $1.5\times 10^{11}$ M$_\odot$), and a central SMBH ($\MBH=5\times10^7 $ M$_\odot$), which has a self-consistent BH accretion rate and AGN feedback power calculation. Their total disc and gas disc masses are similar to ours. Their baryonic mass resolution ($8\times10^4$ M$_\odot$) is comparable to ours. 
However, unlike our thermal AGN feedback, they applied kinetic feedback by injecting momentum into surrounding gas. The corresponding energy rate is given by $\dot{E}_{\rm BH} = \epsilon_w \dot{M}_{\rm acc}c^2 / (\psi + 1)$, with $\psi = 2\epsilon_w c^2 / v_w^2$. Using their fiducial parameters ($\epsilon_w = 5\times10^{-4}$, $v_w = 10^4$ km s$^{-1}$), the energy rate is $2.6\times10^{-3}/\eta$ times ours for the same accretion rate.

For AGN properties, their BHs grow by a factor of 1.2 (1.4) and release $7\times10^{57}$ ($2\times10^{58}$) erg in the stochastic (continuous) mode, while our models inject $\simeq 3-7\times10^{58}$ erg over the same period (\Fig{AGNprop_mass2}, $\MBH=10^8 $ M$_\odot$). Their initial SFR is slightly lower than ours, reflecting a smaller cold gas fraction. After $\sim$1.2 Gyr, their stochastic mode shows a steady SFR decline, whereas our coupling efficiency models reach even lower levels, indicating stronger suppression from higher AGN energy input. Their continuous mode maintains higher SFRs and even multiple starbursts. Despite these trends, the stellar mass growth is similar ($\sim$2 \% change in stellar mass by the final time compared to the initial value in their runs). The quenching timescale in their stochastic mode is $\sim$1 Gyr, longer than our simulations, while their continuous mode remains actively star-forming.

For gas mass outflow rates measured at 50 kpc, both our runs and their stochastic mode show an early peak followed by a decline, with our values slightly higher (\Fig{galaxyprop_mass2}). Their stochastic mode maintains a more stable kinetic energy outflow rate, whereas ours declines more rapidly after a higher initial peak. Both show episodic fluctuations, indicating a galactic fountain process. Their continuous mode instead shows a steady decline in both mass and energy outflow rates from the beginning.

Mass outflow profiles at 0.35 Gyr in both our and their simulations peak at CGM scales before declining (\Fig{outflow_prof_mass}). Our rates are higher, but their outflows extend farther ($\sim$75 kpc), suggesting more effective gas transport to large radii. Their isotropic thermal model injects more AGN energy but produces lower outflow rates and weaker SFR suppression than their stochastic kinetic model, highlighting the importance of the feedback scheme rather than energy input alone. The greater reach of their outflows may also result from their CGM initial density profile that drops steeply beyond $r\gtrsim20$ kpc.

In summary, 
\cite{Farcy25} found that kinetic feedback with high-velocity collimated injection in the polar direction is more effective at suppressing star formation and driving CGM-scale outflows than other schemes. However, our thermal feedback models, particularly those using variable coupling efficiencies, tend to sustain gas reservoirs and influence galaxy properties more gradually. Consistent with previous studies \citep[e.g.][]{Torrey20,Costa20}, stronger energy injection generally produces stronger impacts, but the outcomes are also sensitive to ICs, particularly the gas fraction and CGM structure, making direct comparisons between models difficult.

\subsection{The self-regulation of AGN feedback}
We emphasize that a key feature of our new model is stronger BH self-regulation, which is enhanced by a coupling efficiency that varies with the Eddington ratio. The tendency of AGN feedback to self-regulate in certain circumstances was already found in early cosmological simulations by \citet{Booth09,Booth10}. Similarly to our simulations, \cite{Booth09} assume that BHs accrete gas at a modified Bondi-Hoyle rate, and inject AGN feedback energy thermally, but with a coupling  efficiency that is constant, independent of the BH mass or accretion rate.
Varying the coupling efficiency $\eta$ over the range from $10^{-4}$ to 1, they found that the change in black hole mass scales as $\Delta\MBH\propto \eta^{-1}$, implying constant AGN energy injection \citep{Booth10}. 
They argue that the amount of AGN energy injection in the self-regulated state is controlled by scales that are much larger than those on which the gravity of the BH is important. Although the BH mass varies with $\eta$ due to self-regulation, they find that other global properties such as galaxy SFRs 
remain largely unchanged \citep{Booth09}.
In our idealized galaxy simulations with varying coupling efficiency, we also find similar AGN energy injection when $\eta$ is varied. For example, in \Fig{AGNprop_vary}, the final $E_{\rm AGN}$ varies by only a factor 2 even when the normalization of the variable feedback efficiency formulation varies by a factor 300. Additionally, the differences in the SFR and mass outflow rate among runs with different coupling efficiencies but the same BH mass are small (e.g. \Fig{Galaxyprop_vary}).
 In our work, we also find that BH mass growth is negatively correlated with the coupling efficiency $\eta$ (e.g. \Fig{AGNprop_vary}), but we do not expect a simple $\eta^{-1}$ scaling of BH mass in our case, since our $\eta$ is variable.


\begin{figure*}
    \centering
    \includegraphics[width=0.8\textwidth]{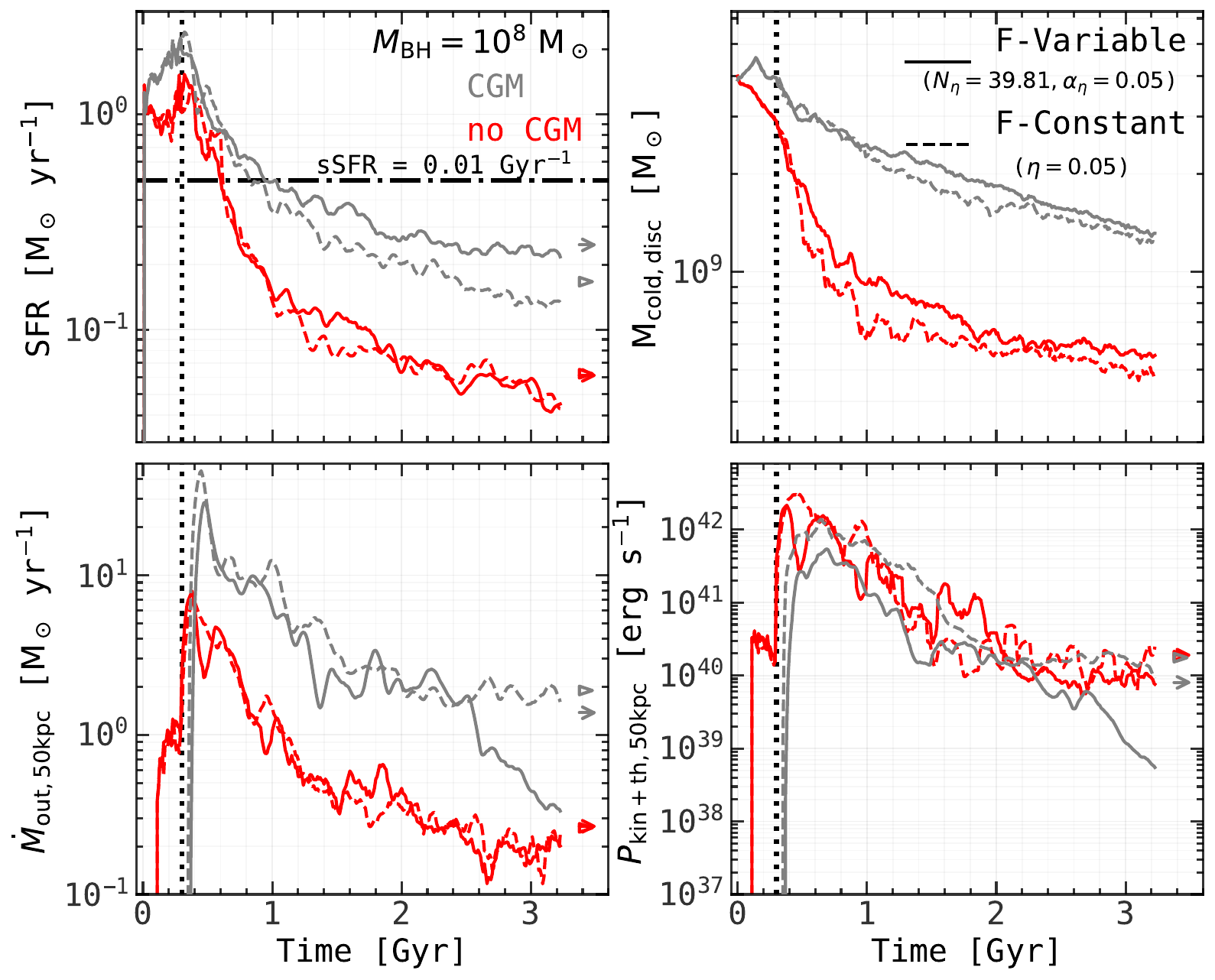}
    \caption{Similar to \Fig{AGNprop_noCGM}, but showing the galaxy SFR, cold gas mass in the disc, mass outflow rate and energy outflow rate at 50 kpc. 
    }
    \label{fig:galaxyprop_noCGM}
\end{figure*}

\subsection{The impact of the circumgalactic medium}
\label{sec:CGMfree}

To assess the impact of including the CGM on our results, we perform simulations with $\MBH = 10^8$ M$_\odot$ using either the fiducial variable coupling efficiency model or the fiducial constant coupling efficiency model, but without including the CGM, i.e. using only the ICs from \cite{Nobels24}. The resulting AGN and galaxy properties from these no-CGM runs are presented in \Fig{AGNprop_noCGM} and \Fig{galaxyprop_noCGM}, respectively.

As shown in \Fig{AGNprop_noCGM}, significant differences between the CGM and no-CGM runs begin to arise approximately 1.7 Gyr after we insert the BH. At this point, for the no-CGM runs, $\eta$ and $P_{\rm AGN}$ decline sharply, while $\MBH$ stops increasing, as the initial cold gas reservoir is depleted. In contrast, when the CGM is present, it continuously supplies gas to the disc gas reservoir through cooling and inflow, sustaining more prolonged BH accretion and feedback. 

In \Fig{galaxyprop_noCGM}, the cold gas disc mass declines rapidly in the no-CGM runs despite identical initial conditions, highlighting the role of CGM cooling in replenishing gas. Consequently, the SFR decreases and stellar mass growth stops after $\sim0.3$ Gyr (not shown here).  At 50 kpc, the no-CGM runs show much lower mass outflow rates, while energy outflow rates are initially higher but becomes comparable by the end. 
Notably, the differences in galaxy properties emerge earlier than the differences in AGN properties.

These tests indicate that including the CGM is important in idealised MW-mass disc galaxies in terms of sustaining SFR and AGN feedback, especially given that the MW is observed to have a CGM.

\subsection{Implications for high redshift supermassive black hole growth and JWST observations}

While we have enhanced BH self-regulation by implementing a coupling
efficiency that varies with the Eddington ratio as a power law, the
outcome in cosmological simulations, or in idealised elliptical or dwarf
galaxies, could differ from what we find here. Our idealised runs remain
in the low-$\dot{m}$ regime, where the variable efficiency model gives weaker
coupling efficiencies than the constant efficiency case, so that BH growth is somewhat
faster (\Fig{AGNprop_mass1}) while galaxy properties are left largely
unchanged. We caution, however, that this may not carry over to a full
cosmological context, and that the result may also depend on our
assumption of a constant heating temperature or on the system of interest.

In particular, the Eddington ratio is predicted to typically be larger at high redshift
\citep[e.g.][]{Husko25,Chaikin26, Pizzati26}, and for $\dot{m}\gtrsim0.08$ the fiducial variable
coupling efficiency exceeds the fiducial constant value of 0.05 (\Fig{eta_mdot}). In
this regime the variable
coupling efficiency model would instead strengthen feedback, which could slow
BH growth and require higher seed masses to match observed BH mass
functions. This is the opposite of what our low-$\dot{m}$ results might naively
suggest, unless the gas supply at high $z$ is high enough. Whether the
rapid growth of the overmassive SMBHs now revealed by JWST is helped or
hindered by this prescription therefore cannot be determined from our
simulations.

A viable model would need to sustain rapid growth at high redshift while
still converging to the observed $\MBH$--$M_*$ relation by $z=0$. This may
require the normalisation and/or slope of the variable coupling efficiency
to depend on redshift or on other BH or galaxy properties, or the desired
behaviour may emerge naturally from self-regulation in a cosmological
setting. Cosmological simulations using this model, calibrated against
local observations, are required to determine which scenario emerges, and
we defer this to future work.

\subsection{Caveats to this work}
In this section, several caveats to our work are discussed, connected to simplifications in the initial conditions for our idealised disc galaxies. The first oversimplification is that our ICs do not include a central bulge. A bulge tends to stabilise the disc against gravitational instabilities, thereby reducing angular momentum loss and making gas inflow less efficient \citep[e.g.][]{Lapiner23}. As a result, a smaller amount of gas would be fed toward the BH, leading to lower accretion rates and slower growth. This decreased fuelling results in weaker feedback. The bulge also deepens the central gravitational potential, and this increased central concentration may allow AGN feedback to couple more effectively with the surrounding medium, although the overall impact depends on the gas distribution and geometry \citep[e.g.][]{Martig09}. Finally, the suppression of star formation becomes more pronounced, particularly in the inner regions of the galaxy, due to either stronger local feedback or reduced gas inflow.

Another caveat is that we did not vary the properties of the CGM, e.g. its total mass, scale radius and central temperature. These properties directly influence BH growth and AGN feedback. The total CGM mass determines the available gas supply for the BH. A more massive CGM can supply more gas to the galaxy disc, potentially increasing the BH accretion rate and triggering stronger AGN feedback, while a lower-mass CGM reduces both BH fueling and star formation by limiting gas replenishment. The extreme case is the absence of a CGM, as discussed in \se{CGMfree}. The scale radius influences the density gradient of the CGM, particularly near the galaxy centre. A more compact CGM with a smaller scale radius produces higher central densities, facilitating faster BH growth, whereas a more extended CGM yields lower central densities and reduces the accretion rate. Finally, the central temperature sets the thermal pressure of the CGM. Lower central temperatures lead to lower central pressures but similar central densities, and therefore faster gas cooling onto the galaxy.

\section{Summary and Conclusions}\label{sec:conclusion}
In this work, we have presented a model relating the coupling efficiency of AGN thermal feedback (the fraction of AGN bolometric luminosity that is converted into AGN feedback energy)  to the Eddington-normalised BH accretion rate as a power law ($\eta=N_\eta\dot{m}^{\alpha_\eta}$), motivated by theoretical predictions for UV line-driven winds from AGN accretion discs \citep{Quera-Bofarull23}. We implemented this model into the hydrodynamical and gravity code \texttt{SWIFT} \citep{Schaller24} and tested it by running a series of idealised MW-mass galaxies with the COLIBRE subgrid physics model \citep{Schaye25}. We assume an analytical DM gravitational potential, and initialise a cold gas disc, a stellar disc and a hot CGM in dynamical equilibrium by extending the methods from \cite{Nobels22} (\Fig{CGM_init}). Our simulations include gas cooling, chemical enrichment, dust modelling, star formation, early stellar feedback, and AGN and SN feedback. Gas is assumed to accrete onto the BH at a rate given by a modified
version of the Bondi-Hoyle formula \Eqb{acc} modified for turbulence and vorticity. We vary both the black hole mass ($\MBH=10^6, 4\times10^6, 10^7, 10^8, 10^9$ M$_\odot$), and the normalisation and slope in the coupling efficiency (\Fig{eta_mdot}) to explore the influence of this model on AGN feedback and galaxy properties. Our main conclusions are:
\begin{enumerate}
    \item The differences in AGN and galaxy properties under variations of the normalisation or slope in the variable coupling efficiency model are relatively small, although some trends can be observed. In contrast, BH mass is the most important quantity in determining the strength of effects of AGN feedback for both the constant and variable coupling efficiencies. 
    \item Increasing the normalisation or decreasing the slope in the variable coupling efficiency model generally leads to higher coupling efficiencies, and consequently,  
    higher cumulative AGN energy injection  (\Fig{AGNprop_vary}). This in turn suppresses star formation and results in greater gas mass outflow rates at CGM scales. However, these differences are small 
    (\Fig{Galaxyprop_vary}). 
    The outflow rate profiles and gas density profiles are also similar (\Fig{outflow_prof_vary} and \Fig{netoutflow_prof_vary}).
    
    \item Higher initial BH mass leads to lower Eddington ratios and coupling efficiencies in our fiducial variable coupling efficiency model (\Fig{AGNprop_mass1}) but higher cumulative AGN energy input (\Fig{AGNprop_mass2}). This suppresses star formation, reduces the mass of the cold gas disc, and slows down stellar mass growth (\Fig{galaxyprop_mass1}). 
    Higher BH mass also increases outflow rates and gas densities at CGM scales and early times ($t=0.35$ Gyr), but reduces outflow rates and central gas densities at late times ($t=2.0$ Gyr, \Fig{galaxyprop_mass2}, \Fig{outflow_prof_mass} and \Fig{netoutflow_prof_mass}).
    
    \item Comparing the fiducial variable coupling efficiency model ($N_\eta=39.81$, $\alpha_\eta=2.6$) with the fiducial constant efficiency model ($\eta=0.05$), we find that, overall, the variable model produces less AGN energy (\Fig{AGNprop_mass2}), shows lower coupling efficiencies, but higher accretion rates and faster BH growth (\Fig{AGNprop_mass1}). Despite these differences, the star formation rate, cold gas disc mass, and stellar mass growth remain similar between the two models (\Fig{galaxyprop_mass1}). Compared to the constant efficiency model, the fiducial variable efficiency model produces lower mass and energy outflow rates at the CGM scale and early times ($t=0.35$ Gyr), but higher rates close the galactic centre and late times ($t=2.0$ Gyr, \Fig{galaxyprop_mass2}, \Fig{outflow_prof_mass} and \Fig{netoutflow_prof_mass}). 
    Additionally, at late times, the fiducial variable coupling efficiency model shows higher gas densities and smaller cavities at the galaxy centre for high-mass BHs (\Fig{outflow_prof_mass}). 

    \item The SN feedback power dominates the total feedback power in low-mass BH cases  ($M_{\rm BH} \lesssim
4\times 10^{6}~\mathrm{M_\odot}$), while AGN and SN feedback powers become comparable for high-mass BHs at early times (\Fig{powers}). The outflow power at a radius of 50 kpc is dominated by the thermal component in runs both with and without AGN feedback. For runs with a BH, the thermal outflow power is a roughly constant fraction $\sim0.7$–$0.8$ of the total outflow power for runs with BH masses up to $10^8$ M$_\odot$. 
    The ratio of total outflow power to total feedback injection power increases from very low values $\sim 0.03-0.07$ for $M_{\rm BH}=10^6$ M$_\odot$ up to a roughly constant value $\sim 0.5-0.6$ for $M_{\rm BH} \gtrsim 10^7$~M$_\odot$, indicating the importance of AGN feedback for driving outflows on CGM scales (\Fig{powers_ratios}).

    \item Variable coupling efficiency models with different parameters enhance self-regulation. This is evident from the smaller differences in cumulative AGN energy injection (\Fig{AGNprop_mass2}) compared to the fiducial constant coupling efficiency model, driven by a balance between coupling efficiency and accretion rate (or Eddington ratio and BH mass), and from higher central gas densities at late times ($t=2.0$ Gyr, \Fig{outflow_prof_mass}), indicating sustained feedback.

    \item AGN-driven outflows at the CGM scale arise in both the polar and equatorial directions (\Fig{netoutflow_prof_mass} and \Fig{netoutflow_prof_vary}). Near the galactic centre, outflows are mainly bipolar, while equatorial regions are dominated by inflows. Galactic fountain features are reflected in radial fluctuations of the net mass outflow rate at small radii.
    
    \item The BH mass accretion history and corresponding AGN feedback power, as well as the star formation history of the galaxy, are notably sensitive to the initial galaxy setup. The contributions from the CGM are important, especially for sustaining SFR and AGN feedback in long-term evolution (\Fig{AGNprop_noCGM} and \Fig{galaxyprop_noCGM}).
    
\end{enumerate}

Overall, the results of this work suggest that implementing a variable coupling efficiency model in AGN thermal feedback produces rather similar feedback compared to a constant coupling efficiency model, while achieving improved BH self-regulation. These conclusions are drawn from simulations of idealised Milky Way-mass galaxies using the COLIBRE model, and may be sensitive to the chosen initial conditions and subgrid physics. 
Further investigation of our coupling efficiency model in a fully cosmological context will be pursued in future work.

\section*{Acknowledgements}
We thank the COLIBRE team for their help in providing and supporting the use of the COLIBRE subgrid model. We acknowledge the use of \texttt{swiftsimio} \citep{Borrow20} in our
analysis. JL thank Fangzhou Jiang and Luis C. Ho for helpful discussions. JL acknowledges the support of Science and Technology Facilities Council (STFC) studentship (ST/Y509346/1). CGL acknowledges support from STFC consolidated grants ST/T000244/1 and ST/X001075/1.  EC acknowledges support from STFC consolidated grant ST/X001075/1. SB is supported by the UKRI Future Leaders Fellowship [grant numbers MR/V023381/1 and UKRI2044]. This project has received funding from the Netherlands Organization for Scientific Research (NWO) through research programme Athena 184.034.002. This work used the DiRAC@Durham facility managed by the Institute for Computational Cosmology on behalf of the STFC DiRAC HPC Facility (www.dirac.ac.uk). The equipment was funded by BEIS capital funding via STFC capital grants ST/K00042X/1, ST/P002293/1, ST/R002371/1 and ST/S002502/1, Durham University and STFC operations grant ST/R000832/1. DiRAC is part of the National e-Infrastructure.

\section*{Data Availability}

 The data supporting the plots within this article are available on reasonable request to the lead author. A public version of the SWIFT code \citep{Schaller24} is available at \href{www.swiftsim.com}{www.swiftsim.com}. The COLIBRE modules implemented in SWIFT will be made publicly available after the public release of the simulation data at \href{http://www.colibre-simulations.org/}{http://www.colibre-simulations.org/}.



\bibliographystyle{mnras}
\bibliography{example} 




\appendix

\section{AGN feedback and galaxy properties in variation models with lower black hole masses}\label{app:AGN_galaxy_prop}
\begin{figure}
    \centering
    \includegraphics[width=0.75\columnwidth]{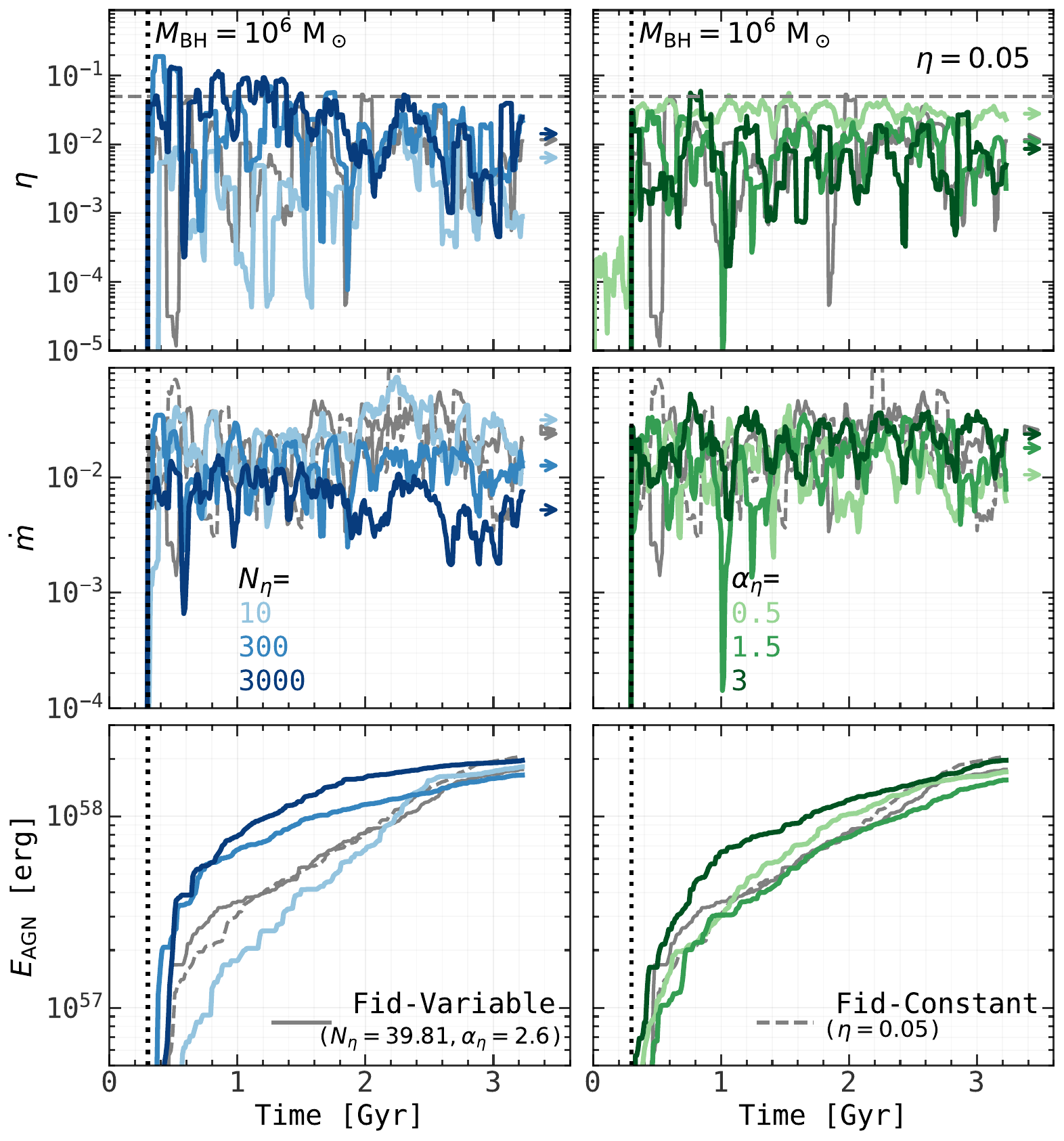}
    \caption{AGN feedback properties versus time for runs with $\MBH=10^6 $ M$_\odot$ using variable coupling efficiency model with varying normalisations (first column) and varying slopes (second column). Colours and line-styles remain the same as \Fig{AGNprop_vary}.}
    \label{fig:AGNprop_vary_1e6}
\end{figure}

\begin{figure}
    \centering
    \includegraphics[width=0.75\columnwidth]{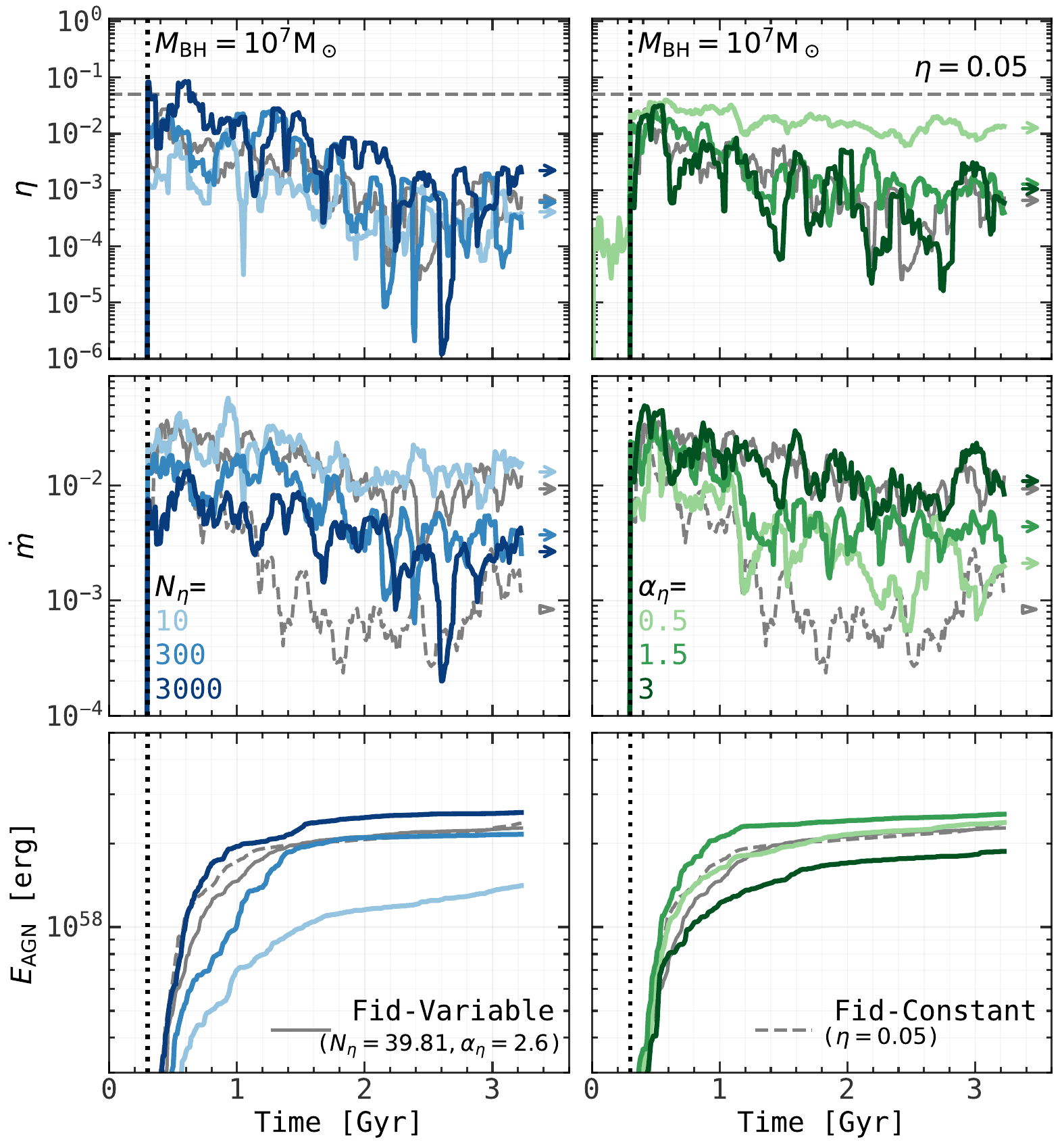}
    \caption{Similar to \Fig{AGNprop_vary_1e6} but showing results with $\MBH=10^7 $ M$_\odot$.}
    \label{fig:AGNprop_vary_1e7}
\end{figure}

\begin{figure}
    \centering
    \includegraphics[width=0.75\columnwidth]{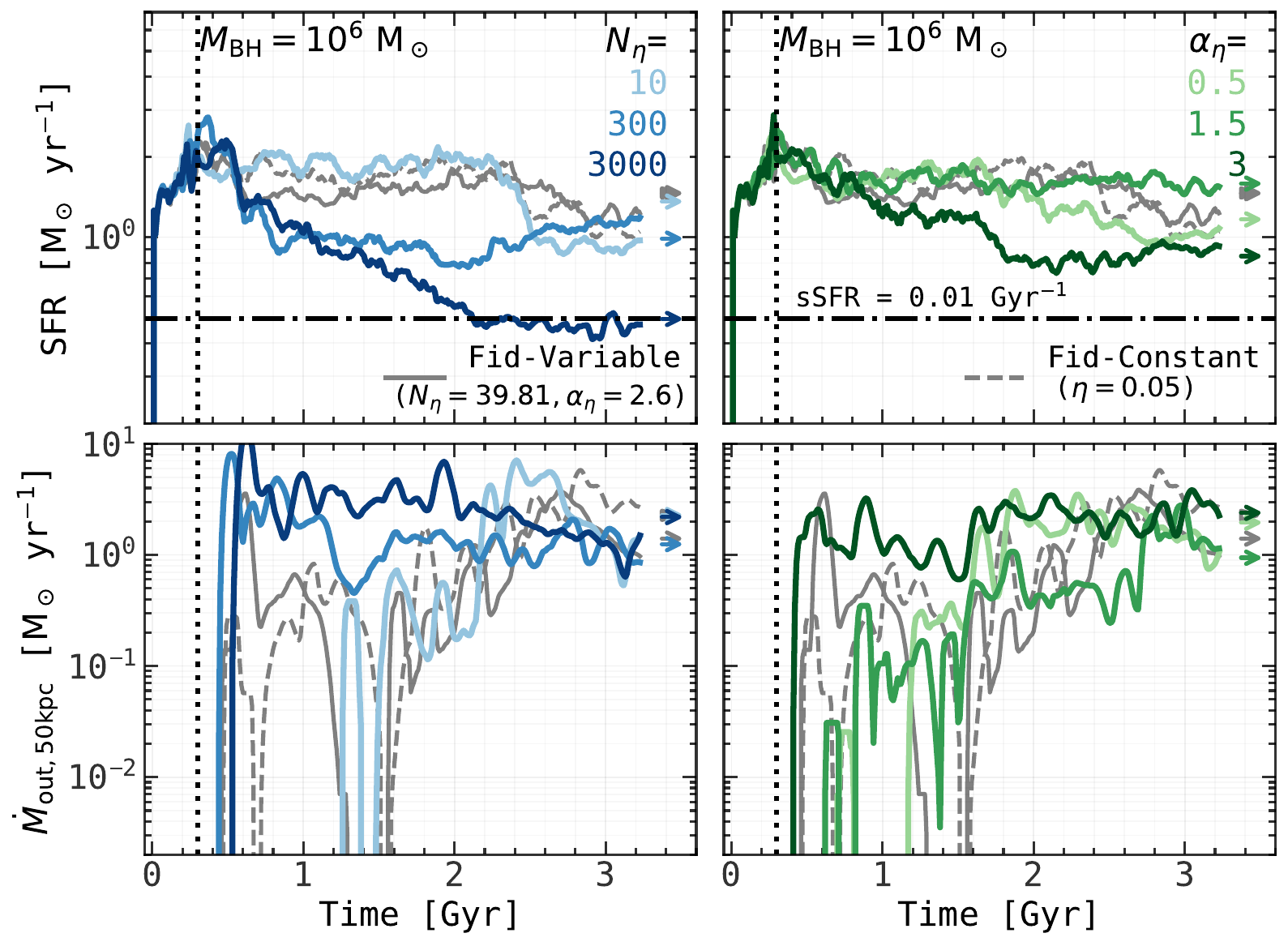}
    \caption{Galaxy properties versus time for runs with $\MBH=10^6 $ M$_\odot$ using variable coupling efficiency model with varying normalisations (first column) and varying slopes (second column). Colours and line-styles remain the same as \Fig{Galaxyprop_vary}.}
    \label{fig:galaxyprop_vary_1e6}
\end{figure}

\begin{figure}
    \centering
    \includegraphics[width=0.75\columnwidth]{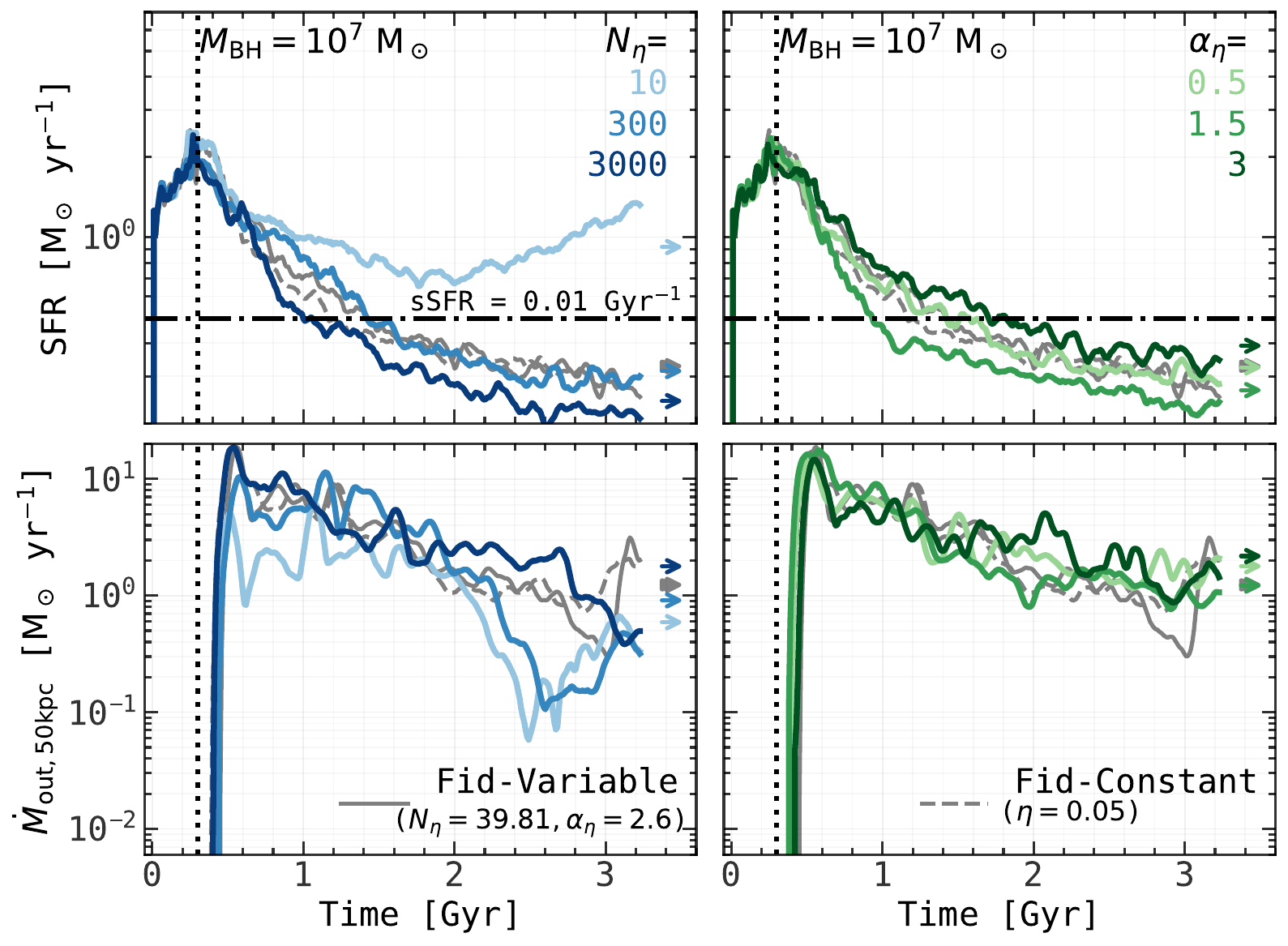}
    \caption{Similar to \Fig{galaxyprop_vary_1e6} but showing results with $\MBH=10^7 $ M$_\odot$.}
    \label{fig:galaxyprop_vary_1e7}
\end{figure}

In this Appendix, we show the AGN (\Fig{AGNprop_vary_1e6}-\Fig{AGNprop_vary_1e7}) and galaxy properties (\Fig{galaxyprop_vary_1e6}-\Fig{galaxyprop_vary_1e7}) at different times for lower BH masses ($\MBH=10^6,10^7 $ M$_\odot$) than the value presented in the main text ($\MBH=10^8$ M$_\odot$), using the variable coupling efficiency model with varying normalisations and varying slopes. These figures show that the trends with $N_\eta$, $\alpha_\eta$ and consequent $\eta$ that we find for high-mass BHs in the main body of the paper also largely hold for lower mass BHs. Specifically, both \Fig{AGNprop_vary_1e6} and \Fig{AGNprop_vary_1e7} demonstrate that a higher normalisation or a lower slope results in a higher coupling efficiency and smaller Eddington ratio. Regarding the AGN energy injection, \Fig{AGNprop_vary_1e6} (for $\MBH=10^6 $ M$_\odot$) does not show a clear trend with normalization or slope, although the differences are small. For $\MBH=10^7 $ M$_\odot$, \Fig{AGNprop_vary_1e7} indicates that a higher $N_\eta$ or a lower $\alpha_\eta$ leads to larger values. Notably, for these low-mass BH cases, the corresponding $\eta$ can exceed that of the fiducial constant model in some cases.

For galaxy properties, \Fig{galaxyprop_vary_1e7} exhibits a clear trend in which a higher $N_\eta$ suppresses the SFR and slightly increases the mass outflow rate measured at a radius of 50 kpc. However, the trend with $\alpha_\eta$ is less clear. The quenching timescales range from 0.7 (0.6) to 1.1 (1.4) Gyr for different $N_\eta$ ($\alpha_\eta$). Note that the case with $N_\eta=10$ is not quenched in 3.3 Gyr. Similarly, \Fig{galaxyprop_vary_1e6} shows a clear trend with $N_\eta$ but the trend with $\alpha_\eta$ is less clear. Additionally, some cases display large fluctuations in the mass outflow rates. All cases with $\MBH = 10^6 \mathrm{M}_\odot$ remain star-forming throughout the entire simulation except the one with $N_\eta=3000$, which has a quenching timescale of 1.8 Gyr.

The differences between cases with $\MBH=10^6$ M$_\odot$ and $\MBH=10^7$ M$_\odot$ shown in this appendix indicate that when AGN feedback is very weak due to a small BH mass, the effects of the variable coupling efficiency model on AGN and galaxy properties become less predictable, as other processes may dominate or interact more strongly with the AGN feedback.

\bsp	
\label{lastpage}
\end{document}